\documentclass[11pt,a4paper]{article}
\usepackage{siunitx}
\usepackage{amsmath}
\numberwithin{equation}{section}
\usepackage{amssymb}
\usepackage{dsfont}
\usepackage{graphicx}
\usepackage{xspace}
\usepackage[multiple]{footmisc}
\usepackage{color}
\usepackage{epsfig,amsthm}
\usepackage{soul}
\usepackage[makeroom]{cancel}
\usepackage{mathrsfs}
\usepackage[utf8]{inputenc}
\usepackage{slashed}
\usepackage{graphicx}
\usepackage{tikz}
\usetikzlibrary{calc, patterns}
\usepackage{pgfplots}
\pgfplotsset{compat=1.17} 
\usepackage{float}
\usepackage[percent]{overpic}
\usepackage{multirow}
\usepackage{physics}
\usepackage [autostyle, english = american]{csquotes}
\usepackage{bbm}
\usepackage{todonotes}
\usepackage{indentfirst}
\MakeOuterQuote{"}
\usepackage{listings}
\usepackage{ctable} 
\usepackage{longtable}
\usepackage[backend=biber, style=numeric-comp, sorting=none, sortcites=none]{biblatex}
\usepackage{hyperref}
\hypersetup{colorlinks=true,citecolor=blue,linkcolor=blue,urlcolor=blue}
\usepackage{fixfoot}
\usepackage{placeins}
\usepackage{enumitem}
\usepackage[usestackEOL]{stackengine}
\usepackage[normalem]{ulem}
\usepackage{tabularray}
\usepackage{makecell}
\usepackage{booktabs}

\makeatletter
\gdef\tshortstack{\@ifnextchar[\@tshortstack{\@tshortstack[c]}}
\gdef\@tshortstack[#1]{%
  \leavevmode
  \vtop\bgroup
    \baselineskip-\p@\lineskip 3\p@
    \let\mb@l\hss\let\mb@r\hss
    \expandafter\let\csname mb@#1\endcsname\relax
    \let\\\@stackcr
    \@ishortstack}
\makeatother

\usepackage{scrextend}
\addtokomafont{labelinglabel}{\sffamily\bfseries}

\usepackage{contour}
\usepackage{ulem}

\contourlength{0.8pt}
\newcommand{\myuline}[1]{\uline{\phantom{#1}}\llap{\contour{white}{#1}}}

\DeclareUnicodeCharacter{2212}{\ensuremath{-}}

\definecolor{codegreen}{rgb}{0,0.6,0}
\definecolor{codegray}{rgb}{0.5,0.5,0.5}
\definecolor{codepurple}{rgb}{0.58,0,0.82}
\definecolor{backcolour}{rgb}{0.95,0.95,0.92}

\newenvironment{psmallmatrix}
  {\left(\begin{smallmatrix}}
  {\end{smallmatrix}\right)}

\lstdefinestyle{mystyle}{
    backgroundcolor=\color{backcolour},   
    commentstyle=\color{codegreen},
    keywordstyle=\color{blue},
    numberstyle=\tiny\color{codegray},
    stringstyle=\color{codepurple},
    basicstyle=\ttfamily\footnotesize,
    breakatwhitespace=false,         
    breaklines=true,                 
    captionpos=b,                    
    keepspaces=true,                 
    numbers=left,                    
    numbersep=5pt,                  
    showspaces=false,                
    showstringspaces=false,
    showtabs=false,                  
    tabsize=2
}

\def\lsim{\mathrel{\rlap{\lower4pt\hbox{\hskip1pt$\sim$}}
    \raise1pt\hbox{$<$}}}                
\def\gsim{\mathrel{\rlap{\lower4pt\hbox{\hskip1pt$\sim$}}
    \raise1pt\hbox{$>$}}}                

\newcommand{\beq}{\begin{eqnarray}}
\newcommand{\eeq}{\end{eqnarray}}
\newcommand{\ba}{\begin{eqnarray}}
\newcommand{\ea}{\end{eqnarray}}
\newcommand{\be}{\begin{equation}}
\newcommand{\ee}{\end{equation}}
\newcommand{\bpmatrix}{\begin{pmatrix}}
\newcommand{\epmatrix}{\end{pmatrix}}

\newcommand{\h}[1]{\ensuremath{\hat{#1}}}

\newcommand{\download}{\url{https://github.com/phbasler/BSMPT}}

\newcommand{\comment}[1]{\ignorespaces}

\newcommand{\s}{\newline \vspace*{-3.5mm}}

\makeatletter
\def\blfootnote{\xdef\@thefnmark{}\@footnotetext}
\makeatother

\allowdisplaybreaks

\begin{document}

\title{
	\vspace*{-3cm}
	\phantom{h} \hfill\mbox{\small KA-TP-08-2024}
	\vspace*{0.7cm}
\\[-1.1cm]
	\vspace{15mm}   
\textbf{BSMPT v3\\A Tool for Phase Transitions and \\ Primordial
  Gravitational Waves \\ in Extended Higgs Sectors\\[4mm]}
} 
\date{}

\author{
    Philipp Basler\footnote{E-mail: \texttt{philipp.basler@alumni.kit.edu}}$\,\,\,^{\bigstar}$,
    Lisa Biermann$^{1\,}$\footnote{E-mail: \texttt{lisa.biermann@kit.edu}}$\,\,\,^{\bigstar}$,
    Margarete M\"{u}hlleitner$^{1\,}$\footnote{E-mail: \texttt{margarete.muehlleitner@kit.edu}}$\,\,\,^{\bigstar}$,
    Jonas M\"uller$\,^{\bigstar}$,\\
    Rui Santos$^{2,3\,}$\footnote{E-mail: \texttt{rasantos@fc.ul.pt}}$\,\,\,^{\bigstar}$,
    Jo\~ao Viana$^{2\,}$\footnote{E-mail: \texttt{jfvvchico@hotmail.com}}$\,\,\,^{\bigstar}$
\\[9mm]
{\small\it
$^1$Institute for Theoretical Physics, Karlsruhe Institute of Technology,} \\
{\small\it Wolfgang-Gaede-Str. 1, 76131 Karlsruhe, Germany}\\[3mm]
{\small\it
$^2$Centro de F\'{\i}sica Te\'{o}rica e Computacional,
    Faculdade de Ci\^{e}ncias,} \\
{\small \it    Universidade de Lisboa, Campo Grande, Edif\'{\i}cio C8
  1749-016 Lisboa, Portugal} \\[3mm]
{\small\it
$^3$ISEL -
 Instituto Superior de Engenharia de Lisboa,} \\
{\small \it   Instituto Polit\'ecnico de Lisboa
 1959-007 Lisboa, Portugal} \\[3mm]
}

\maketitle

\vspace*{-1.0cm}

\begin{abstract}
\blfootnote{${}^{\bigstar}$All authors can be reached via: \texttt{bsmpt@lists.kit.edu}}
  Strong first-order phase transitions (SFOPT) during the evolution of
  the Higgs potential in the early universe not only allow for the
  dynamical generation of the observed matter-antimatter asymmetry,
  they can also source a stochastic gravitational wave (GW) background
  possibly detectable with future space-based gravitational waves
  interferometers. As SFOPTs are phenomenologically incompatible with
  the Standard Model (SM) Higgs sector, the observation of GWs from
  SFOPTs provides an exciting interplay between cosmology and particle
  physics in the search for new physics. With the {\tt C++} code {\tt
    BSMPTv3}, we present for the first time a tool that performs the
  whole chain from the particle physics model to the gravitational
  wave spectrum. Extending the previous versions {\tt BSMPTv1} and
  {\tt v2}, it traces the phases of beyond-SM (BSM) Higgs potentials
  and is capable of treating multiple vacuum directions and multi-step
  phase transitions. During the tracing, it checks for discrete
  symmetries,  flat directions, and electroweak symmetry restoration,
  and finally reports the transition history. The transition
  probability from the false to the true vacuum is obtained from the
  solution of the bounce equation which allows for the calculation
    of the nucleation, percolation and completion temperatures. The
    amplitude and characteristic 
    frequencies of the GWs originating from bubble collisions and
    highly relativistic fluid shells, sound waves and turbulence, 
  are evaluated after the calculation of the thermal parameters at the
  transition temperature, and finally the signal-to-noise ratio at
  {\tt LISA} is provided. The code {\tt BSMPTv3} is a powerful
  self-contained tool that comes more than timely and will be of
  great benefit for investigations of the vacuum structure of the
  early universe of not only simple but also complicated Higgs
  potentials involving several vacuum directions, with exciting
  applications in the search for new physics.  
  
\end{abstract}



{\bf PROGRAM SUMMARY/NEW VERSION PROGRAM SUMMARY}

  \begin{small}
    \noindent
    {\em Program Title:} {\tt BSMPT} - Beyond the Standard Model Phase
    Transitions: A {\tt C++} package for the computation of phase transitions 
    and of gravitational waves sourced by first-order phase transitions (FOPT) in 
    beyond-Standard-Model (BSM) theories with extended Higgs sectors. \\
    {\em CPC Library link to program files:} (to be added by Technical Editor) \\
    {\em Developer's repository link:} \url{https://github.com/phbasler/BSMPT} \\
    {\em Code Ocean capsule:} (to be added by Technical Editor)\\
    {\em Licensing provisions(please choose one):} GPLv3  \\
    {\em Programming language:} C++17                             \\
    {\em Journal reference of previous version:} \href{https://www.sciencedirect.com/science/article/abs/pii/S0010465521002368}{10.17632/sjtp7bb33t.1}                  \\
    {\em Does the new version supersede the previous version?:} Yes   \\
    {\em Reasons for the new version:} Add algorithms for multi-step phase tracking, 
    the computation of the bounce solution, of thermal parameters 
    and of the spectrum of gravitational waves sourced by FOPTs.\\
    {\em Summary of revisions:} \texttt{BSMPTv3} can track multiple
    non-global and global minima of the effective potential as a function
    of the temperature, solve the bounce equation to calculate the
    tunnelling rate and compute the spectrum of the primordial
    gravitational waves sourced by FOPT as well as their
    signal-to-noise ratio at {\tt LISA}, a future space-based gravitational waves
    observatory.\\
    {\em Nature of problem:} Scalar extensions are promising models
    capable of providing a strong first-order electroweak phase
    transition, one (i.e.~departure from thermal equilibrium) of
    the three Sakharov conditions for successful
    electroweak baryogenesis. Furthermore, they can provide addtional sources of CP
    violation, which is another Sakharov condition. Such FOPT can source
    gravitational waves that are possibly detectable at future GW
    observatories. In order to decide if and which phase transitions take
    place during the evolution of the universe, the effective potential of
    the model under consideration has to be computed at higher orders
    including the thermal mass corrections. Its non-global and global
    minima then have to be traced as function of the temperature. For
    overlapping minima, the tunnelling rate from the false to the true
    minimum needs to be calculated. For this the bounce equation has to be
    solved. For the phase transitions that take place, the thermal
    parameters are then determined. They are required for the computation
    of the generated GW spectrum. Finally, the signal-to-noise ratio is
    computed for the future GW observatory {\tt LISA}.\\
    {\em Solution method:} The minima are tracked across the 
    temperature range by using a Newton-Raphson method with 
    seed points determined by the \texttt{BSMPTv2} routines.
    For the compuation of the tunnelling rate,
    the bounce equation is solved by the path deformation
    algorithm. This consists of splitting the bounce equation into a
    tangential and a perpendicular component, and then solve it
    iteratively until convergence is reached. With the tunnelling
    rate as a function of the temperature, the nucleation, percolation
    and completion temperatures are calculated as well as the other
    thermal parameters (released latent heat, inverse time scale) required for the computation of the GW spectrum. 
    The computations of the spectrum and its respective
    signal-to-noise ratio are performed using results from numerical
    simulations and fits available online.\\ 
    {\em Additional comments including restrictions and unusual features:}\\
    The wall velocity is considered an input parameter. It can be set
    to a specific constant value or alternatively chosen between two
    different approximations. If none of these is chosen, the wall velocity
    is set by default to 0.95. Models with a spontaneously broken discrete
    symmetry can lead to domain walls. The code does not take into account
    the possible existence of different domains separated by domain
    walls. In this case, only phase transitions with the shortest path
    between false and true vacuum are considered. The code assumes
    only one transition taking place between one pair of false and true
    vacuum phases. The code calculates GWs originating from bubble collisions and highly relativistic fluid shells, sound/shock
    waves and magneto-hydrodynamic turbulence. \\
    \end{small}

\thispagestyle{empty}
\vfill
\newpage

\tableofcontents

\section{Introduction}
\label{sec:intro}

Despite the success of the Standard Model (SM), which has been
structurally completed with the discovery of the Higgs boson
\cite{ATLAS:2012yve,CMS:2012qbp} and 
tested to great accuracy, there are open questions which cannot be
answered by the SM. The investigation of the development of the vacuum
structure during the evolution of the universe allows us to get
exciting insights, which may help to find answers to some 
long-standing open problems. Among these is the question of the
generation of the observed baryon asymmetry \cite{WMAP:2012fli} of the universe. A
dynamical explanation is given by electroweak baryogenesis (EWBG) \cite{Kuzmin:1985mm,Cohen:1990it,Cohen:1993nk,Quiros:1994dr,Rubakov:1996vz,Funakubo:1996dw,Trodden:1998ym,Bernreuther:2002uj,Morrissey:2012db}, provided
the three Sakharov conditions \cite{Sakharov:1967dj} are fulfilled. While these could in
principle be met by the SM, for the measured value of the SM-like
Higgs mass of \SI{125}{GeV}, there is a smooth cross-over
\cite{Kajantie:1996mn,Csikor:1998eu}. 
Since EWBG requires a strong first-order electroweak phase transition
(SFOEWPT), compatibility with Higgs phenomenology hence leads to the
investigation of beyond-the-SM (BSM) Higgs sectors.  The requirement
of an SFOEWPT restricts the allowed parameter space of BSM models entailing observable
consequences at collider experiments, like the Large Hadron Collider
(LHC), and provides us with an indirect probe of new physics
scenarios. \s

The first direct observation of gravitational waves (GWs) reported by
the LIGO Collaboration in 2016 \cite{LIGOScientific:2016aoc}, was awarded
the Nobel Prize for physics in 2017 and initiated a new era of multi-messenger
astronomy. Future GW observatories with increased sensitivity provide
new exciting avenues with unprecedented opportunities for the
exploration of particle physics. Thus, first-order phase transitions (FOPTs) can source a stochastic
gravitational wave background that can be detectable with future
space-based gravitational waves interferometers. This would not only provide us
with the exciting possibility to directly probe the echo of a
cosmological FOPT, but it would also amount to the discovery of
physics beyond the SM. \s

The combination of collider phenomenology and cosmological
observations is hence an exceptional opportunity to get insights
into the true physics that underlies nature, spanning a large range of
energy scales. For this to be meaningful, we have to consistently
combine information from collider observables and gravitational wave
observation. Specific new physics models, that fulfil all relevant
theoretical and experimental constraints, are tested w.r.t.~to their
ability to induce an FOPT, and if the 
gravitational wave spectrum that is generated during this
transition may be detectable at future GW observatories. This program
has to be performed at the highest possible 
precision taking into account all available state-of-the-art
information. There exist several codes that are publicly available,
which trace the minima of extended Higgs sector potentials at
non-zero temperature some of which also 
determine the bounce action. None of these codes, however,
performs the whole chain from testing an arbitrary extended
  Higgs sector model w.r.t.~to its
constraints via tracing the minima of its scalar potential at non-zero temperatures,
the determination of the bounce solution and the possible phase transitions, to the
computation of the gravitational wave spectrum originating from these 
first-order phase transitions, in a self-contained
  way. Moreover, some codes become very
slow and even fail when it comes to the determination of the various
vacuum phases of involved potentials with multiple field
directions. Last but not least, the widely used code {\tt
CosmoTransitions} \cite{Wainwright:2011kj} is not publicly maintained any
more. \s

It is hence timely and imperative to develop a new
self-contained code being able to
perform the whole chain of calculations starting from a particle
physics model considering all relevant constraints to the spectrum of
gravitational waves from FOPTs, 
implementing the state-of-the-art approaches at highest available
precision, which are updated 
constantly when new developments appear. This is the aim of the
\texttt{C++} code {\tt BSMPT} and the here presented extension {\tt BSMPTv3}. 
With {\tt BSMPTv1/v2}
\cite{Basler:2018cwe,Basler:2020nrq} and the link to {\tt  
  ScannerS} \cite{Coimbra:2013qq,Ferreira:2014dya,Costa:2015llh,Muhlleitner:2016mzt,Muhlleitner:2020wwk} the
check of new physics models w.r.t.~to their potential 
of generating an FOPT while simultaneously being compatible with all
relevant theoretical and experimental constraints is possible. In this
paper, we present the new release \texttt{BSMPTv3}, a \texttt{C++} code that extends
the previous versions substantially by the ability to  
\begin{itemize}
    \item track temperature-dependent coexisting minimum phases over
      arbitrary temperature intervals; 
    \item trace multi-step phase transitions;
    \item deal with discrete symmetries;
    \item deal with flat directions;
    \item test electroweak symmetry restoration at high temperature; 
    \item calculate the bounce solution for regions of coexisting minima;
    \item determine the critical temperature, the nucleation
        temperature, the percolation temperature, the
        completion temperature, and the reheating temperature;
    \item calculate the released latent heat $\alpha$ and the inverse
      time scale $\beta/H$ of the phase transition;
  \item derive the amplitude and characteristic frequencies of gravitational waves,
        with bubble collisions and highly relativistic fluid shells, sound waves
        and turbulence as their possible origins;
    \item calculate the signal-to-noise ratio at {\tt LISA};
\end{itemize}
for all models implemented in {\tt BSMPT} and those included by the
user. \s

The outline of the paper is as follows. We start with a description of
the state-of-the-art and the new features of {\tt BSMPTv3} in
Sec.~\ref{sec:stateoftheart}. In Sec.~\ref{sec:codedescription}, we
describe in great detail the program. After giving details on download and installation in
Sec.~\ref{sec:installation}, we describe the structure and new algorithms
of the code, before moving on to the presentation of the newly implemented classes {\tt
  MinimumTracer}, {\tt BounceSolution}, {\tt GravitationalWave}, and
{\tt TransitionTracer} in
Secs.~\ref{subsec:mintracer}-\ref{subsec:transtracer}. Subsequently,
we describe the new executables {\tt MinimaTracer}, {\tt CalcTemps},
{\tt CalcGW} and {\tt PotPlotter} in
Secs.~\ref{sec:MinimaTracer}-\ref{sec:PotPlotter}, as well as new
stand-alone features in Sec.~\ref{sec:standalone}. 
We finish the program section in Sec.~\ref{sec:status_codes} with
the summary on the status codes that are given out. Section
\ref{sec:examples} contains example runs, the discussion of 
their results as
well as comparisons with the existing code {\tt
  CosmoTransitions}. Our conclusions are given in Sec.~\ref{sec:conclusions}. The appendix details the improvement on 
the bosonic thermal function used in {\tt BSMPT} that we implemented
together with this new version of the code.

\section{State-Of-The-Art and New Features \label{sec:stateoftheart}}
The vacuum structure of the universe is theoretically described by the
effective Higgs potential at non-zero temperature. As the universe
cools down and expands, its vacuum structure changes. At a certain temperature a new 
minimum evolves which may eventually become degenerate with the 
existing one, and finally become the global 
minimum. The degenerate situation defines the critical 
temperature $T_c$ and the corresponding critical vacuum expectation
value (VEV) $v_c$, but it
does not guarantee that actually a phase transition from the false to
the true vacuum takes place. First-order phase transitions from the false to
the new true vacuum proceed via bubble nucleation. The bubbles with
non-zero VEV evolve and expand 
until they dominate the universe. These bubbles generate GWs through
friction with the thermal plasma and through collisions (see
e.g.~Refs.~\cite{Binetruy:2012ze,Caprini:2015zlo,Weir:2017wfa,Caprini:2019egz,Hindmarsh:2020hop}
for reviews). \s 

The crucial quantity for the phase transition is the decay rate of the
false vacuum, or in other words the tunnelling rate for the transition
from the false vacuum to the true vacuum. The probability for the
phase transition to take place at finite temperature is computed via minimisation of the
$O(3)$-symmetric Euclidean action $S_3$, the bounce 
action, of the scalar field. For the bounce
solution the trajectory in field space connecting the
local minima needs to be found, which minimises the Euclidean action. This is
technically very challenging and is complicated by the fact that
the vacuum structure of extended Higgs sector potentials is very involved
and changes when loop and temperature effects are included. The
behaviour of the ground state of the universe as it cools down is 
hence highly non-trivial and requires tracing the ground
states, given by the minima of the effective potential, as a function
of the temperature. Through the decay rate and the Euclidean action
the thermal parameters characterising FOPTs are obtained. These are
the transition temperature, the inverse duration of the PT and the transition
strength given by the latent heat released during the PT. Together
with the bubble wall velocity, they ultimately determine the
characteristic frequencies and the amplitude of the gravitational wave
spectrum. \s

There are several codes on the market, besides {\tt BSMPT}, which
trace the minima of involved scalar potentials and some of them also provide bounce
solutions. They are briefly reviewed here:
\begin{itemize}
\item \texttt{CosmoTransitions} \cite{Wainwright:2011kj}: traces minima upwards
  and downwards in the temperature, initiating the tracing with a collection of starting
  points that are then optimized locally using a Nelder-Mead-type
  algorithm. 
  It contains \texttt{Python} modules to calculate the bounce solution via path
  deformation. The nucleation temperature is obtained using the
  approximation $S_3(T)/T\lesssim\num{140}$.
\item In \texttt{Vevacious} \cite{Camargo-Molina:2013qva} homotopy
  continuation is exploited 
  to find all extrema of the tree-level potential. These are then used
  as starting points for gradient-based minimisation of the one-loop
  effective potential. Tunnelling times are obtained by using {\tt
    CosmoTransitions}.
\item \texttt{VevaciousPlusPlus}
  \cite{Camargo-Molina:2013qva,Camargo-Molina:2014pwa} has no
  new implementation of the bounce solution calculation, but a
  \texttt{C++} code wrapper of \texttt{CosmoTransitions} interfaced
  with models in the framework of \texttt{SARAH}
  \cite{Staub:2009bi,Staub:2010jh,Staub:2012pb,Staub:2013tta}.
\item \texttt{AnyBubble} \cite{Masoumi:2017trx} is a \texttt{Mathematica}
  code for finding bubble nucleation instantons via a multiple
  shooting algorithm.
%
\item \texttt{EVADE} \cite{Hollik:2018wrr,Ferreira:2019iqb,EVADE}
  performs the minimisation of the scalar potential through polynomial
  homotopy continuation and estimates the decay rate of the false
  vacuum in a multi-scalar theory by using the straight-path approximation.
\item \texttt{BubbleProfiler} \cite{Athron:2019nbd} is a \texttt{C++}
  is library for finding the bounce solution via a semi-analytic
  algorithm formulated in \cite{Akula:2016gpl}.
%
\item The {\tt C++}  code \texttt{PhaseTracer} \cite{Athron:2020sbe} tracks phases and identifies critical temperatures using an algorithm
    that is similar to the one used in {\tt CosmoTransitions}, but faster. It
  handles discrete symmetries and can be linked to potentials
  implemented in 
{\tt FlexibleSUSY} \cite{Athron:2014yba,Athron:2017fvs} and {\tt
  BSMPT}.
\item \texttt{SimpleBounce} \cite{Sato:2019wpo} applies the gradient
    flow method from \cite{Sato:2019axv} to calculate the bounce solution.  
\item \texttt{FindBounce} \cite{Guada:2020xnz} finds the bounce
  solution via the polygonal multi-field method described in \cite{Guada:2018jek}.
\item \texttt{OptiBounce} \cite{Bardsley:2021lmq} obtains the bounce
  solution via solving the `reduced' minimisation problem \cite{Coleman:1977th}. 
\item \texttt{TransitionListener} \cite{Ertas:2021xeh,Bringmann:2023iuz} extends {\tt CosmoTransitions} by functionalities to study the case of dark sector phase transitions and also derive e.g.~stochastic GW spectra and signal-to-noise ratios.
\end{itemize}

The {\tt C++} code {\tt BSMPTv1} \cite{Basler:2018cwe} has been
developed to compute the loop-corrected daisy-resummed effective
potential of BSM Higgs sectors 
at non-zero temperature, applying an on-shell (OS) renormalization scheme.
It checks for absolute vacuum stability
requiring the electroweak vacuum to be the global minimum of the
one-loop corrected effective potential at zero temperature. 
This ensures that we live in a stable vacuum today, however, it
excludes the valid parameter region of points with metastable vacuum
configurations.\footnote{The electroweak vacuum of the SM at zero
  temperature is metastable
\cite{Degrassi:2012ry,Bednyakov:2015sca}, its quartic coupling
$\lambda$ is negative for scales $\gsim 
10^{10}$~GeV and a lifetime larger than the age of the universe.} 
Our code \texttt{BSMPTv1/v2} traces the position of the global minimum
for temperatures $T\in\{0,\,300\}\,\si{GeV}$, looking for a discontinuity in
the electroweak VEV as the order parameter of the phase transition
between the high-temperature symmetric and the electroweak vacuum at
$T=\SI{0}{GeV}$. However, this approach only shows the possible
coexistence of two minima. It does not guarantee that the transition
actually takes place. It can
also not reveal the possible existence of multiple phases during the
evolution of the universe. In {\tt BSMPTv2} \cite{Basler:2020nrq}
the code was extended to
the computation of the generated 
baryon asymmetry in the CP-violating 2-Higgs-Doublet Model (C2HDM),
and included a new model, the Complex Singlet Extension of the SM
(CxSM). A detailed description of {\tt BSMPTv1} and {\tt v2} is given 
in the two corresponding manuals \cite{Basler:2018cwe} and
\cite{Basler:2020nrq}, respectively, 
phenomenological investigations using the code can be found in
\cite{Basler:2016obg,Basler:2017uxn,Basler:2018dac,Wang:2018hnw,Wang:2019pet,Basler:2019nas,Basler:2019iuu,Basler:2019ici,Glaus:2020ihj,Abdussalam:2020ssl,Su:2020pjw,Gabelmann:2021ohf,Basler:2021kgq,Chaudhuri:2021agl,Biekotter:2021ysx,Chaudhuri:2021rwt,Atkinson:2021eox,Chaudhuri:2021ibc,Chaudhuri:2021vdi,Muller:2021hrn,Egle:2022wmq,Anisha:2022hgv,Biermann:2022meg,Atkinson:2022pcn,Song:2022xts,AbdusSalam:2022idz,Biekotter:2022kgf,OleaRomacho:2022crq,Chang:2022psj,Su:2022upm,Biermann:2023owb,Egle:2023pbm,Aoki:2023lbz,
  Goncalves:2023svb,Heinemeyer:2024vqw,Heinemeyer:2024hxa}. \s

The here presented version
\texttt{BSMPTv3}\footnote{For a phenomenological study, an early
  version of the code was used in \cite{Anisha:2023vvu}.} extends
\texttt{BSMPT} into a capable single and 
multi-step phase transition finder. Starting from an
absolutely stable electroweak vacuum at zero temperature, it traces the
electroweak and all emerging global minima as functions of the temperature
and calculates all possible found transitions and key parameters
(temperature scales, released latent heat, inverse time scale etc.) as well
as the resulting gravitational wave spectra. We also implement a
framework of status codes that report e.g.~on electroweak symmetry 
non-restoration at high temperature\footnote{The possibility and
  consequences of electroweak symmetry non-restoration were studied in
  e.g. \cite{Biekotter:2021ysx,Biekotter:2022kgf,Meade:2018saz,Baldes:2018nel,Matsedonskyi:2020mlz,Carena:2021onl}.},
as well as vacuum trapping\footnote{Vacuum trapping has been studied
  in e.g. \cite{Biekotter:2021ysx,Biekotter:2022kgf,Baum:2020vfl}.}. \s

The code {\tt BSMPT} with its extension to {\tt v3} goes beyond
existing codes in the following sense. First of all, the innovations related to the
versions {\tt v1} and {\tt v2} are: 
\begin{itemize}
\item {\tt BSMPT} was the first code to implement an on-shell renormalization
scheme \cite{Basler:2016obg}, where, at zero temperature, the
loop-corrected masses and mixing angles are renormalized to their
corresponding leading-order (LO) values in the minimum of the potential. 
Crucial for phenomenology, this allows for directly checking the relevant theoretical and
  experimental constraints of the investigated model without resorting
  to an involved time-consuming iterative procedure.
  In this on-shell renormalization scheme UV-finite
counterterms are added after the potential has been renormalized in the
$\overline{\mbox{MS}}$ scheme. We have fixed the renormalization
scale in the $\overline{\mbox{MS}}$ scheme to the  value of the VEV, 
$v=246$~GeV. Since this fixing does not allow to study the renormalisation scale dependence of the obtained results, we provided the option to change the value of the renormalisation scale in the model file. This allows the users to use their own preferred value for the renormalisation scale. Furthermore, it provides the possibility to study the impact of the change of the renormalisation scale on the results. This residual scale dependence originates from the fact that the on-shell renormalisation conditions for the masses and mixing angles are derived for the minimum of the zero-temperature potential and that moreover the Higgs self-interactions are not renormalised on-shell.
 \item The code can be easily linked to {\tt ScannerS} \cite{Coimbra:2013qq,Ferreira:2014dya,Costa:2015llh,Muhlleitner:2016mzt,Muhlleitner:2020wwk}, so that extensive scans in the
parameter spaces of the investigated models can be
performed checking for the relevant theoretical constraints,
implemented in {\tt ScannerS}, and (through
    the links to {\tt HiggsBounds} \cite{Bechtle:2008jh,Bechtle:2011sb,Bechtle:2012lvg,Bechtle:2013wla,Bechtle:2015pma,Bechtle:2020pkv,Bahl:2021yhk} and {\tt HiggSignals} \cite{Bechtle:2013xfa,Stal:2013hwa,Bechtle:2014ewa,Bechtle:2020uwn},
which have been recently merged into the new package {\tt HiggsTools} \cite{Bahl:2022igd}, and {\tt MicrOMEGAs} \cite{Belanger:2001fz,Belanger:2004yn,Belanger:2006is,Belanger:2008sj,Belanger:2010gh,Belanger:2013oya,Belanger:2014vza,Barducci:2016pcb,Belanger:2018ccd,Alguero:2023zol})
for the experimental collider and DM constraints. {\tt ScannerS} also checks
  for flavour constraints, and, in CP-violating
models, tests the compatibility with results from the electric
dipole moments. 
\item Several models are already pre-implemented, namely the complex singlet
  extension of the SM (CxSM) \cite{Basler:2020nrq}, the CP-conserving, i.e.~real,
  2-Higgs-Doublet Model 
  (R2HDM) \cite{Basler:2018cwe} and its CP-violating version C2HDM \cite{Basler:2018cwe}, the next-to-minimal
  2HDM (N2HDM) \cite{Basler:2018cwe}, and the model `CP in the Dark' \cite{Biermann:2022meg,Biermann:2023owb}. New
  models can easily be added following the prescription in the manual of {\tt BSMPTv2}
  \cite{Basler:2020nrq}. Also, consult the {\tt README.md}-file on how
  to use our {\tt SymPy}~\cite{10.7717/peerj-cs.103} as well as {\tt
        Maple}~\cite{maple} and {\tt Mathematica}~\cite{mathematica} interfaces for model implementation. In {\tt
    BSMPTv3}, we furthermore added an exemplary model file for the SM.
\item  The code allows to calculate the loop-corrected trilinear
    Higgs self-couplings of the pre-implemented models and any model
  provided and implemented by the user, from the effective potential
  calculated in the code, at zero temperature.
\item In {\tt BSMPTv2} the computation of the baryon asymmetry for
  the model C2HDM was implemented. 
\end{itemize}
The new version {\tt v3} presented here surpasses the existing codes
because of the following features:
\begin{itemize}
\item Our algorithms for tracing minima and calculating the bounce solutions
  are more stable than the ones in 
  {\tt CosmoTransitions}, in particular for complicated potentials with
  numerous field directions.  It is for most scenarios faster than
  {\tt CosmoTransitions}. Importantly, it finds phase transitions,
  where {\tt CosmoTransitions} fails to identify them. We will discuss
  the comparison with {\tt CosmoTransitions} in Sec.~\ref{sec:examples}.
\item {\tt BSMPTv3} allows to check for symmetry restoration at high
  temperature, and it can treat potentials with discrete symmetries
    and potentials with flat directions.
\item In the derivation of the nucleation temperature, we do not only
  rely on the approximation applied in {\tt CosmoTransitions}
  $S_3(T)/T\lesssim\num{140}$, to get the nucleation temperature, but
  also derive it from the condition that the tunnelling decay rate per
  Hubble volume matches the Hubble rate.
\item Unlike existing codes, {\tt BSMPTv3} calculates the
  nucleation, the percolation,
        the completion and the reheating
  temperature. The user can optionally define the values of the false
  vacuum fractions to be applied for the percolation and completion
  temperature, respectively. 
\item The user can select the characteristic temperature scale at
        which the thermal parameters relevant for the gravitational wave
  spectrum are calculated. 
\item Contrary to the above listed codes, {\tt BSMPTv3} computes the
    gravitational wave spectrum originating from bubble collisions and
    highly relativistic fluid shells, from sound waves and 
  magneto-hydrodynamic turbulence. 
\item {\tt BSMPTv3} computes the related signal-to-noise ratio at
    {\tt LISA}.  
\end{itemize}
The code {\tt BSMPTv3} hence provides the whole chain from tracing the
phases of extended Higgs sectors, calculating the bounce action, the
transition rate, the strength of the phase transition to the
gravitational wave spectrum (and it calculates also the baryon
asymmetry in case of the 
C2HDM), in a self-contained framework applying 
on-shell renormalization in the effective potential. \s

Recently, an excellent overview has been published in 
\cite{Athron:2023xlk}, which reviews comprehensively the path from a particle
physics model to GWs.
The review also briefly comments on
  lattice methods. Lattice simulations allow to complete the
  incomplete perturbative approach to the study of EWPTs. They are
  computationally demanding, however, and not able to explore in
  detail many beyond-SM extensions. For lattice treatments of the
  effective potential
  cf.~e.g.~\cite{Kainulainen:2019kyp,Niemi:2020hto,Gould:2021dzl,Gould:2022ran},
  for works on lattice simulations of gravitational waves,
  cf.~e.g.~\cite{Hindmarsh:2015qta}.
An algorithm for the construction of an effective, dimensionally
  reduced, high-temperature field theory for generic models has been
  implemented in {\tt DRalgo} \cite{Ekstedt:2022bff}, which allows to
  better describe infrared effects \cite{Linde:1980ts} that can only be
  treated properly by lattice simulations
  \cite{Braaten:1994na}. The approximation through leading-order
  perturbation theory, which is widely used, contains large
  theoretical uncertainties because it converges only slowly, as has
  been pointed out in \cite{Arnold:1992rz,Farakos:1994kx, Croon:2020cgk,Gould:2021oba}. In \cite{Athron:2023xlk}, also
  gauge-independent approaches are reviewed which address the problem
  of gauge dependence of the effective potential
  \cite{Jackiw:1974cv,Patel:2011th}. For recent developments,
  cf.~e.g.~\cite{Niemi:2020hto,Croon:2020cgk,Hirvonen:2021zej,Lofgren:2021ogg,Schicho:2022wty,Athron:2022jyi,Qin:2024dfp}.
The review introduces all relevant quantities, discusses
related obstacles and open problems, reviews the state-of-the-art and
related literature, and provides useful formula. While we restrict
here to a minimal description for the introduction of our notation needed for the
presentation of our code and its new 
features, without a discussion of pros and cons of different
approaches\footnote{Where appropriate, we provide flags that allow the
user to choose between approaches.}, we
refer the reader for further background information to Ref.~\cite{Athron:2023xlk}.

\section{Program Description \label{sec:codedescription}}
In the following, we describe how to install and use {\tt BSMPTv3} focusing on the new executables and describing the new structure.

\subsection{Download and Installation}\label{sec:installation}
The program was developed and tested on {\tt Arch Linux}, {\tt OpenSuse 15.3} and {\tt macOS 13.2.1}
with a range of compilers: {\tt GNU
  7.5.0-13.2.1}\footnote{\url{https://gcc.gnu.org/}} and {\tt Clang
  14.0.3}\footnote{\url{https://clang.llvm.org/}}\,\footnote{We
  furthermore continuously ensure that {\tt BSMPT} compiles and passes
  all unit tests under the latest versions of {\tt
  macOS}, {\tt Windows} and {\tt Ubuntu} as well as {\tt Ubuntu-20.09}}.
The here presented version {\tt v3}, as well as all previously released versions, can be obtained from:
\begin{center}
    \download\,.
\end{center}
The code is structured into the following directories:
\begin{labeling}{standalone\quad}
    \item[{\tt example}] example input and output files for all executables and models 
    \item[{\tt include}] header files
    \item[{\tt manual}] manuals 
    \item[{\tt profiles}] {\tt Conan} profiles
    \item[{\tt sh}] {\tt Python} files for converting input data files to required format
    \item[{\tt src}] source files
    \item[{\tt standalone}] stand-alone example codes that allow users
      to directly use selected algorithms 
    \item[{\tt tests}] input and source files used for the unit tests
    \item[{\tt tools}] a {\tt SymPy}~\cite{10.7717/peerj-cs.103} as well as a {\tt Maple}~\cite{maple} and a {\tt Mathematica}~\cite{mathematica} interface to calculate all necessary input needed to implement a new model, as well as configurations for the installation with {\tt CMake}
\end{labeling}
\noindent
The directory {\tt src} contains the following:
\begin{labeling}{src/gravitational\_waves \quad}
    \item[{\tt src/prog}] executable source code
    \item[{\tt src/models}] implemented  models and SM parameters
    \item[{\tt src/minimiser}] minimisation routines
    \item[{\tt src/ThermalFunctions}] thermal integrals
    \item[{\tt src/WallThickness}] calculation of the wall thickness
    \item[{\tt src/Kfactors}] calculation and interpolation of the
      thermal transport coefficients  
    \item[{\tt src/baryo\_calculations}] {\tt VIA} and {\tt FH} approach to
calculate the baryon asymmetry of the universe\footnote{We used the
  {\tt FH}
\cite{Cline:1997vk,Kainulainen:2001cn,Fromme:2006wx,Fromme:2006cm} and
the {\tt VIA} \cite{Riotto:1995hh,Riotto:1997vy,Lee:2004we,Chung:2009qs} approach to
compute the baryon asymmetry of the universe in the C2HDM. Recently,
it was argued, however, that the source term in the {\tt VIA} method  vanishes at
leading order \cite{Postma:2022dbr}, which would have consequences for the derived
 baryon asymmetry in this method.}
    \item[{\tt src/minimum\_tracer}] (multi-step) phase tracing and
      identification of coexisting phase pairs
    \item[{\tt src/bounce\_solution}] bounce solution and characteristic temperatures
    \item[{\tt src/gravitational\_waves}] derivation of GW spectrum parameters
    \item[{\tt src/transition\_tracer}] transition history evaluator
      that operates the classes \\ {\tt minimum\_tracer}, {\tt bounce\_solution}, {\tt gravitational\_waves} for the new executables
\end{labeling}

{\tt BSMPTv3} requires a {\tt C} and {\tt C++} compiler that supports
the language standard \num{17}, as well as an installation of {\tt
  CMake}\footnote{\url{https://cmake.org/}} and {\tt
  Conan}\footnote{\url{https://conan.io/}}. The latter two can e.g.~be
installed with the {\tt Python} package manager {\tt pip}~\cite{pip}
through the command {\tt pip3 install cmake conan}.\s

For a default installation of {\tt BSMPTv3}, our {\tt Build.py} script can be used. 
This script installs the necessary {\tt Conan} profiles for the operating system, handles the dependencies and compiles {\tt BSMPT} with its default settings.
It is executed via the command (from within the main directory of {\tt BSMPTv3}):
\begin{lstlisting}{language=bash}
  python3 Build.py
\end{lstlisting}

If the installation is successful, a new directory {\tt build} is created and the following new executables are built in {\tt \$BSMPT/build/[operating-system-specific-name]/bin}:
\begin{labeling}{MinimaTracer\quad}
\item[{\tt MinimaTracer}] Tracing of minima as function of the temperature (Sec.~\ref{sec:MinimaTracer})
\item[{\tt CalcTemps}] Calculation of the bounce solution and characteristic temperatures for first-order phase transitions between pairs of coexisting phases (Sec.~\ref{sec:CalcTemps})
\item[{\tt CalcGW}] Calculation of the gravitational wave spectrum sourced by first-order phase transitions (Sec.~\ref{sec:CalcGW})
\item[{\tt PotPlotter}] Visualization of multi-dimensional potential contours (Sec.~\ref{sec:PotPlotter})
\end{labeling}
By default, also the {\tt BSMPTv2}-executables {\tt BSMPT}, {\tt
  CalcCT}, {\tt NLOVEV}, {\tt TripleHiggsCouplingsNLO}, \\
{\tt Test}, {\tt VEVEVO} are built. 
To build the baryogenesis executables, {\tt  CalculateEWBG}, {\tt PlotEWBG\_nL} and {\tt PlotEWBG\_vw}, one needs
to set {\tt CompileBaryo=True} when installing {\tt BSMPT} via our
{\tt Setup.py} script, as will be described below. Before doing so, we
first comment on the dependencies that are used by {\tt Conan}: 
\begin{itemize}
    \item The library {\tt GSL}~\cite{GSL} is required
  for its routines for numerical derivation, integration,
  interpolation as well as minimisation and its mathematical
  algorithms.
    \item {\tt Eigen3}\footnote{\url{http://eigen.tuxfamily.org/} and
        \url{https://gitlab.com/libeigen/eigen}} is used for matrix-
      and vector-manipulations.
  \item {\tt nlohmann\_json}~\cite{nlohmann-json} is used for the option to supply input to the executables in the form of {\tt json}-files as further described below.
  \item {\tt Catch}~\cite{catch} and {\tt benchmarks}~\cite{benchmark}
    are used for unit tests. If the unit tests should not be compiled,
    the option {\tt EnableTests=False} must be set when using the
    detailed installation method via the {\tt Setup.py} script, as further described below. 
    \item {\tt Boost}\footnote{\url{https://www.boost.org/}} is
      optional and only required for the calculations
      related to baryogenesis. In order to compile the baryogenesis
      calculation, the option {\tt CompileBaryo=True} must be set in
      the detailed installation method, as described below.
\end{itemize}
In addition to {\tt GSL}, at least one of the following minimisation libraries should be used. By default, both are installed:
\begin{itemize}
    \item {\tt libcmaes}~\cite{libcmaes} is a {\tt C++} implementation
      of the Covariance Matrix Adaptation Evolution Strategy 
      algorithm. If an installation is not wanted, {\tt UseLibCMAES=False} must be specified when using {\tt Setup.py} for installation, as will be explained below. 
  \item {\tt NLopt}~\cite{NLopt} uses the {\tt DIRECT\_L}~\cite{DIRECT_L} algorithm. If the user does not want to install {\tt NLopt}, the option {\tt UseNLopt=False} can be specified analogously.
\end{itemize}

We provide the script {\tt Setup.py} which allows for a customized installation. It can take several optional arguments, e.g.~all above listed options and more.
All possible optional arguments can be viewed by running {\tt python3 Setup.py -h} or {\tt python3 Setup.py --help}\footnote{Note, that if a compiled version of {\tt BSMPT} is distributed to other machines, which do not share the same or related CPUs, it is advisable to disable vectorization for its compilation.
This can be done by setting {\tt UseVectorization=False}.}.
A complete installation of {\tt BSMPTv3} using our {\tt Setup.py} script looks like:
\begin{lstlisting}{language=bash}
  python3 Setup.py [optional arguments]
  cmake --preset ${profile}  
  cmake --build --preset ${profile} -j  
  cmake --build --preset ${profile} -j -t doc
\end{lstlisting}
The {\tt -t doc} uses {\tt
  doxygen}\footnote{\url{https://www.doxygen.nl/index.html}} to create
the documentation in the local {\tt build} directory\footnote{The
  documentation for {\tt BSMPT} can also be found online at \url{https://phbasler.github.io/BSMPT/documentation/}.}.
The {\tt \$\{profile\}} parameter depends on the operating system. After
running {\tt Setup.py}, {\tt cmake --list-presets} gives a list of all selectable profiles.\s

When {\tt BSMPTv3} is successfully compiled (with option {\tt EnableTests=True}), unit tests can be run by calling (in the main directory):
\begin{lstlisting}{language=bash}
  ctest --preset ${profile} -j
\end{lstlisting}
These tests should be extended if the user implements a new model.

\subsection{Structure and Description of the Algorithms of {\tt BSMPTv3}}\label{sec:structure}
\begin{figure}
      \centering
      \begin{tikzpicture}
          \node[draw,thick,text width=8cm,align=center] (h0) at (0,18) {
              {\tt BSMPTv1/v2}\par
              one-loop daisy-resummed finite-temperature effective potential
          };
          \node[draw,text width=8cm] (a) at (0,12.5) {
              \myuline{class {\tt MinimumTracer}}\par
              \vspace*{-.5em}
              \begin{itemize}
                  \item[-] derivation of (finite temperature) phase structure in the
                      temperature range $T\in[T_\text{low}=\SI{0}{GeV},\,T_\text{high}]$
                  \item[-] identification of coexisting phase pairs and their critical temperatures $T_c$
              \end{itemize}
          };
          \node[draw,text width=8cm] (b) at (0,6) {
              \myuline{class {\tt BounceSolution}}\par
              \vspace*{-.5em}
              \begin{itemize}
                  \item[-] calculation of the bounce solution as a function of temperature
                  \item[-] finding the nucleation temperature $T_n$ through matching the tunnelling rate with the Hubble rate
                  \item[-] derivation of the percolation $T_p$ and completion temperature $T_f$ via solving the integral of the false vacuum fraction
              \end{itemize}
          };
          \node[draw,text width=8cm] (c) at (0,2) {
              \myuline{class {\tt GravitationalWave}}\par
              calculation of all parameters of the GW spectrum,
              e.g. $\alpha$, $\beta/H$, $\kappa$,
              $c_s$, \& the GW spectrum 
          };
          \node[text width=8cm] (d) at (0,15) {
              \myuline{class {\tt TransitionTracer}} transition history evaluater, interfacing with executables
          };
          \draw[black] ($(a.north west)+(-0.3,1.25)$) rectangle ($(a.south east)+(0.3,-9.7)$);
          \node[draw, text width=3.2cm,align=center] (e) at (7,12.5) {\tt MinimaTracer.cpp};
          \node[draw, text width=3.2cm,align=center] (f) at (7,6) {\tt CalcTemps.cpp};
          \node[draw, text width=3.2cm,align=center] (g) at (7,2) {\tt CalcGW.cpp};
          \node[draw, text width=3.2cm,align=center] (h) at (7,18) {\tt PotPlotter.cpp};
          \draw[-latex] (a) -- (b) node[midway,right,text width=4cm] () {for each phase pair with $T_c$};
          \draw[-latex] (b) -- (c);
          \draw[-latex] (a) -- (e);
          \draw[-latex] (b) -- (f);
          \draw[-latex] (c) -- (g);
          \draw[-latex] (h0) -- (h);
          \draw[-latex] (h0) -- (d) node[pos=.5,right,text width=4cm] () {expanded in {\tt BSMPTv3} by};
      \end{tikzpicture}
      \caption{Structure and dependencies of the algorithm of {\tt
          BSMPTv3}. The four classes are compiled as libraries. The
        class {\tt TransitionTracer} acts as a logic interface between the
        executables and the three subclasses that contain the steps of
        the calculation and, based on the results, reports on the
        transition history.}\label{fig:structure}

  \end{figure}
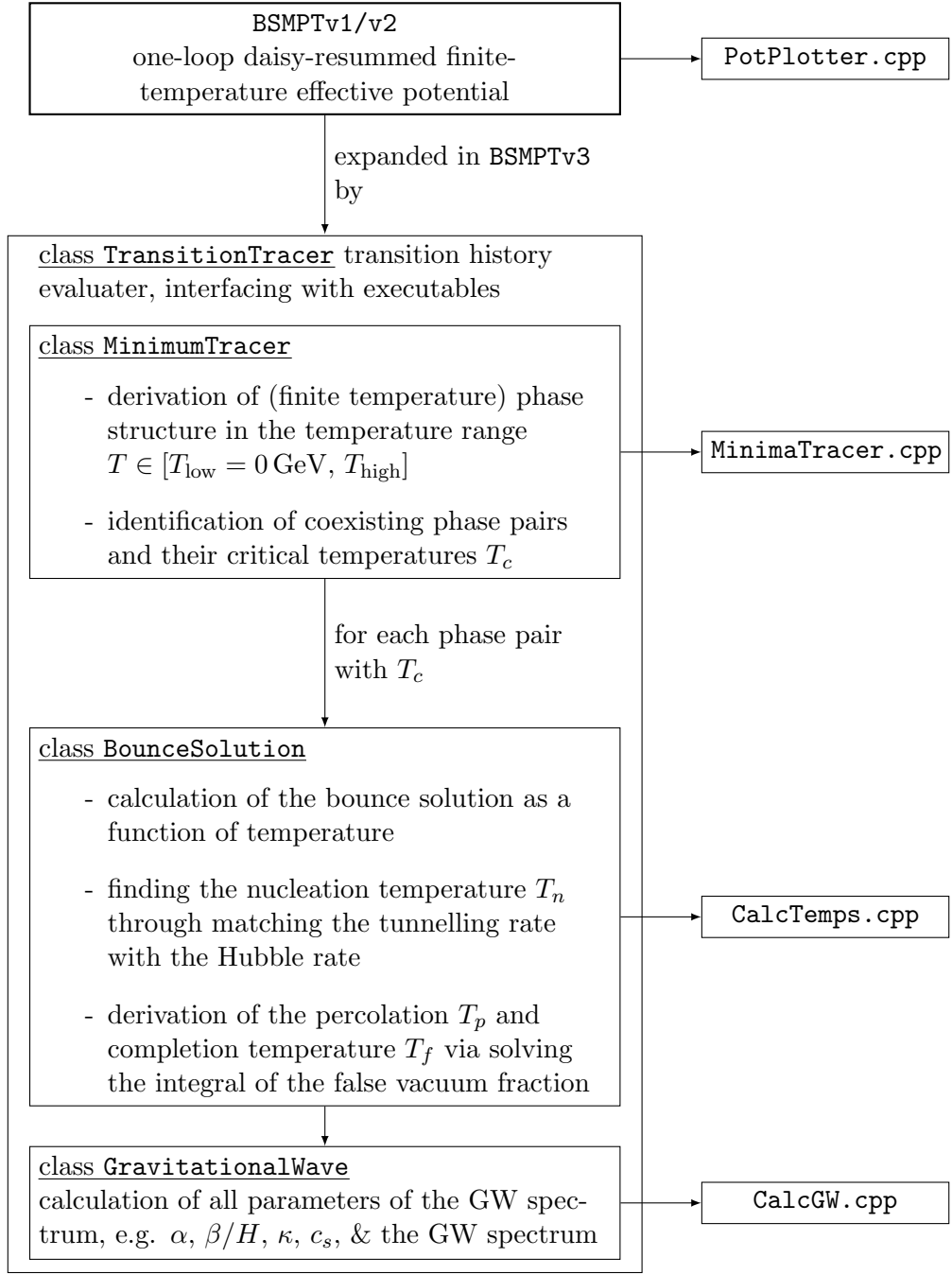

The main objective of the code {\tt BSMPTv3} is the tracing of (multiple)
phases as a function of the temperature and the calculation of the
transition probability from the respective false to the true vacuum, the 
computation of the relevant thermal parameters 
and the determination of the gravitational wave spectrum from a
FOPT. The solution is divided into three steps: $(i)$ Construction of the
loop-corrected effective potential (evaluated in
the Landau gauge) including thermal masses and
applying the on-shell renormalization scheme; tracing the minima of
this potential and identification of pairs of coexisting phases. 
$(ii)$ Determination of the bounce solution for each of the
found pairs of coexisting phases; calculation of the tunnelling rate from the
false to the true vacuum of the phase pair; computation
of the critical, nucleation, percolation, and completion
temperatures. $(iii)$ For the found FOPT, computation of the
gravitational wave spectrum based on the transition
temperature (percolation temperature by default) and the bounce
solution determined in step $(ii)$. These three steps are performed in the corresponding
three classes {\tt MinimumTracer}, {\tt BounceSolution}, and {\tt
  GravitationalWave} and organized by a fourth class, {\tt TransitionTracer}, cf.~Fig.~\ref{fig:structure}. 
  The user interface to extracting the results is given by four
executables, namely  
{\tt MinimaTracer.cpp} (reports on all found minima as functions of the
temperature), {\tt CalcTemps.cpp} (gives out characteristic temperatures for all
found coexisting phase pairs), {\tt CalcGW.cpp} (reports on characteristic
temperatures and GW parameters for all found coexisting phase pairs) and {\tt PotPlotter} (calculates potential contours useful for visualization).
  
In the following four sections
\ref{subsec:mintracer}-\ref{subsec:transtracer}, we will describe the 
four classes with our applied solutions and the relevant formulae. 
In the four subsequent sections \ref{sec:MinimaTracer}-\ref{sec:PotPlotter}, the four executables will be explained together with the flags that can
be applied. In this context, we will also describe various
algorithms that can be chosen by the user through the flags. In
  Sec.~\ref{sec:standalone} we collect functions that the user might
  want to use for specific computations. The last section
  \ref{sec:status_codes} finally is devoted to the summary of  
the given out status codes and their explanation.

\subsection{The Class {\tt MinimumTracer} \label{subsec:mintracer}}
The computation of the effective potential in the on-shell
renormalization scheme for an already implemented model or a new model
implemented by the user was described in the {\tt BSMPT}
manuals of {\tt v1} and {\tt v2}, to which we refer 
the user for details. Here we describe the newly implemented algorithm
for the tracing of (possibly multiple) coexisting phases as function
of the temperature. \s

Between a user-defined high temperature $T_\text{high}$ and the low
temperature $T_\text{low}=\SI{0}{GeV}$, {\tt MinimumTracer} traces
phases using found global minima at high and low temperature as well
as the zero-temperature electroweak minimum as seed points.  
We start with the definitions of phases and phase transitions in
  Sec.~\ref{sec:expressionsPT} and then give in
Sec.~\ref{sec:phasetracker} details on the
algorithm for tracing one phase. In Sec.~\ref{sec:symmetries} we describe
how we identify symmetries of the potential and find the closest
distinct phases in field space. Section~\ref{sec:flat_directions}
shows how {\tt BSMPTv3} deals with flat directions in potential
space. More details on how {\tt BSMPTv3} traces
  landscapes with possibly multiple coexisting phases are given
below in Sec.~\ref{sec:multistepmode}.

\subsubsection{Phases, First and Second Order Phase Transitions}\label{sec:expressionsPT}
Multi-scalar Higgs potentials exhibit complicated cosmological
histories accompanied by possibly a multitude of phase transitions
between different vacuum states. For the sake of clarity, we have to define
the meaning of the expressions {\it phase} as well as {\it
  first-order} and {\it second-order phase transition} 
as they are used in our code for the derivation of the 
gravitational waves related to a first-order phase transition. The
{\it phase} of a multi-scalar potential
is defined by the values of its temperature-dependent
vacuum expection values of the
complete set of scalar fields. Different phases differ in the scalar
field directions in which they exhibit a non-zero VEV. 
With decreasing temperature during the
evolution of the universe, the temperature-dependent effective
potential changes and different minima, maxima, and saddle points at
different locations in field space and with different potential values
evolve. Starting from a global minimum at high temperature, with
decreasing temperature at some moment a second minimum starts
evolving, which may become degenerate with the existing global minimum
at the critical temperature $T_c$, however, separated by a barrier,
so that we then have a discontinuity in the VEVs of
the two degenerate global minima. Finally, the second minimum becomes the global one
and then, if the tunnelling rate is sufficiently large, a {\it first-order phase transition} 
from the false to the true vacuum takes place. 
In a {\it second-order phase transition}, on the other hand, we have a
continuous change of the VEV as a function of the
temperature. In {\tt BSMPTv3}, we only consider GWs sourced by FOPTs.


\subsubsection{Phase Tracker}\label{sec:phasetracker}
Our goal is to find the transition rate between two distinct phases as
a function of the temperature $T$. Thus, before we do any calculation,
it is of utmost importance to have an accurate description of the
vacuum structure of a particular BSM model. Usually, models are too
complex to allow for an analytical description of their vacuum
so that we need to employ numerical methods to find the phases and track them across the whole temperature range. \s

The location of the global minimum in \texttt{BSMPTv2} is searched for using 
algorithms of the libraries \texttt{GSL}, \texttt{libcmaes} or
\texttt{NLopt}. These gradient-free methods only use potential values
to locate the minimum, so that they are rather fast.
The available precision provided in \texttt{BSMPTv2} was considered 
sufficient in previous iterations. However, for the identification of
the bounce solution in \texttt{BSMPTv3}, significantly higher precision is 
required, primarily as a result of the complexity of the boundary
conditions. In {\tt BSMPTv3}, we still
use these gradient-free methods to find the seed points from where we start the tracing. \s 

To find the minimum $\vec{\phi}$ of the potential we have to search
for points with vanishing gradient, $\nabla V(\vec{\phi}) =
\vec{0}$. The method that we settled for is the Newton–Raphson method 
which uses the  
gradient and the Hessian matrix to take an educated step in the right
direction. To understand the core of the method let us Taylor expand the
gradient around a point $\vec{\phi}$, 
\begin{align}
    \nabla V(\vec{\phi} + \vec{\varepsilon}) = \nabla V(\vec{\phi}) +
  H(\vec{\phi})\vec{\varepsilon} + \mathcal{O}(\vec{\varepsilon}^{\, 2})\,,
\end{align}
where $H(\vec{\phi})$ is the Hessian matrix calculated at
$\vec{\phi}$. Suppose that this small step takes us to the minimum of
the potential, i.e. $\nabla V(\vec{\phi} + \vec{\varepsilon}) = 0$,
then we can invert the equation to find the step $\vec{\varepsilon}$,  
\begin{align}
	\vec{\varepsilon} = -H(\vec{\phi})^{-1} \nabla V(\vec{\phi}) \,.
\end{align}
In our code, we start with an initial guess $\vec{\phi}_0$ provided by
the gradient-free minimisers and take educated steps $\vec{\phi}_{i+1}
= \vec{\phi}_{i} -H(\vec{\phi}_i)^{-1} \nabla V(\vec{\phi}_i)$ until the
gradient vanishes. This method is the equivalent of locally
approximating the potential with a multivariate second-degree
polynomial and finding the minimum with a single iteration. Although
each iteration is computationally costly, as one needs to numerically
calculate the gradient and the Hessian and its inverse, the convergence is so fast
that near the minimum only a few iterations are needed to find the
minimum. \s

We then want to track the phases across the whole temperature range, so
after finding the minimum for some temperature $T$ we slightly change
the temperature $T\rightarrow T + \delta T$ and rerun the algorithm
starting at the minimum found at $T$. If the minimum is now a saddle
point (which can also be found by the method) then we decrease $\delta
T$. If the Hessian matrix is singular this method will not work
because the Hessian matrix will have no inverse. This might be a
problematic scenario because the Hessian matrix eigenvalues coincide
with the masses of scalar particles and massless particles are a
possibility. To circumvent this scenario we add a small constant to
the diagonal elements of the Hessian, thereby 
shifting all eigenvalues by this constant value. This allows us to have
zero, and even small negative eigenvalues to account for numerical
errors, in the Hessian matrix without destroying the convergence. \s 

This method uses gradients and Hessian matrices that, in our case,
have to be calculated numerically. A useful trick that stabilises the
numerical derivatives is to rescale the potential as 
\begin{align}
	V(\vec{\phi}) \rightarrow \frac{V(\vec{\phi})}{1\,\,
  \text{GeV}^2 + T^2}\,. \label{eq:potrescal}
\end{align}
Obviously, a minimum in the rescaled potential is also a minimum in the non-rescaled potential. The advantage comes from the fact that at high $T$ the potential behaves as
\begin{align}
	\lim_{T\rightarrow\infty} V(\vec{\phi},T) \sim T^2 \widetilde{V} (\vec{\phi}) + T^4 \times(\text{field-independent constants})\,,
	\end{align}
  where $\widetilde{V}(\vec{\phi})$ is the part of the potential
  proportional to $T^2$ and is $T$-independent.\footnote{There are two
    different approaches to implement the temperature-corrected masses
    in the effective potential, referred to as `Arnold-Espinosa' \cite{Arnold:1992rz} and
    `Parwani' \cite{Parwani:1991gq} approach. For details,
    cf.~e.g.~\cite{Basler:2016obg}. In \texttt{BSMPT}, the default
    option is the Arnold-Espinos approach, which
    consistently implements the thermal masses at one-loop level in
    the high-temperature expansion. The scaling with $T^2$ is found
    for the Arnold-Espinosa approach, for the Parwani 
    approach it is $T^2 \log T^2$. The numerical advantage holds for
both methods. \label{footnote:methods}} The $T^4$ term 
  is irrelevant for the derivatives, the other term scales as $T^2$ so
  that 
  the rescaled potential becomes $T$-independent at high $T$. Because of
  this, it is much easier to choose a good step size for the numerical
  derivatives. We chose to  rescale the potential with
  $1\,\,\text{GeV}^2 + T^2$ and not just $T^2$ in order not to run into
  $\frac{1}{0}$ at low $T$. \s
  
A short comparison with \texttt{CosmoTransitions} is in
order. While we use the Newton-Raphson method to find the minimum,
    \texttt{CosmoTransitions} first uses a Newton-Raphson step combined
    with a gradient descent step in temperature from the previous
    temperature iteration to find a good approximation; next it uses
    the Nelder-Mead downhill simplex method \cite{Nelder:1965zz} until it
    finds the minimum. While, as said above, our method is computationally more
    expensive, as we have to compute and invert the Hessian matrix, we
    found that it converges much faster than the Nelder-Mead simplex
    method, in particular if the temperature step is well-chosen.

\subsubsection{Discrete Symmetries}\label{sec:symmetries}
Some of the models may exhibit discrete $\mathbb{Z}_2$ symmetries in
the scalar sector or $\mathbb{Z}_2$ subgroups of the gauge groups. As
these symmetries increase the number of possible 
minima, it is important to know if two particular minima can be
transformed from one to the other. By knowing the symmetries,
\texttt{BSMPTv3} does not trace the same minimum twice which reduces the
computational time. Another important issue is that, although minima
which are related through symmetries have the same physics, they may
have different transition rates to other minima. Let us consider a
model with a symmetry transformation $\vec\phi \to \vec{\widetilde\phi}$,
and with a true vacuum $\vec{\phi}_t$ and a false vacuum
$\vec{\phi}_f$ that cannot transform into each other or themselves
applying the symmetry transformation. We hence have four distinct
minima. Obviously, we have for the Euclidean action $S_3$ that
$S_3(\vec{\phi}_f \to \vec{\phi}_t) = S_3(\vec{\widetilde\phi}_f \to
\vec{\widetilde\phi}_t)$. But we also have other possibilities,
$S_3(\vec{\phi}_f \to \vec{\widetilde\phi}_t) = S_3(\vec{\widetilde\phi}_f \to
\vec{\phi}_t$) etc., which might produce different transition rates. If this is not taken into account, the code might miss
the transition with the lowest action. This is precisely what
was noticed and discussed in
Ref.~\cite{Branchina2018} in the 2HDM and which alerted us for such scenarios. \s
 
The user does not need to provide the symmetries. \texttt{BSMPTv3}
deals with this scenario by first computing the group $G$ of all
$\mathbb{Z}_2$ symmetries that the potential can have. The general
group is given by the following direct product
 \begin{align}
 	G = \prod_i^n \mathbb{Z}^{(i)}_2
 \end{align}
 where $\mathbb{Z}^{(i)}_2 = \{ e,\, z^{(i)}  \}$ is the symmetry
 group that affects the sign of the $i$-th component, i.e. $e$ is the
 identity and $\{\phi_1,\dotsb,\phi_i,\dotsb,\phi_n)\}
 \overset{z^{(i)}}{\to} \{\phi_1,\dotsb,-\phi_i,\dotsb,\phi_n\}$. The
 order of the group $G$ is $2^n$ where $n$ is the dimension of the
 field space. 
 As an example, we consider the 2HDM. The eight group generators of
 its symmetry group in the field basis 
 $\{\omega_\text{CB},\,\omega_1,\,\omega_2,\,\omega_\text{CP}\}^T$ 
are given in Tab.~\ref{table:discretesymmetries}. The
     indices `CB, 1, 2, CP' denote the charge-breaking, the two
     CP-even neutral and the CP-breaking VEV directions,
     respectively. In general, we denote VEV directions at arbitrary temperature by
 $\omega_i$ and at zero temperature by $v_i$.
After generating the group elements, the code verifies
 which of the symmetries keeps the potential invariant and saves this
 information to be used later. \s 

 We also introduce the notion of `principal quadrant'.\footnote{While
   depending on the number of VEVs connected through discrete
   symmetries, geometrically this is not necessarily a quadrant, for
   the sake of simplicity we still keep this expression.}
 Its definition takes into account our
 preference for positive VEVs (which of course is
   an arbitrary choice) and makes sure, by comparing two
 elements, that we have all VEVs in the same quadrant so that we do
 not follow the same VEV twice. We apply the symmetries such that we
 get the largest number of positive VEVs in the upper components of
 the field vector. This means, given an arbitrary field configuration $\vec{\phi}$ we apply the group element $g_i$ that maximizes the  measure $M(g_i\vec{\phi})$ given by
 \begin{align}
 	M(g_i\vec{\phi}) = \left(\vec\theta( g_i\vec{\phi}) \right)_2 = \sum_i^n 2^i \left\{\vec\theta( g_i\vec{\phi}) \right\}_i,
 \end{align}
 where $\vec{\theta}(\vec x) \equiv \{ \vec{\theta}(\vec x)_i =
 \theta(x_i) \}$ is the vectorised Heaviside step function, and the
 subscript $2$ indicates that we should interpret the components of
 the vector as a binary number. It is best to think of this measure as
 mapping the field space on binary numbers. Let us consider two field
 configurations $g_a \vec\phi$ and $g_b \vec\phi$ such 
 that $g_a\vec\phi \neq g_b\vec\phi$. This means that the measures are
 different, i.e.~$M(g_a \vec\phi)\neq M(g_b \vec\phi)$, because they
 have different binary representations. With this we showed that given
 a field configuration $\vec\phi$ there is a single\footnote{This is
   not a injective mapping, in particular if there are zeros in the
   components. In that case we can have two symmetries $g_i$ and $g_j$
   that maximize the measure but then they are equal,
   i.e. $g_i\vec\phi=g_j\vec\phi$.} $g_i\vec\phi$ that maximizes
 $M(g_i\vec\phi)$. We choose the set of field configurations that
 maximize $M(g_i\vec\phi)$ under the symmetries, to be the principal
 quadrant. \s

To give some context to this measure, we apply all group elements for
 an arbitrarily chosen 2HDM field configuration given by
 $\{-10,5,-20,0\}^T$, cf.~Tab.~\ref{table:discretesymmetries}.
 From the second and the sixth row of the table, we
 can conclude that there are two group elements that produce the same
 measure. This is not an issue, however, as both symmetries transform the
 initial field configuration into the same configuration
 $\{10,5,-20,0\}^T$. \s
 
\begin{table}[h!]
\renewcommand{\arraystretch}{2}
\centering
\begin{tblr}{ccc}
\textbf{Group Element} & \textbf{Measure in binary} & \textbf{Measure in decimal} \\\hline
$\begin{psmallmatrix}1 & 0 & 0 & 0\\0 & 1 & 0 & 0\\0 & 0 & 1 & 0\\0 & 0 & 0 & 1\end{psmallmatrix}$     &  $0101$                 &  $5$                  \\ \hline
$\begin{psmallmatrix}-1 & 0 & 0 & 0\\0 & 1 & 0 & 0\\0 & 0 & 1 & 0\\0 & 0 & 0 & 1\end{psmallmatrix}$    &  $1101$                 &  $13$                  \\ \hline
$\begin{psmallmatrix}1 & 0 & 0 & 0\\0 & -1 & 0 & 0\\0 & 0 & -1 & 0\\0 & 0 & 0 & 1\end{psmallmatrix}$   &  $0011$                 &  $3$                  \\ \hline
$\begin{psmallmatrix}1 & 0 & 0 & 0\\0 & 1 & 0 & 0\\0 & 0 & 1 & 0\\0 & 0 & 0 & -1\end{psmallmatrix}$    &  $0101$                 &  $5$                  \\ \hline
$\begin{psmallmatrix}-1 & 0 & 0 & 0\\0 & -1 & 0 & 0\\0 & 0 & -1 & 0\\0 & 0 & 0 & 1\end{psmallmatrix}$  &  $1011$                 &  $11$                  \\ \hline
$\begin{psmallmatrix}-1 & 0 & 0 & 0\\0 & 1 & 0 & 0\\0 & 0 & 1 & 0\\0 & 0 & 0 & -1\end{psmallmatrix}$   &  $1101$                 &  $13$                  \\ \hline
$\begin{psmallmatrix}1 & 0 & 0 & 0\\0 & -1 & 0 & 0\\0 & 0 & -1 & 0\\0 & 0 & 0 & -1\end{psmallmatrix}$  &  $0011$                 &  $3$                  \\ \hline
$\begin{psmallmatrix}-1 & 0 & 0 & 0\\0 & -1 & 0 & 0\\0 & 0 & -1 & 0\\0 & 0 & 0 & -1\end{psmallmatrix}$ &  $1011$                 &  $11$                 
\end{tblr}
\caption{2HDM group elements applied on the initial field
  configuration $\{-10,5,-20,0\}^T$ and the resulting measure
  $M(g_i\vec\phi)$ in binary and decimal numbers. We can see that $\{10,5,-20,0\}^T$ is the field configuration mapped into the principal quadrant.}
\label{table:discretesymmetries}
\end{table}

The method has one caveat. Models with a spontaneously broken
discrete symmetry give rise to domain walls, which are a topological defect
\cite{Zeldovich:1974uw,Kibble:1976sj,Zurek:1985qw}. If domain walls
were to exist they would dominate the energy density of the universe at
some late time \cite{Zeldovich:1974uw,Larsson:1996sp,Press:1989yh} and
be in contradiction with observation, which 
is also known as the domain wall problem. In this case, constraints
would have to be placed on models that can lead to the formation of domain
walls, such that the domain wall domination does not occur
\cite{Lazanu:2015fua}, or at least not until today.\footnote{For
recent works on domain walls in the 2HDM, see \cite{Battye:2011jj,Battye:2020sxy,Law:2021ing, Sassi:2023cqp,Yang:2024bys}.} In our code, we do
not take into account the possible existence of different domains
separated by domain walls. As an approximation, we only consider phase transitions with
  the shortest path between false and true vacuum. The users have to make sure
themselves not to apply models with unphysical domain walls,
respectively, else be aware that the existence of domain walls is not taken
into account by {\tt BMSPTv3}. 
In case of explicitly (softly) broken discrete symmetries the vacuum configurations of
the related minima have different energies, such that domain walls
are unstable and the domain with the higher energy eventually decays
into the lower energy configuration. Such decays can lead to gravitational waves
\cite{Hiramatsu:2010yz}. The 
fact that the domain walls exist by some time, may furthermore influence the
energy of the universe and hence also the cosmological history of the
universe. Again such effects are not described by our code. In
summary, while the impact of topological defects may play an
interesting role in the dynamics of phase transitions and electroweak
baryogenesis, this is beyond the present goal of our code and is left
for future work. 

\subsubsection{Flat Directions}\label{sec:flat_directions}
Multi-dimensional scalar potentials can exhibit flat directions
resulting in an effective sub-dimensional minimisation problem that is
notoriously difficult to be dealt with numerically. In {\tt BSMPTv3} we
identify flat directions, i.e. in the one-dimensional case when the
potential is invariant in one field direction $\omega_i$ with
$\Delta \omega_i \gg \omega_i$,  
\begin{align}
    V(\omega_i, \dots) = V(\omega_i + \Delta \omega_i, \dots)\,,
\end{align}
or in the two-dimensional case when the potential is invariant in
$\omega_i^2+\omega_j^2$, $i\neq j$, with
\begin{align}
    V(\omega_i, \omega_j, \dots) = V(\omega_i + \Delta\omega_i,
  \omega_j + \Delta \omega_j, \dots)\quad\text{with}\quad \sum_{a
  \in \{i,j\}} \omega_a^2 = \sum_{a\in \{i,j\}} (\omega_a + \Delta \omega_a)^2\,,
\end{align}
and in the three-dimensional case, checking for invariance in $\omega_i^2+\omega_j^2+\omega_k^2$, $i\neq j\neq k,\, i\neq k$, analogously to the two-dimensional case above.
In order to catch the largest possible flat dimension first, we check subsequently for three-dimensional, two-dimensional and one-dimensional flat directions.
If an $n-$dimensional flat direction is encountered with
$n\in\{1,2,3\}$, we set the last $(n-1)$ VEV directions in the
model-specific {\tt VevOrder} to zero and use the respective first one 
only to report on found phases and transitions.\footnote{Furthermore, a message is printed on the screen if the corresponding {\tt logginglevel} is enabled with {\tt --logginglevel::mintracerdetailed=true}. More details on useful diagnosing output, managed by all implemented {\tt logginglevels}, can be found in
Sec.~\ref{sec:MinimaTracer}.}

\subsection{The Class {\tt BounceSolution} \label{subsec:bouncesol}}

In the following, we describe the newly implemented algorithm for the
determination of the bounce solution, which is needed for the
computation of the transition probability from a false to a true
vacuum, the characteristic temperatures and the gravitational wave
spectrum. 

\subsubsection{Bounce Equation}
Our starting point is the Lagrangian
\begin{align}
	\mathcal{L} = \frac{1}{2}\left(\partial_\mu \vec{\phi}\right)\left(\partial^\mu \vec{\phi}\right) - V(\vec{\phi})\,,
\end{align}
where $\vec{\phi}$ is the vector of scalar fields of some particular
theory and $V(\vec{\phi})$ is the effective potential. 
As shown by Coleman~\cite{FateOfFalseVacuum} based on the WKB approximation 
developed by Banks, Bender and Wu in~\cite{Banks1973}, the transition
rate per unit volume of the false vacuum $\vec{\phi}_f$ into the true
vacuum $\vec{\phi}_t$ is obtained by 
\begin{align}
	\Gamma(\vec{\phi}_f \rightarrow \vec{\phi}_t) \equiv \Gamma =
  A(T)\,e^{-S_E} \;,
\end{align}
where $S_E$ is the Euclidean action of the classical path given by
\begin{align}
	S_E(T) = \int d\tau d^3x  \left[  \frac{1}{2}\left(\partial_\mu \vec{\phi}\right)\left(\partial^\mu \vec{\phi}\right) + V(\vec{\phi}) \right]\,,
\end{align}
  and $A(T)$ is a temperature-dependent prefactor that will be
  discussed shortly. At $T = 0$, the lowest 
  Euclidean action is $O(4)$-symmetric~\cite{FateOfFalseVacuum},
  therefore, one can make the following change of variables
$\rho = \sqrt{\tau^2 + \vec{x}^2}$ that simplifies the calculation of the action to 
 \begin{align}
	S_4(T) =2\pi^2\int_0^\infty d\rho\, \rho^{3}\left[  \frac{1}{2}\left(\frac{d\vec{\phi}}{d\rho}\right)^2 + V(\vec{\phi}) \right]\,.
\end{align}
At finite $T$, the statistics of bosons is periodic in the imaginary
time $\tau$ direction with period $\frac{1}{T}$. That allows us to 
combine all contributions into an $O(3)$-symmetric Euclidean action,
with the imaginary time integration giving a factor
$\frac{1}{T}$~\cite{LINDE198137}, i.e. we have the replacement $S_4(T)
\rightarrow \frac{S_3(T)}{T}$. Thus, we can make the change of
variables $\rho = \sqrt{\sum_{i\leq3}x_i^2}$, simplifying the action
to 
 \begin{align}
	S_3(T) =4\pi\int_0^\infty d\rho\, \rho^{2}\left[  \frac{1}{2}\left(\frac{d\vec{\phi}}{d\rho}\right)^2 + V(\vec{\phi}) \right]\,.
\end{align}
The Euler-Lagrange equations are given by
\begin{align} 
  \frac{d^2\vec{\phi}}{d\rho^2} +
    \frac{D-1}{\rho}\frac{d\vec{\phi}}{d\rho} = \nabla V(\vec{\phi}) ,&
    & \text{w/ the boundary conditions :} &
    &\vec{\phi}(\rho)\big|_{\rho\rightarrow\infty}= \vec{\phi}_f, & &
                                                                      \frac{d\vec{\phi}}{d\rho}\Big|_{\rho=0}=0\,, 
  \label{eq:euler-lagrange}
  \end{align}
  where $D = 4$ at zero temperature and $D = 3$ at finite
  temperature. The boundary conditions state that far away from the
  true vacuum bubble the false vacuum remains undisturbed, and that the
  transition happens at $\rho = 0$, which can be chosen without loss of
  generality. For $D = 3$, the prefactor $A(T)$ can be well approximated
  \cite{LINDE198137,FateOfFalseVacuumII} by 
  \begin{align}
A(T) \simeq
T^4\left(\frac{S_3}{2\pi T}\right)^\frac{3}{2}\quad &\text{if} \quad  T > 0\,.
\end{align}

We can cast the two expressions of the spherically symmetric action,
$S_3$ and $S_4$, into a single expression ($D=3,4$)
\begin{align}
	S_D(T) = A_{D}\int_0^\infty  d\rho\, \rho^{D-1}\left[  \frac{1}{2}\left(\frac{d\vec{\phi}}{d\rho}\right)^2 + V(\vec{\phi}) \right]=S^K_D(T) + S^V_D(T) \label{eq:action_kinetic_potential}\,,
\end{align}
where $A_{D}$ denotes the area of the $D$-dimensional unitary sphere,
and $S^K_D(T)$ and $S^V_D(T)$ are the contributions to the action
coming from $\frac{d\vec{\phi}}{d\rho}$ and $V(\vec{\phi})$,
respectively. We can draw a relation between these two contributions
to the action~\cite{Derrick:1964ww}. To see that, let us assume that
$\vec{\phi}(\rho)$ is a solution to the bounce equation. Making the
ansatz $\vec{\phi}(\lambda \rho)$ for the solution, where $\lambda$ is a
real number, the action can be written as 
\begin{align}
	S_D(T) =S^K_D(T) \lambda^{2-D} + S^V_D(T) \lambda^{-D}\,.
\end{align}
The action must be stationary at $\lambda=1$ so that we must have 
\begin{align}
	\frac{d S_D(T)}{d\lambda}\Bigg|_{\lambda=1} = (2-D) S^K_D(T) -
  D S^V_D(T)= 0 \implies S^K_D(T) &= \frac{D}{2-D} S^V_D(T) \,,
\end{align}
which allows us to write the action as
\begin{align}
	 S_D(T) &= S^K_D(T) + S^V_D(T) \nonumber \\
	 &= \frac{2}{D} S^K_D(T)\label{eq:action_kinetic} \\
	 &= \frac{2}{2-D} S^V_D(T)\label{eq:action_potential}\,.
\end{align}
This will be later used as a consistency check of the results: We
  compute $S_D^K(T)$ and $S_D^V(T)$, the kinetic and
  potential part of the action, respectively, and use
  Eqs.~(\ref{eq:action_kinetic}) and (\ref{eq:action_potential}) to
  verify that they are consistent with the original expression for the action,
  Eq.~(\ref{eq:action_kinetic_potential}). Let us
  note also that we assume that the finite-temperature transition rate
  is the dominant contribution for phase transitions taking place at
  finite temperature. We neglect zero-temperature contributions to the
  tunnelling rate. \s

The solution of Eq.~(\ref{eq:euler-lagrange}) is highly
non-trivial. Before diving into its detailed derivation, let us first
discuss if the differential equation has a solution or 
  not. In \cite{FateOfFalseVacuum} it was shown that the one-dimensional
  version of the bounce action always has a solution. We shortly repeat
  this proof. If we take a look at Eq.~(\ref{eq:euler-lagrange}), we can see that it
  resembles the equation of motion of a particle in an upturned
  potential that starts at rest in a position $\phi_0$ at $\rho = 0$ and
  ends up at the false vacuum $\phi_f$ in the limit of $\rho \rightarrow
  \infty$. The solution of the bounce equation can be uniquely
  characterised by the starting position $\phi_0$ so, to complete the
  proof, we must show that there is a starting position that makes the
  system end up at $\phi_f$. \s

  \begin{figure}[h]
    \centering
    \includegraphics[width=0.6\textwidth]{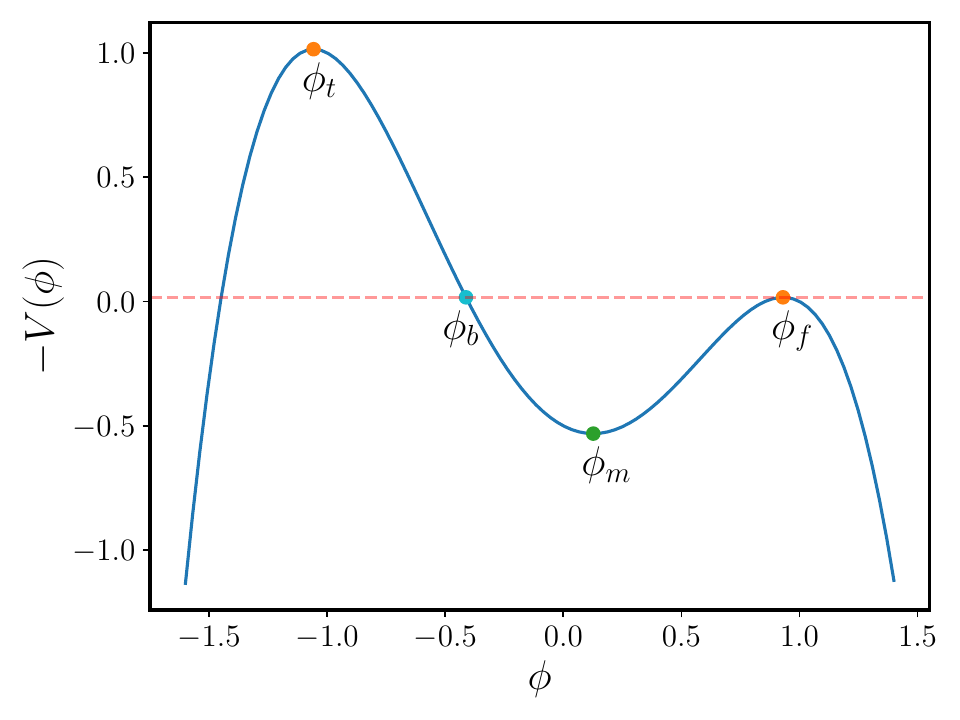}
    \caption{Plot of a generic upturned potential. The orange points $\phi_f$ and $\phi_t$ are the false and true vacuum, respectively, the green point $\phi_m$ is the potential barrier, the cyan point $\phi_b$ is the point with the same potential value as the false vacuum, the dashed red line is the potential value of the false vacuum.}
    \label{fig:1dinvertedpotential}
    \end{figure}

In the particle analogy, the variation of the energy as a function of
$\rho$ is given by \cite{FateOfFalseVacuum}
\begin{align}
	\frac{d}{d\rho}\left[
  \frac{1}{2}\left(\frac{d\phi}{d\rho}\right)^2- V(\phi) \right] =
  -\frac{D-1}{\rho}\left(\frac{d\phi}{d\rho}\right)^2 \leq 0 \;,
\end{align}
which means that energy is lost due to the drag term. For this reason,
if the starting position $\phi_0$ is between $\phi_b$ and $\phi_f$ in
Fig.~\ref{fig:1dinvertedpotential}, then it will never have enough
energy to reach $\phi_f$. We call this an undershoot. In the case of
thin-walled situations, the particle stays close to $\phi_t$ across a
large range of $\rho$. In 
this case, we can neglect the drag term, as it is proportional to
$\rho^{-1}$, making the particle go over $\phi_f$ without stopping. We
call this an overshoot. Therefore, by continuity, between these two
situations, there must exist a $\rho_0$ which solves the ODE, as to
one side we overshoot and to the other we undershoot.  \s

The bounce equation (\ref{eq:euler-lagrange}) can only be solved 
numerically, as will be described below, except for
  $\rho = 0$. Here an analytical approach is required,  because of the 
$\frac{D-1}{\rho}\frac{d\vec{\phi}}{d\rho}$ term. This term has an 
$\infty \times 0$ indetermination. Another benefit of analytically
integrating at $\rho = 0$ is
that it allows for a quick integration over very thin-walled solutions,
i.e. solutions that stay close to $\phi_t$ across a large range of
$\rho$. \s

From now on, we use a different parametrisation
for the equations of motion, similar to the one described in Refs.~\cite{Wainwright:2011kj,PhysRevD.60.105035}. We parameterize the tunnelling path as
\begin{align}
	\vec{\phi}(\rho) \rightarrow \vec{\phi}(l(\rho))\,,
\end{align}
where we impose unitary velocity, i.e.
\begin{align}
	\left| \frac{d\vec{\phi}(l)}{dl} \right| = 1\,,
\end{align}
which allows us to interpret $l$ as length along the tunnelling
path. This parametrisation simplifies the Euler-Lagrange
equation (\ref{eq:euler-lagrange}) into
\begin{align}
\frac{d^2l}{d\rho^2} + \frac{D-1}{\rho} \frac{dl}{d\rho} & =
                                                           \frac{dV(\vec{\phi})}{dl}\,,\label{eq:1D-equation} 
\end{align}
with the boundary conditions
\begin{align}
	\frac{dl}{d\rho}\Big|_{\rho = 0}= 0 \;, \quad l(\rho = \infty)
  = l_f \,,
\end{align}
where $l_f$ is the total length of the tunnelling path. \s

The analytical solutions are found by approximating the potential by a
quadratic function and writing it as a function of the spline
parameter $l$, with the starting position $l_0\equiv l(\rho=0)$ and
the spline parameter of the true vacuum $l_t$.
We made one further assumption, namely that the
smallest bounce solution is monotonic in $\rho$. In the particle
analogy this means that it never moves backwards, equivalent to
$\frac{dl}{d\rho} \geq0$. The solutions at $\rho = 0$ depend on
$D$, and we are going to provide the solutions for the $D =
\{3,\,4\}$ cases specifically. We distinguish between the
  following solutions:

  \begin{itemize}
    \item \textbf{Solutions starting near the true minimum}\\
    In this scenario we start very close to the true vacuum at a  position
    $l_t$. The potential is then approximated by $V(l) \approx
    \frac{1}{2}H (l-l_t)^2$, where $H \equiv
    \frac{d^2V}{dl^2}\big|_{\rho=0}$. The solutions start at
    $l(\rho=0) = l_0$ and are given by 
    \begin{align}
    l(\rho) = \begin{cases}
    l_t-\frac{(l_t - l_0) \sinh(\rho\sqrt{H})}{\rho\sqrt{H}}\quad &\text{for} \quad D = 3\\
    l_t-\frac{2(l_t - l_0) I_1\left(\rho\sqrt{H} \right)}{\rho\sqrt{H}}\quad &\text{for} \quad D = 4
    \end{cases} \;,
    \label{eq:thinwalled1d}
    \end{align}
    where $\sinh(x)$ is the hyperbolic sine function and $I_1(x)$ is the
    modified Bessel function of the first kind. 
\item \textbf{Solution starting where $\mathbf{H>0}$}\\
    In this scenario we still start near a local minimum but with a non-negligible 
    potential gradient $G \equiv \frac{dV}{dl}\big|_{\rho = 0}$,
    which is taken to be positive to ensure the assumption that
    $\frac{dl}{d\rho} \geq0$. Therefore, the potential is approximated by
    $V(l)\approx G(l-l_0) + \frac{1}{2}H (l-l_0)^2$ and the solutions are
    given by 
    \begin{align}
    l(\rho) = \begin{cases}
    l_0 - \frac{G}{H} + \frac{G \sinh(\rho \sqrt{H})}{\rho H^\frac{3}{2}}\quad &\text{for} \quad D = 3\\
    l_0 - \frac{G}{H} + \frac{2\,G I_1(\rho \sqrt{H})}{\rho H^\frac{3}{2}}\quad &\text{for} \quad D = 4
    \end{cases} \;.
    \label{eq:thickwalled1d}
    \end{align}
\item \textbf{Solution starting where $\mathbf{H<0}$}\\
    In this scenario we start near the potential barrier with a non-zero gradient $G \equiv \frac{dV}{dl}\big|_{\rho = 0}$, which is taken to be positive to ensure the assumption that $\frac{dl}{d\rho} \geq0$. The potential is approximated by $V(l)\approx G(l-l_0) + \frac{1}{2}H (l-l_0)^2$ and the solutions are given by
    \begin{align}
    l(\rho) = \begin{cases}
    l_0 - \frac{G}{H} - \frac{G \sin(\rho \sqrt{-H})}{\rho (-H)^\frac{3}{2}}\quad &\text{for} \quad D = 3\\
    l_0 - \frac{G}{H} - \frac{2\,G J_1(\rho \sqrt{-H})}{\rho (-H)^\frac{3}{2}}\quad &\text{for} \quad D = 4
    \end{cases} \;.
    \label{eq:ultrathickwalled1d}
    \end{align}
The $J_1(x)$ is the Bessel function of the first kind. 
  \end{itemize}
There are two analytical solutions that can be used when $H>0$,
Eq. ~(\ref{eq:thinwalled1d}) and Eq.~(\ref{eq:thickwalled1d}). To have
a smooth transition between these two branches, we first calculate the
branch switching point $l_\text{threshold}$ which minimizes the
relative error on $\rho$  between the two expressions. Secondly, given
a starting point $l_0$, we compute both solutions and combine them in
a logistic function weighted average based on the distance to
$l_\text{threshold}$, such that the transition between the two branches is continuous.
  
\subsubsection{Numerical Solution of the Bounce Equation}
The bounce equation that we need to solve to find the tunnelling rate
has no analytical solution for a generic potential $V(\vec{\phi})$,
so that we have to employ numerical methods to find an
approximate solution, which we will sketch in the following. \s

The ordinary differential equation (ODE), presented in the last
section, is difficult to solve. The main issue is that the boundary
conditions are not applied at the same point in $\rho$ and, since we
are working with the inverted potential, i.e.~with a potential with the
opposite sign, the integration of the ODE is highly 
susceptible to small changes in
$\vec{\phi}_0\equiv\vec{\phi}(\rho=0)$, which is exactly what  
we want to find. Once $\vec{\phi}_0$ is known, it is trivial to
integrate the ODE. There is some bias for $\vec{\phi}_0$ to be around
$\vec{\phi}_t$ but, apart from that, $\vec{\phi}_0$ lives in a space
with the same number of dimensions as the number of fields that can
acquire a VEV. From a computational point of view, this problem is
very difficult to solve since, as mentioned above, small
variations of the starting position $\vec\phi_0$ will produce very
different solutions to the bounce equation. For this reason, trying to
guess the initial position is very inefficient, requiring an
alternative approach. \s

The approach of
  \texttt{CosmoTransitions}~\cite{Wainwright:2011kj} is to start with a
  guess path, a straight line between $\vec{\phi}_t$ and
  $\vec{\phi}_f$, and solve Eq.~(\ref{eq:1D-equation}), which is
  computationally feasible. Then one deforms the path such
  that
\beq  
\frac{d^2\vec{\phi}}{dl^2}\left(\frac{dl}{d\rho} \right)^2 &= \nabla_\perp V(\vec{\phi})\label{eq:perpendicular_forces}\,,
\eeq
where
\begin{align}
\nabla_\perp V(\vec{\phi}) = \nabla V(\vec{\phi}) - \left(\nabla V(\vec{\phi})\cdot \frac{d\vec{\phi}}{dl}\right)\frac{d\vec{\phi}}{dl}\,,
	\end{align}
is fulfilled. This equation imposes that the curvature of the
  path matches the perpendicular forces originating from the
  potential. This is done
  by selecting points on the guess path and applying a small (rescaled)
  force to them,
\begin{align}
	\vec{N} =
  \frac{d^2\vec{\phi}}{d^2l}\left(\frac{dl}{d\rho}\right)^2-\nabla_\perp
  V(\vec{\phi})\,,  \label{eq:nforce}
\end{align} 
until they converge and this produces a new, hopefully better,
path. Then one goes back to Eq.~(\ref{eq:1D-equation}), solves it, and
deforms the path again. This process is repeated 
until convergence is achieved, i.e. $\vec{N} = 0$. 
\s

  While there are numerous methods to
  numerically solve the problem (cf.~e.g.~the algorithm of the codes
  described Sec.~\ref{sec:stateoftheart} as well as \cite{Bardsley:2021lmq} and references
  [28-36] therein), we chose in {\tt BSMPTv3} an algorithm fulfilling 
  our needs, that is very
  similar to the one used by {\tt CosmoTransitions} described here
  above, but with a few differences. \texttt{CosmoTransitions} relies on $B$-splines
  to describe the tunnelling path, whereas we use cubic-splines with
  not-a-knot boundary conditions to describe
  it. This allows for a more general tunnelling path as $B$-splines
  have a finite resolution. During path deformation, however, we also use
  $B$-splines to remove slight numerical instabilities.
  The step size for the path deformation is
  calculated in the same way as in {\tt CosmoTransitions}. i.e. we
  start with a small step size. If we are constantly deforming in the
  same direction we increase the step size; if the direction inverts
  we reduce the step size. 
For the computation of the relevant thermal quantities, it is
necessary to have the Euclidean action $S_3/T$ as a
function of the temperature for a wide temperature range. For this reason,
after solving the bounce equation for a given temperature $T$ we
solve the bounce equation for a temperature $T+\delta T$
by slightly warping the path from the solution previously found. This
increases the reliability of the computation and considerably decreases the computational time. \s

 To find the
solution we apply bisection until the desired resolution is
reached. Here, we initially perform the binary search on a linear scale but if a thin-walled solution is detected we switch to a log scale. Note, however, that the solution needs not be unique. Furthermore, for
more than one field a solution does not always exist
\cite{Mukhanov:2021kat}. This does not mean that the false vacuum is
stable, but rather that the decay rate must be computed differently
\cite{Espinosa:2019hbm,Mukhanov:2020pau,Mukhanov:2020wim,Espinosa:2021qeo}. \s

After providing the one-dimensional solution for the bounce equation
using the overshoot/undershoot method as described above, we 
deform the path with a force perpendicular to the path and
proportional to $\vec{N}$, cf.~Eq.~(\ref{eq:nforce}). 
Since the normal force depends on the one-dimensional solution, it
only makes sense  to deform the path
from the initial position at 
$\vec{\phi}_0 = \vec{\phi}(\rho = 0)$ up until $\vec{\phi}_f(\rho =
\infty)$, so that the path from $\vec{\phi}_t$ to $\vec{\phi}_0$ is
thrown away during each path deformation 
iteration. This can be problematic because we cannot guarantee that
the new path can solve the bounce equation. If e.g.~the new path is
longer or has a steeper starting position, then it can dissipate more
energy than the previous iteration, making it impossible to solve the
one-dimensional bounce equation. For this reason, it is necessary to
add path to the beginning of the current guess. We do that by using
the spline to extrapolate the beginning of the path until we reach
through this backwards propagation (`bp') a
point $\vec{\phi}_{\text{bp}}$ with
$\frac{dV\left(\vec{\phi}(l)\right)}{dl}\Big|_{\vec{\phi}=\vec{\phi}_{\text{bp}}}=0$. This
condition ensures that, since $-V(\vec{\phi}_{\text{bp}}) > -V(\phi_f)$,
the one-dimensional bounce equation has a solution for the same
reasons presented above. 

\subsubsection{Characteristic Temperature
  Scales}\label{sec:characteristic_temperature_scales}
 
The effective potential and hence also the vacuum decay rates are
temperature-dependent. Depending on the temperature, the transition
rate may eventually become large enough to make the
universe go from one to the other vacuum with the cosmological FOPT
taking place in a certain temperature interval.
It is hence useful to consider certain characteristic moments of the
transition as the decay rate grows, which, instead of time, is done
using the temperature of the universe. We will present here these
characteristic temperatures and how they are 
calculated in {\tt BSMPTv3}. They will be used in the calculation of the
spectrum of the gravitational waves presented below. 
\begin{itemize}
\item \textbf{Critical Temperature} - ${T_c}$ - This is the
  temperature, where the effective potential has two
  degenerate minima and, consequently, the transition from the false
  vacuum to the true vacuum may start via quantum tunnelling.
\item \textbf{Nucleation Temperature}  - ${T_n}$ - This is the temperature at which
the tunnelling decay rate per Hubble volume matches the Hubble rate, 
\begin{align}
    \frac{\Gamma(T_n)}{H^4(T_n)} = 1 \,, \label{eq:nucl_exact}
\end{align}
which can be further approximated as (cf.~e.g.~\cite{Wainwright:2011kj})
\begin{align}
    \frac{S_3(T_n)}{T_n} \sim 140\,.\label{eq:nucl_approx}
\end{align}
We note that {\tt BSMPTv3} calculates and outputs the nucleation
temperature calculated via Eq.~(\ref{eq:nucl_exact}) as well as
Eq.~(\ref{eq:nucl_approx}) in order to compare to e.g. {\tt
  CosmoTransitions} which uses this approximation. 
\item \textbf{Percolation Temperature} - ${T_p}$ - This is the
    temperature at which at least 29\% of the false vacuum has
  tunnelled into the true vacuum or, equivalently, the probability of
  finding a point still in the false vacuum is \SI{71}{\%}
  \cite{LorenzZiff:2001,LIN2018299,LI2020112815}. This condition 
  imposes that at the percolation temperature there is a large
  connected structure of true vacuum that spans the whole universe,
  and that is stable and cannot collapse back to the false
  vacuum. This large structure is known as the \textit{percolating
    cluster}. 
The probability of finding a point in the false vacuum is given by  
\begin{align}
    P_p = P(T= T_p) = e^{-I(T= T_p)} = \epsilon_{\text{p}}, & & I(T) = \frac{4\pi v_w^3}{3}
\int_T^{T_c} \frac{\Gamma(T')dT'}{T'^4 H(T')}\left(\int_T^{T'}\frac{d\tilde{T}}{H(\tilde{T})}\right)^3\label{eq:perc} \;,
\end{align}
where $v_w$ denotes the wall velocity.
To find the percolation temperature one has to solve $I(T_p) = 0.34$
or, equivalently, $P(T_p) = 0.71$. This is the default set in {\tt
    BSMPTv3}. The user has the possibility, however, to set
  $\epsilon_p$ through the input.
\item \textbf{Completion Temperature} - ${T_f}$ - This is the temperature, at
  which the transition completes, no finite regions in the universe in
  the false vacuum are left. It is obtained from demanding the
  probability of finding a false vacuum to be
  \begin{align}
    P_f = P(T = T_f) = \epsilon_f \;, \label{eq:compl}
  \end{align}
  with the default setting $\epsilon_f=0.01$. Again, the user has the
  possibility to choose a different value of $\epsilon_f$ through the input.
\end{itemize}

\subsection{The Class {\tt GravitationalWave} \label{subsec:GravWave}}
In this class both the gravitational wave spectrum as well as the
signal-to-noise ratio at the Laser Interferometer Space Antenna
(\texttt{LISA}) \cite{Caprini:2019egz,LISA,LISACosmologyWorkingGroup:2022jok}
are calculated as will be described in the following.

\subsubsection{Gravitational Wave Spectrum}\label{sec:GW_spectrum}
Thermal parameters, like the transition temperature, the transition
strength, the characteristic length scale, the bubble wall velocity,
not only characterise FOPTs, they are also relevant for the
gravitational wave predictions. Thermal parameters are evaluated,
however, only at a single temperature. Gravitational waves on the
other hand are produced during the whole phase transition. The question is hence,
which temperature to apply when evaluating the thermal parameters used
in gravitational wave predictions. In the following, we generically
call this transition temperature and denote it by $T_*$. A common choice
for $T_*$ is the nucleation temperature $T_n$. Since GWs originate from
bubble collisions and sound shells and the following turbulence, it
might be more appropriate to chose the percolation temperature
$T_p$. Another choice might be the completion temperature. In a plasma
reheating to a homogeneous temperature $T_{\text{reh}}$ after the
completion of the transition, the redshift
of the GWs should be calculated from $T_{\text{reh}}$, see Eq.~\eqref{eq:Treh} below, instead from $T_*$
\cite{Cai:2017tmh}, but not the characteristic length scale and the
energy available for the production of gravitational waves, as they
take place before reheating. 
In the following, we will give the formulae for the relevant thermal
parameters of the gravitational waves at the transition temperature
$T_*$. In \texttt{BSMPTv3} the default setting is
\begin{eqnarray}
\mathrm{Default\, in\, \texttt{BSMPTv3}:} \; T_* = T_p \;.
\end{eqnarray}
The user can also choose other settings using the input
flags. \s

The second key parameter is the strength of the phase transition,
which is measured by the parameter $\alpha$.
It can be, and most commonly is, defined by the trace
anomaly~\cite{Hindmarsh:2015qta,Hindmarsh:2017gnf} as 
\begin{equation}
    \alpha = \frac{1}{\rho_\gamma(T_*)} \Big[ V(\vec{\phi}_f) - V(\vec{\phi}_t) - \dfrac{T}{4} \Big( \frac{\partial V(\vec{\phi}_f)}{\partial T} - 
    \frac{\partial V(\vec{\phi}_t )}{\partial T} \Big) \Big]_{T = T_*}\,,
\label{alpha}
\end{equation}
where $\rho_\gamma(T_*)$ is the energy density of a radiation dominated
universe at $T_*$, written as a function of the effective number of relativistic degrees
of freedom $g(T)$, which we derive as a
  function of the temperature in App.~\ref{sec:gstar}.
The $\rho_\gamma$ is given by 
\begin{equation}
    \rho_\gamma(T) = g(T) \frac{\pi^2}{30} T^4 \;.\label{eq:rho_gamma}
\end{equation}
The parameter $\alpha$ measures the energy budget available for the
production of GWs. A common classification is that $\alpha \sim {\cal
  O}(0.01)$ corresponds to weak transitions, $\alpha \sim {\cal
  O}(0.1)$ to intermediate transitions, and $\alpha \gsim {\cal O}
(1)$ to strong transitions. Strong phase transitions not only may source preferrably GWs through bubble collisions, but may also lead to an early
onset of turbulence
\cite{Caprini:2015zlo,Hindmarsh:2015qta,Hindmarsh:2017gnf,Ellis:2018mja}, which we account for by introducing the GW source lifetime factor $\Upsilon$, see Eq.~\eqref{eq:lifetime_factor} below. Note,
that some studies use in the nominator of the right-hand side of Eq.~(\ref{alpha}) 
the latent heat released during the transition, which does not have
the factor 1/4 in the second term. The definition used here, has been
shown to describe the energy budget better, however, 
\cite{Giese:2020rtr}. \s

We take the occasion to comment on the notion of
  the strength of the 
phase transition. One of the three Sakharov conditions
\cite{Sakharov:1967dj} for successful electroweak baryogenesis is a strong
first-order EWPT. It is quantified by the sphaleron suppression
criterion which classifies a PT as of strong first-order if the ratio $\xi_c$
of the critical VEV $v_c$ and the critical temperature $T_c$ is larger
than one \cite{Quiros:1994dr,Moore:1998swa},
\begin{align}
\xi_c = \frac{v_c}{T_c} > 1 \;.
\end{align}
Here $v_c$ is the vacuum expectation value at the critical temperature
$T_c$ taking into account only the doublet VEV values. Non-zero
singlet VEVs in models where they are available, are not
included here. There is hence some ambiguity in the definition of the
strength of the PT, so that it has to be made clear, where necessary,
which definition is used. Note, also that a value of $\xi_c > 1$ does not
guarantee that an SFOPT actually takes place. The universe could also be
trapped in the wrong vacuum,
cf.~e.g.~\cite{Biekotter:2021ysx,Biekotter:2022kgf,Baum:2020vfl}. This
can only be decided by 
applying the criterion for the nucleation temperature,
Eq.~(\ref{eq:nucl_exact}). \s

The third important thermal parameter is the inverse
duration of the phase transition in Hubble units, denoted as
$\beta/H$, and defined as (cf.~e.g.~\cite{Ellis:2018mja})
\begin{equation}
\frac{\beta}{H_*} = T_*  \left. \frac{d}{d T} \left(\frac{S_3(T)}{T}\right) \right|_{T_*}\,,
\label{betaH}
\end{equation}
with $H_*$ being the Hubble rate at the transition temperature $T_*$, $H_* =
  H(T=T_*)$. \s

The Hubble rate as a function of temperature is derived taking into
account contributions from the radiation energy density
$\rho_\gamma$, as well as from the difference in the vacuum energy
between the false and the true vacuum, $\Delta V = V_\mathrm{false} - V_\mathrm{true}$, cf.~\cite{Kierkla:2022odc,Ellis:2022lft,Goncalves:2024lrk},
\begin{align}
    H^2(T) = \frac{1}{3 \widetilde{M}_\mathrm{Pl}^2} \left(\rho_\gamma(T) + \Delta V(T)\right) = \frac{1}{3\widetilde{M}_\mathrm{Pl}^2} \left(\frac{g(T) \pi^2}{30}T^4 + \Delta V(T)\right)\,,
\end{align}
where $\widetilde{M}_{Pl} \approx 2.4\cdot 10^{18}\,\text{GeV}$ is
the reduced Planck mass. The inverse time scale $\beta/H$ is
  obtained from a linear approximation of the action in time,
  i.e. $S_3(t)/T(t)\approx S_3(t_*)/T(t_*) - \beta (t-t_*)$, assuming an
  adiabatic expansion of the universe with $dT/dt = -T
  H(T)$~\cite{Athron:2023xlk}, which leads to an additional
  factor $H_* T_*$. This expansion allows us to write the tunnelling rate as 
\begin{align}
\Gamma \approx e^{-\frac{S_3(t_*)}{T(t_*)}+\beta (t-t_*)} =
  e^{-\frac{S_3(T_*)}{T_*}-\frac{\beta}{H_* T_*} (T-T_*)}. 
\end{align}
Although there is no physical reason to impose any lower bound on
$\beta/H_*$, it has been argued in
Refs.~\cite{Athron:2023xlk,Giblin2014TheDO,Konstandin:2011dr} that
$\beta/H_* < 1$ would constitute a GW wavelength which is larger than
the Hubble-horizon and would lead to difficulties with causality
bounds on the amplitude~\cite{Athron:2023xlk,Caprini:2006jb}. We therefore
only consider GW signals with $\beta/H_* \ge 1$ to be
realistic. Furthermore, within our internal testing we even found GWs
point with negative $\beta/H_*$ which are not necessarily a
mistake. These points might have an action that plateaus around $140$
and the percolation temperature is only reached when the action is
increasing again as the temperature lowers.
For these cases, we still consider that the universe transitions
into a different phase, but do not compute the gravitational waves
spectrum. In such an instance, the characteristic time scale of the
transition would have to be determined differently, which is
beyond the current scope of the code. 
\s
 
The fourth parameter that determines the GW spectrum is the bubble
  wall velocity $v_w$. This parameter is extremely complicated to
determine. The bubble walls start at rest and accelerates due to the
difference of pressure between the phases, so that the wall velocity
is a time-dependent quantity. The
wall velocity is subject to a lot of ongoing activity and (also
controversial) discussions.\footnote{For a recent summary of the
  discussion, cf.~\cite{Ai:2023see}, which provides a model-independent
  determination of bubble wall velocities in local thermal equilibrium.} We therefore treat the (terminal) wall
velocity as input parameter. If it is not given by the user it is by
default set to
\begin{eqnarray}
\mbox{default wall velocity: }  v_w = 0.95 \;.
\end{eqnarray}  
Furthermore, the user can choose to
use the approximate expression ($\alpha$ and $\rho_\gamma$ evaluated
at $T_*$)~\cite{Lewicki2022,Ellis2023} 
\begin{align}
	v_w \simeq \begin{cases}
        \sqrt{\frac{\Delta V}{\alpha \rho_\gamma}} \quad &\text{if} \quad \sqrt{\frac{\Delta V}{\alpha \rho_\gamma}} < v_\mathrm{CJ}\,,\\
        1 \quad &\text{if} \quad \sqrt{\frac{\Delta V}{\alpha \rho_\gamma}} > v_\mathrm{CJ}\,,\label{eq:vwallapprox}
	\end{cases}
\end{align}
or the formula given in~\cite{Ai:2023see},
\begin{align}
  v_w=\left(\left|\frac{3 \alpha+\Psi-1}{2\left(2-3 \Psi+\Psi^3\right)}\right|^{\frac{p}{2}}+\left|v_{\mathrm{CJ}}\left(1-a \frac{(1-\Psi)^b}{\alpha}\right)\right|^{\frac{p}{2}}\right)^{\frac{1}{p}}\,,\label{eq:vwallupperbound}
\end{align}
where $a = 0.2233,\, b = 1.704
,\, p = -3.433$ are numerically fitted values, $\Psi =
\omega_t/\omega_f$ is the ratio of enthalpies $\omega_i = - T
\frac{dV(\vec{\phi}_i)}{dT}$ at the transition temperature $T_*$ and
$v_\mathrm{CJ}$ is the Chapman-Jouguet velocity, for general sound speeds $c_s$ defined as~\cite{Giese:2020rtr}
\begin{equation}
    v_\mathrm{CJ} = \dfrac{1 + \sqrt{3 \alpha\left(1 - c_{s,\mathrm{f}}^2 + 3 c_{s,\mathrm{f}}^2 \alpha\right)}}{1/c_{s,\mathrm{f}} + 3 c_{s,\mathrm{f}} \alpha}\,,\label{eq:vJ}
\end{equation}
where $c_{s,\mathrm{f}}$ denotes the sound speed in the false vacuum.
The sound speeds in the true and false vacuum, respectively, are calculated as~\cite{Giese:2020znk}
\begin{align}
    c^2_{s,\mathrm{t}} &=
               \frac{1}{T_*}\frac{V'(\vec\omega_\mathrm{t},T_*)}{V''(\vec\omega_\mathrm{t},T_*)}
    \;, & c^2_{s,\mathrm{f}} &= \frac{1}{T_*}\frac{V'(\vec\omega_\mathrm{f},T_*)}{V''(\vec\omega_\mathrm{f},T_*)}\label{eq:sound_speed}
\end{align}
where the prime denotes the partial derivative with
  respect to the temperature. In the derivation of both expressions for
  the velocity, the authors assumed local thermal equilibrium, and for
  the result (\ref{eq:vwallapprox}) a constant temperature accross the bubble was
  assumed. As argued by the authors of \cite{Ai:2023see}, their result can be
  interpreted as an upper bound on the wall velocity, as
  out-of-equilibrium effects always slow down the wall velocity
  compared to the equilibrium case. Alternatively, the user can
  calculate the bubble wall velocity with the recently published
code {\tt WallGo} \cite{Ekstedt:2024fyq}.
\s

The primordial GW signals produced in such violent out-of-equilibrium
cosmological processes as given by the FOPTs, are redshifted by the cosmological
expansion and look today like a cosmic gravitational stochastic
background. The corresponding power
spectrum~\cite{Grojean:2006bp,Leitao:2015fmj,Caprini:2001nb,Figueroa:2012kw,Hindmarsh:2016lnk}
of the GW is given by 
\begin{equation}
h^2 \Omega_{\rm GW}(f) \equiv \frac{h^2}{\rho_c} \frac{\partial \rho_{\rm GW}}{\partial \log f}\,,
\end{equation}
where $\rho_c$ is the critical energy density today and $h= 0.674 \pm
0.005$ is the reduced Hubble constant~\cite{planck2018}. It can be split
into three contributions,
\begin{align}
    h^2\Omega_{\rm GW}(f) = h^2\Omega_{\rm coll}(f) + h^2\Omega_{\rm
    sw}(f) + h^2\Omega_{\rm turb}(f) \,.
\end{align}
In our analysis we consider GWs originating from the collision of bubbles and highly relativistic fluid shells (coll), as well as from sound/shock waves
(sw) and from magnetohydrodynamic turbulence (turb)
\cite{Caprini:2019egz,Caprini:2024hue} which are generated by breaking the spherical
symmetry through the process of rapid expansion of the bubble wall in, 
and especially through its interaction with, the surrounding plasma in the
early universe. For non-runaway bubbles with $\alpha<1$, the shock wave is the
contribution that dominates the GW spectrum \cite{Caprini:2019egz}.  
\s

We implement the spectra of \cite{Caprini:2024hue} for gravitational waves sourced by first-order phase transitions.
The spectrum for gravitational waves sourced by the collision of
bubbles and by highly relativistic fluid shells, is modelled by a broken
power law \cite{Caprini:2024hue} as
\begin{align}
    \Omega_\mathrm{coll}(f) = \Omega^\mathrm{coll}_b \left(\frac{f}{f^\mathrm{coll}_b}\right)^{n_1}\left[\frac{1}{2} + \frac{1}{2}\left(\frac{f}{f^\mathrm{coll}_b}\right)^{a_1}\right]^\frac{n_2-n_1}{a_1}\,,
\end{align}
with $n_1=\num{2.4}$, $n_2=\num{-2.4}$, $a_1=\num{1.2}$ and the
amplitude and characteristic frequency given by \cite{Lewicki:2022pdb,Caprini:2024hue}
\begin{align}
    \Omega_b^\mathrm{coll} &\simeq \num{0.05} \, F_{\mathrm{GW},0}  \, K_\mathrm{coll}^2 \left(\frac{\beta}{H(T_*)}\right)^{-2}\,,\label{eq:Omegab}\\
    f_b^\mathrm{coll} &\simeq \num{.11} \, H_{*,0} \frac{\beta}{H(T_*)}\,.\label{eq:fb}
\end{align}
The redshift of the Hubble rate, $H_{*,0}$, and the redshift of the fractional energy density
$K_\mathrm{coll}$, $F_{\mathrm{GW},0}$, are expressed as \cite{Caprini:2024hue}
\begin{align}
    H_{*,0} &= \SI{1.65e-5}{Hz} \left(\frac{g_*}{100}\right)^{1/6}\left(\frac{T_\mathrm{reh}}{\SI{100}{GeV}}\right)\,,\\
    h^2\,F_{\mathrm{GW},0} &= \num{1.64e-5}\left(\frac{100}{g_*}\right)^{1/3}\,,
\end{align}
with the \textit{reheating} temperature $T_\mathrm{reh}\geq T_*$ approximated by
\begin{align}
    T_\mathrm{reh} = \begin{cases} T_* & ,\,\alpha(T_*) < 1 \\  T_* \left[ 1 + \alpha(T_*) \right]^{1/4}& ,\, \mathrm{else} \end{cases}\,,\label{eq:Treh}
\end{align}
assuming that the universe enters radiation domination almost instantaneously, cf.~\cite{Goncalves:2024lrk}.
Note, that this simplified prescription is only valid for large enough
bubble wall velocities in the case of strong enough phase transitions, cf.~\cite{Athron:2023xlk} and references within. 
We therefore print a warning if for $\alpha(T_*) > 1$, $v_w < v_\mathrm{CJ}$, with $v_\mathrm{CJ}$ as defined in Eq.~\eqref{eq:vJ}.
In contrast to \cite{Caprini:2024hue}, we introduce the
model-dependent efficiency factor $\kappa_\mathrm{coll}$ of
\cite{Ellis:2019oqb}, as also used in \cite{Goncalves:2024lrk}, in the
definition of the fractional energy density, $K_\mathrm{coll}$, of the collsion GW source
\begin{align}
    K_\mathrm{coll}=\frac{\kappa_\mathrm{coll}\,\alpha}{1+\alpha}\,. \label{eq:enfraction}
\end{align}
The efficiency factor $\kappa_{\mathrm{coll}}$ for
bubble collisions is defined as the fraction of the vacuum energy that is stored in the bubble walls. We implement the updated prediction \cite{Ellis:2019oqb,Ellis:2020nnr,Lewicki:2022pdb,Kierkla:2022odc} of
\begin{align}
  \kappa_{\mathrm{coll}}=\left(1-{\frac{\alpha_{\infty}}{\alpha}}\right)\left(1-{\frac{1}{\gamma_{\mathrm{eq}}^{n}}}\right){\frac{R_{\mathrm{eq}}}{R_{*}}}{\frac{\gamma_{\mathrm{*}}}{\gamma_{\mathrm{eq}}}}\,,\label{eq:kappa_coll}
\end{align}
where $\alpha_\infty$ defines the weakest strength of the phase
transition for which Eq.~\eqref{eq:DeltaP} (see below) is valid if the next-to-leading order friction contribution $P_{1\rightarrow N}$ is neglected \cite{Ellis:2019oqb,Ellis:2020nnr}, and it is given by
\begin{align}
    \alpha_\infty = \frac{P_{1\rightarrow 1}}{\rho_\gamma}\,,
\end{align}
where $\rho_\gamma$ is the energy density given by
Eq.~\eqref{eq:rho_gamma} and $P_{1\to 1}$ is the
  leading-order friction given below in Eq.~(\ref{eq:pnlo1}).
The Lorentz factor $\gamma_*$ in the absence of plasma friction at the moment of collisions is derived from
\begin{align}
    \gamma_* = \min(\gamma_\mathrm{eq},\gamma_\mathrm{run-away})\,,
\end{align}
with \cite{Gouttenoire:2021kjv}
\begin{align}
  \gamma_\mathrm{run-away} = R_*/(3 R_0) \quad\mbox{ and } \quad
  R_\mathrm{eq} = 3 R_0 \gamma_\mathrm{eq} \;.
\end{align}
This definition takes into account whether the bubbles were still accelerating at
collision or if they reached their terminal velocity. The initial
bubble radius $R_0$~\cite{Kierkla:2022odc} is given by
\begin{align}
	R_0 = \left[\frac{3 S_3}{8 \pi \Delta V}\right]^\frac{1}{3}\,,
\end{align}
and the average bubble radius at the transition temperature $R_*$ \cite{Guth:1979bh,Guth:1981uk,Enqvist:1991xw,Turner:1992tz,Kierkla:2022odc,Goncalves:2024lrk,Bertenstam:2025jvd} is given by
\begin{align}
    R_* = \left[T_*^3 \int_{T_*}^{T_c} \frac{dT'}{T'^4} \frac{\Gamma(T')}{H(T')} \mathrm{e}^{-I(T')} \right]^{-1/3}\,.
\end{align}
The above derivation of the collision efficiency factor is valid in the regime where the bubble wall is accelerated by the pressure difference
\begin{align}
    \Delta P = \Delta V - P_{1\rightarrow 1} - P^{(n)}_{1\rightarrow N}\label{eq:DeltaP}
\end{align}
between vacuum pressure $\Delta V$ and the plasma friction terms, $P_{1\rightarrow 1}$  and $P^{(n)}_{1\rightarrow N}$ . The leading-order pressure contribution~\cite{Bodeker:2009qy} is given by
\begin{align}
    P_{1\rightarrow 1} \simeq \frac{1}{24} \,\Delta m^2 \,T_*^2\;,
\end{align}
where
\begin{align}
    \Delta m^2 = \sum_i k_i c_i (m_{i, \mathrm{t}}^2 - m_{i, \mathrm{f}}^2 )
\end{align}
with $c_i = 1\,(1/2)$ for bosons (fermions) and $k_i$ denoting the
internal number of degrees of freedom.
The masses of the particles $i$ in the true and the 
false vacuum are denoted by $m_{i, \mathrm{t}}$ and $m_{i, \mathrm{f}}$, respectively.
For the next-to-leading order (NLO)
pressure contribution there are two conflicting
results~\cite{Bodeker:2017cim,Ellis:2019oqb,Gouttenoire:2021kjv,Hoche:2020ysm}
that have different scalings with the Lorentz $\gamma$ and will, in
general, produce a different terminal Lorentz factor
$\gamma_\mathrm{eq}$. To the best of our knowledge, the correct result
is an open question so that we include both results in our
implementation. The NLO 
pressure~\cite{Bodeker:2017cim, Ellis:2019oqb, Gouttenoire:2021kjv}
result with linear $\gamma$-scaling is given by
\begin{align}
    P^{(1)}_{1\rightarrow N}&\simeq \,\gamma \sum_i g_i^2 \Delta m_i \,T_*^3\,,\label{eq:pnlo1}
\end{align}
where the sum runs only over the gauge bosons, $g_i$ is the gauge
coupling and $\Delta m_i$ is the difference in the masses of the gauge
bosons in the true vacuum and the false vacuum, respectively. The NLO pressure~\cite{Hoche:2020ysm} result with the $\gamma^2$-scaling is given by
\begin{align}
    P^{(2)}_{1\rightarrow N}&\simeq \,\gamma^2 \sum_i k_i g_i^2 \,T_*^4\,.\label{eq:pnlo2}
\end{align}
The bubble wall accelerates driven by the energy difference until it
reaches at equilibrium pressure $\Delta P = 0$ a terminal velocity
with a Lorentz $\gamma_\mathrm{eq}$, which in general is given by~\cite{Ellis:2020nnr,Kierkla:2022odc}
\begin{align}
    \gamma_\mathrm{eq}=\sqrt[{n}]{\frac{\Delta V - P_{1\rightarrow 1}}{P^{(n)}_{1\rightarrow N}/ \gamma^{n}}}\;,
\end{align}
where $(n)$ denotes the $\gamma^{(n)}$-scaling chosen for the NLO pressure contribution.\s

We set $\kappa_\mathrm{coll}$ as defined in Eq.~\eqref{eq:kappa_coll} to zero if either $\alpha < \alpha_\infty$, or if $\gamma_\mathrm{eq} < 1$.
The former allows us to deal with supercooled and non-supercooled phase transitions simultaneously, the latter excludes situations with an unphysical Lorentz factor $\gamma$.
The spectrum for sound waves and for magnetohydrodynamic turbulence is fitted with a double broken power law template \cite{Caprini:2024hue},
\begin{align}
    \Omega_\mathrm{GW}^\mathrm{DBPL}(f) = \Omega_\mathrm{int} \times S(f) = \Omega_2 \times S_2(f)\,,
\end{align}
with the shape function
\begin{align}
    S(f) = N \left(\frac{f}{f_1}\right)^{n_1}\left[1+\left(\frac{f}{f_1}\right)^{a_1}\right]^{\frac{-n_1+n_2}{a_1}}\left[1+\left(\frac{f}{f_2}\right)^{a_2}\right]^{\frac{-n_2+n_3}{a_2}}\,,
\end{align}
and $\{n_1,\,n_2,\,n_3,\,a_1,\,a_2\}=\{3,1,-3,2,4\}$ for sound waves
and $\left\{3,1,-\frac{8}{3},4,2.15\right\}$ for turbulence,
respectvely, and $S_2(f) = S(f) / S(f_2)$. 
The normalization $N$ of the shape function is then determined via the requirement $S_2(f_2) = 1$.
The characteristic frequency breaks $f_{1,2}$ of the sound wave spectrum are
\begin{align}
    f_1^\mathrm{sw} &\simeq \num{.2} H_{*,0} \left(H_* R_*\right)^{-1} \,, \label{eq:f1_sw}\\
    f_2^\mathrm{sw} &\simeq \num{.5} H_{*,0} \Delta_w^{-1} (H_* R_*)^{-1}\,, \label{eq:f2_sw}
\end{align}
with $\Delta_w = \xi_\mathrm{shell} / \mathrm{max}(v_w,\, c_{s,\mathrm{f}})$ and
the fluid shell thickness estimated as $\xi_\mathrm{shell} = \xi_\mathrm{front} - \xi_\mathrm{rear}$ with \cite{Caprini:2024hue,Caprini:2024gyk}
\begin{align}
    \xi_\mathrm{front} = \begin{cases} \xi_\mathrm{shock}\,, & \text{for}\,\, v_w < v_\mathrm{CJ} \\ v_w\,, & \text{for}\,\, v_w \geq v_\mathrm{CJ}\end{cases}\;, \qquad \xi_\mathrm{rear} = \begin{cases} v_w\,, & \text{for}\,\, v_w < c_{s,\mathrm{t}}\\ c_{s,\mathrm{t}}\,, & \text{for}\,\, v_w \geq c_{s,\mathrm{t}} \end{cases}\,.
\end{align}
We determine the shock velocity $\xi_\mathrm{shock}$ from the fluid
velocity profile that is determined by solving the hydrodynamics
equations using the method of~\cite{Giese:2020rtr,Giese:2020znk}. \s

For sound waves, the integrated amplitude $\Omega_\mathrm{int}$ is derived as \cite{Caprini:2024hue}
\begin{align}
    \Omega_\mathrm{int}^\mathrm{sw} = F_\mathrm{GW,0} A_\mathrm{sw} K_\mathrm{sw}^2 \left(H_* R_* \right) \Upsilon  \,,
\end{align}
where $A_\mathrm{sw} = 0.11$ and with the kinetic energy fraction 
\begin{align}
    K_\mathrm{sw}=\frac{\num{.6}\,\kappa_\mathrm{sw}\,\alpha}{1+\alpha}\;.
\end{align}
We calculate the efficiency factor $\kappa_\mathrm{sw}$ from $\kappa_\mathrm{sw}^{\mu\nu}$ which we obtain with the template model introduced by \cite{Giese:2020rtr,Giese:2020znk},
\begin{align}
    \kappa_\mathrm{sw} = \frac{\alpha_\text{eff}}{\alpha} \kappa_\mathrm{sw}^{\mu\nu}\;,
\end{align}
by a rescaling with the effective strength $\alpha_\text{eff}\equiv \alpha(1-\kappa_\mathrm{coll})$ to account for supercooled transitions, where a large portion of the energy goes into collisions. 
For non-supercooled transitions we have that $\alpha \simeq \alpha_\text{eff} $ as the collision efficiency factor is small, i.e.~$\kappa_\mathrm{coll} \ll 1$. \s

The suppression factor $\Upsilon$ arising from the finite lifetime $\tau_\mathrm{sw}$ of sound waves, is estimated by \cite{Hindmarsh:2015qta,Ellis:2018mja,Ellis:2019oqb,Caprini_2020,Ellis:2020awk,Guo:2020grp}
\begin{align}
    \Upsilon=1-\frac{1}{\sqrt{1+2H_{\ast}\tau_{\mathrm{sw}}}} \mathrm{,\quad where\quad} H_* \tau_\mathrm{sw} = \mathrm{min}\left[\frac{2 H_* R_*}{\sqrt{3 K_\mathrm{sw}}},\, 1\right]\,.\label{eq:lifetime_factor}
\end{align}
The sound wave amplitude $\Omega_2$ at $f_2$ is expressed
in terms of the integrated amplitude $\Omega_\mathrm{int}$ as
\begin{align}
    \Omega_2 = \frac{1}{\pi}\left(\sqrt{2}+\frac{2 f_2/f_1}{1+f_2^2/f_1^2}\right)\,\Omega_\mathrm{int}\;.\label{eq:Omega2_sw}
\end{align}
Magnetohydrodynamic (MHD) turbulence is characterized by
\begin{align}
    f_1 &= \frac{\sqrt{3\Omega_s}}{2 \mathcal{N}} H_{*,0} \left(H_* R_*\right)^{-1}\,,\label{eq:f1_turb}\\
    f_2 &\simeq \num{2.2} H_{*,0} \left(H_* R_*\right)^{-1}\,,\label{eq:f2_turb}\\
    \Omega_2 &= F_{\mathrm{GW},0}\, A_\mathrm{MHD}\, \Omega_s^2\, \left(H_* R_*\right)^2\label{eq:Omega2_turb}\,,
\end{align}
with $A_\mathrm{MHD}=\num{4.37e-3}$ and $\Omega_s = \kappa_\text{turb}
K_\mathrm{sw}$, where $\kappa_\text{turb}$ represents the fraction of
the overall kinetic energy in the bulk motion that is
converted to magnetohydrodynamic turbulence, and $\mathcal{N} \simeq 2$. The turbulence efficiency factor $\kappa_\text{turb}$ is set to
\begin{align}
    \kappa_\mathrm{turb} = \epsilon \, \kappa_\mathrm{sw}\,,\label{eq:kappa_turb}
\end{align}
with the efficiency factor $\epsilon$ that can be set by the user to a value between $0$ and $1$ or to the upper bound given~\cite{Athron:2023xlk} by
\begin{align}
    \epsilon = \sqrt{1-\Upsilon}\,.\label{eq:eps_turb}
\end{align}

\subsubsection{Signal-to-Noise Ratio at LISA}
The stochastic gravitational wave signal produced in an FOPT is in a
frequency range to which the future space-based gravitational wave
observatories like {\tt LISA} 
\cite{Caprini:2019egz,LISA,LISACosmologyWorkingGroup:2022jok} could
potentially be sensitive. The Signal-to-Noise ratio (SNR) of the GWs tells us if a
GW signal from an FOPT can be detected by {\tt LISA}. It can be computed as \cite{Caprini:2019egz} 
\begin{align}
\mathrm{SNR}=\sqrt{\mathcal{T} \int_{f_{\min }}^{f_{\max }} \mathrm{d} f\left[\frac{h^{2} \Omega_{\mathrm{GW}}(f)}{h^{2} \Omega_{\mathrm{Sens}}(f)}\right]^{2}}\,,
\label{eq:SNRLISA}
\end{align}
where $h^{2} \Omega_{\mathrm{GW}}(f)$ is the gravitational wave 
signal, $h ^2\Omega_{\mathrm{Sens}}(f)$ is the nominal
strain sensitivity of {\tt LISA}~\cite{ESA_SciRD}  of
a given {\tt LISA} configuration to stochastic sources,
$\mathcal{T}$ is the experimental acquisition time in seconds, and $f_{\min
}$ and $f_{\max }$ are minimum and maximum frequency, respectively,
to which {\tt LISA} is sensitive. The expected acquisition time of
data for {\tt LISA} is around 4 years with a minimum duty cycle of
$75\%$~\cite{LISA:documents} so that we choose $\mathcal{T} =
3\,\text{years} \cdot 365.25\,\text{days/year}\cdot 86400
\,\text{s/day} = 94672800 \,\text{s}$, and hence
\beq
\text{SNR in {\tt BSMPTv3}: }  \mathrm{SNR}
(3\, \text{years}) \;.
\eeq
In case one wants to
calculate the SNR with an acquisition time of $\mathcal{Y}$ years, the
SNR calculated by \texttt{BSMPT} can be rescaled as 
\begin{align}
\text{SNR}(\mathcal{Y}) =
  \sqrt{\frac{\mathcal{Y}}{3}}\text{SNR}(3\,\text{years}) \;.
\end{align} 
The nominal sensitivity $h ^2\Omega_{\mathrm{Sens}}(f)$ can be written
as a function of the power spectral density $S_h(f)$, given in the
{\tt LISA} mission requirements
\cite{LISA:documents,Caprini_2020,Babak:2021mhe,} as
\begin{align}
  \Omega_{\mathrm{Sens}}(f) = \frac{4\pi^2}{3 H_0^2} f^3 S_h(f) \;,
  \end{align}
  where $H_0 = 67.4 \pm 0.5\,\text{km}/s/\text{Mpc}$ is the Hubble
  constant today \cite{planck2018}. A GW signal is considered to be
  detectable if it gives rise to an SNR $> 10$.

\subsection{The Class {\tt TransitionTracer}}\label{subsec:transtracer}

The class {\tt TransitionTracer} interfaces all previously described classes, {\tt MinimumTracer}, {\tt BounceSolution} and {\tt GravitationalWave}, with the executables. 
It initiates the phase tracking, calling the routines of {\tt
MinimumTracer}, and collects all phases and coexisting phase pairs
with their critical temperatures. 
It then goes through all pairs of coexisting false and true
phases\footnote{With decreasing temperature newly appearing phases are
  first local minima relative to the already existing
  phases. Therefore, in a pair of coexisting phases, the phase which
  is found to exist since a higher temperature is always considered
  the respective false phase.} for which a critical temperature could
be determined\footnote{A critical temperature can be determined if the
  false phase starts as the lower minimum at the highest temperature of the
  overlap and ends as the higher minimum at the lowest temperature of
  the overlap, or if the true phase starts already as the lower
  minimum at the highest temperature of the overlap. In the former
  case, the critical temperature is found in between the lowest and the highest
  temperature of the coexisting temperature region, in the latter case
  the critical temperature is set to the highest temperature of the
  overlap.} and tries to determine a bounce solution using the
algorithms described in {\tt BounceSolution}. If a bounce solution is
successfully determined, it is evaluated in the temperature range of
the overlap region to determine the characteristic temperatures of the
transition, i.e.~the nucleation temperature (cf.~Eq.~(\ref{eq:nucl_exact})), the
percolation temperature (cf.~Eq.~(\ref{eq:perc})), and the completion
temperature (cf.~Eq.~(\ref{eq:compl})). If requested by the user, we then calculate the
gravitational wave signal for each phase pair that was found to have the
transition temperature (as chosen by the user through the input). \s

{\tt BSMPTv3} calculates characteristic temperatures and gravitational
wave signals for all phase pairs that are found. However, some transitions might cosmologically be impossible to realize due to the respective false phase never getting populated by a sufficient fraction of the universe.
Therefore, apart from managing the calculation, the class {\tt TransitionTracer} also reports on the transition history for each point. 
We label found phases and coexisting phase pairs with increasing indices $\{{\tt 0},{\tt 1},\dots\}$ for decreasing upper temperatures $T_\text{high}$. 
After studying all phase pairs with the algorithms of {\tt BounceSolution} and {\tt GravitationalWave} as described above, we collect all phase pairs for which a \textit{completing} transition could be calculated, meaning a completion temperature was reached.
Then, starting from the initial phase which is assumed to be the global minimum at the user-specified highest temperature of the tracing, $T_\text{high}$, phase {\tt 0}, {\tt TransitionTracer} goes through all pairs with false phase {\tt 0} until a first pair with $T_\text{compl}$ is found before any other transition becomes possible.
Then, the old true phase becomes the new false phase and we continue to look for transitions until no transition for the current false phase can be found anymore. 
We then report the transition history for the point in a column {\tt transition\_history} in the form of a string of the following form in the output file, as further described in the following sections for the executables,
\begin{align*}
    {\tt 0} -({\tt i1})\rightarrow {\tt j1} -({\tt i2})\rightarrow {\tt j2} - \dots\,,
\end{align*}
with {\tt i1}, {\tt i2} being placeholders for the phase pair indices and
{\tt j1} and {\tt j2} being placeholders for the phase indices in
the notation described above. In this example, first in the
pair {\tt i1} a transition completes into the true phase {\tt j1} that then is
the false phase of a second transition in the pair {\tt i2} into the
true phase {\tt j2}. 
For examples on how the transition history is reported and how the
output is interpreted, consult Sec.~\ref{sec:examples} where we
illustrate results for benchmark points. \s

Note, that {\tt BSMPTv3} assumes non-overlapping
transitions: The calculation of the percolation and completion
temperatures described in
Sec.~\ref{sec:characteristic_temperature_scales} and the reported
transition history are only valid for one
transition happening between one pair of false and true phases. \s

During the calculation, we report on its intermediate state by
throwing status codes, managed by {\tt TransitionTracer}. In the
sections about the executables,
Secs.~\ref{sec:MinimaTracer}-\ref{sec:CalcGW}, all relevant codes are
introduced, and a complete summary of them is given in
Sec.~\ref{sec:status_codes}. 

\subsection{The Executable \texttt{MinimaTracer}}\label{sec:MinimaTracer}
The minimum tracing algorithm is capable of identifying the
temperature evolution of non-global and global minima in a user-defined temperature
interval  
$T_\text{low}=\SI{0}{GeV} \leq T \leq T_\text{high}$.
Minimum tracing is the first step before we determine the characteristic temperatures and from there calculate the spectrum of gravitational waves.
The executable {\tt MinimaTracer} allows to separately perform the phase tracing for one or more input parameter points and saves all found phases in one output file per point.
Calling the executable without arguments {\tt ./bin/MinimaTracer} or
with the {\tt --help}-flag {\tt ./bin/MinimaTracer --help} prints out
the following menu: 
\begin{lstlisting}
MinimaTracer traces phases in T = [0, Thigh] GeV
it is called by

	./bin/MinimaTracer model input output firstline lastline

or with arguments

	./bin/MinimaTracer [arguments]

with the following arguments, ([*] are required arguments, others are optional):

argument                   default   description
--help                               shows this menu
--model=                             [*] model name
--input=                             [*] input file (in tsv format)
--output=                            [*] output file (in tsv format)
--firstline=                         [*] line number of first line in input file
                                         (expects line 1 to be a legend)
--lastline=                          [*] line number of last line in input file
--thigh=                   300       high temperature [GeV]
--multistepmode=           default   multi-step PT mode
                                     default: default mode
                                     0: single-step PT mode
                                     >0 for multi-step PT modes:
                                     1: tracing coverage
                                     2: global minimum tracing coverage
                                     auto: automatic mode
--num_pts=                 10        intermediate grid-size for default mode
--checkewsr=               on        check for EWSR at high temperature
                                     on: perform check
                                     off: check disabled
--usegsl=                  true      use GSL library for minimization
--usecmaes=                true      use CMAES library  for minimization
--usenlopt=                true      use NLopt library for minimization
--usemultithreading=       false     enable multi-threading for minimizers
--json=                              use a json file instead of cli parameters
\end{lstlisting}
A minimal example call being
\begin{lstlisting}
./bin/MinimaTracing --model=MODEL --input=input.tsv --output=output --firstline=2 --lastline=2
\end{lstlisting}
traces the point of model {\tt MODEL} found in the second line of the
tab-separated input file {\tt input.tsv} in between
$T\in\{0,300\}\,\si{GeV}$. 
Note that the first line of the input file is expected to be a legend.
The temperature range for the tracing can be specified by setting the
optional flag {\tt --thigh} to a user-defined value.
The optional mode for multi-step phase tracing {\tt --multistepmode}
with its optional grid size {\tt --num\_pts} for the default mode is discussed in detail below in Sec.~\ref{sec:multistepmode}. 
The check of electroweak symmetry restoration controlled via {\tt --checkewsr} is discussed in detail below in Sec.~\ref{sec:ewsrcheck}.
The flags {\tt --UseGSL}, {\tt --UseCMAES}, {\tt --UseNLopt} can be used to enable or disable the three implemented minimising libraries separately. By default, all installed and linked libraries are enabled.
Setting {\tt --UseMultithreading=true} enables CPU-parallelization via the {\tt C++-thread} class. 
Additional terminal output for any of the executables can be requested by enabling any or all of the following {\tt logginglevels} of the {\tt Logger} class.
All output of {\tt BSMPT} is channelled through the {\tt Logger} class
since {\tt BSMPTv2.3}, the new release of {\tt BSMPTv3}
extends this by five new {\tt logginglevels} so that we have
\vspace*{-1cm}
\begin{center}
    \setlength{\extrarowheight}{.4em}
    \setlength{\tabcolsep}{.5em}
    \begin{table}[!h]
        \centering
        \begin{longtable}{ll p{8cm}}
            {\tt --logginglevel::}&default& description \\\midrule
            {\tt default=} & {\tt true} & print output enabled by default\\
            {\tt debug=} & {\tt false} & print additional output useful for debugging\\
            {\tt disabled} & - & disable all output\\
            {\tt ewbgdetailed=} & {\tt false} & show additional output  during the calculation of the baryon  asymmetry\footnotemark\\
            {\tt progdetailed=} & {\tt false} & show status messages generated by executables\\
            {\tt minimizerdetailed=} & {\tt false} & show additional minimizer-output \\
            {\tt transitiondetailed=} & {\tt false} & show additional output of the {\tt TransitionTracer} class\\
            {\tt mintracerdetailed=} & {\tt false} & show additional output of the {\tt MinimumTracer} class\\
            {\tt bouncedetailed=} & {\tt false} & show additional output of the {\tt BounceSolution} class\\
            {\tt gwdetailed=} & {\tt false} & show additional output of the {\tt GravitationalWave} class\\
            {\tt complete=} & {\tt false} & enable all {\tt logginglevels} above except {\tt minimizerdetailed}\\
        \end{longtable}
    \end{table}
\end{center}
\footnotetext{We remind the reader, that this is only relevant for the
  C2HDM, as only in this model the baryon asymmetry is calculated.}
\FloatBarrier

The executables also accept input in form of {\tt json} files if the
package \cite{nlohmann-json} was found during installation. Examples
for all executables, on how {\tt json}-files can look like can be
found in {\tt example/JSON}. After the executable ran successfully,
the output is saved in {\tt 
  output\_1.tsv}\footnote{The index refers to the point number,
    for which the output is given.} in tabular-separated form by extending {\tt
  input.tsv} by the status columns (a summary on
  all status codes is presented in Sec.~\ref{sec:status_codes}) 
\begin{labeling}{{\tt status\_nlo\_stability}\quad}
  \item[{\tt status\_nlo\_stability}] Reports {\tt success} if
    the next-to-leading order (NLO)
    zero-temperature global minimum is found to lie at the position of
    the electroweak tree-level minimum and {\tt no\_nlo\_stability} if
    not. Note, that for the {\tt MinimaTracer} executable 
   NLO stability is merely a status, not an error code. 
  \item[{\tt status\_ewsr}] Stores information on the status of the check for electroweak symmetry restoration, more information is found in Sec.~\ref{sec:ewsrcheck}.
  \item[{\tt status\_tracing}] Contains information on the success of the tracing; details on the multi-step phase transition mode can be found in Sec.~\ref{sec:multistepmode}.
\end{labeling}
as well as the following columns for each found and traced phase {\tt i}:
\begin{labeling}{{\tt omega\_X(Temp\_i)}\quad}
    \item[{\tt Temp\_i}] Temperature in [\si{GeV}] of each tracing step in phase {\tt i}.
    \item[{\tt omega\_X(Temp\_i)}] Field value of direction {\tt X} in [\si{GeV}] at temperature {\tt Temp\_i}. The labels of direction {\tt X} are model-specific and defined in {\tt addLegendVEV()} in the respective model file.
    \item[{\tt Veff(Temp\_i)}] Value of the one-loop corrected effective potential in [\si{GeV}] at phase configuration {\tt omega\_X(Temp\_i)} at temperature {\tt Temp\_i}.
\end{labeling}
The last column {\tt runtime} logs the runtime of the code after each
tracing step in seconds. \s

Note, that in addition to the new {\tt MinimaTracer} executable, we continue to ship the {\tt VEVEVO} executable with {\tt BSMPTv3}. 
{\tt VEVEVO} calculates and outputs the location of the global minimum
in a multi-dimensional field space using minimisation routines from
{\tt GSL}, {\tt CMAES} and {\tt NLopt}. For a documentation
consult~\cite{Basler:2020nrq}.  
The new executable {\tt MinimaTracer} is designed to use the new
algorithms of local-minimum tracing enabling {\tt BSMPTv3} to track
the location of global and non-global minima over temperature ranges and to 
look for regions of coexisting phases. \s
 
The next section, Sec.~\ref{sec:multistepmode}, describes in detail
how we manage the tracing of multiple, possibly coexisting, phases and
how the users can customize the tracing method according to their needs.

\subsubsection{Multi-Step Phase Transition Mode}\label{sec:multistepmode}
We trace individual phases using the algorithms described in
Sec.~\ref{subsec:mintracer}. In order to be able to study phase
transition histories with multiple phases that possibly exist in
overlapping temperature regions we make the following assumptions: 
\begin{enumerate}
    \item At the user-defined temperature $T_\text{high}$ and at
      $T_\text{low}=\SI{0}{GeV}$ the universe is realized in the
      global minimum of its one-loop corrected effective potential.
    \item Phases that always remain the non-global minimum over the
      whole temperature range escape our multi-step phase tracing as we, for the
        moment, only use global minima positions as \textit{seeds} for
        the phase tracing. 

        The only exception to this assumption is the electroweak
        minimum with $v=\SI{246}{GeV}$ at $T=\SI{0}{GeV}$, that we
        always use as an additional seed point, as this is at least a
        local minimum due to the choice of our counterterm potential. 
        For the executables {\tt CalcTemps} (see
        Sec.~\ref{sec:CalcTemps}) and {\tt CalcGW} (see
        Sec.~\ref{sec:CalcGW}) the potential is required to be NLO
        stable by default, meaning that unphysical points with a
        one-loop global minimum at zero temperature, which is different
        from the electroweak minimum, are discarded
        immediately. However, the user can switch off this requirement
        with {\tt --checknlo=off}. In this case then also an only non-global electroweak
        minimum gets traced.
    \item Phases start as a non-global minimum when they are first
      found at their highest temperature, they only become the global
      minimum at a lower temperature. This statement assumes that {\tt
        BSMPTv3} is able to trace the phase over the whole temperature
      region in which it exists.  
\end{enumerate}
The user can specify how the minimum tracing algorithm detects possible
multi-step phase transitions by setting the
flag {\tt --multistepmode} to {\tt default}, {\tt 0}, {\tt 1}, {\tt 2}
or {\tt auto}. By default, {\tt mode default} is selected. For most
points, it will provide successful tracing with {\tt status\_tracing =
  success}, while being the most resource-optimized. In addition to
{\tt mode default}, we offer four tracing modes with slightly
different algorithms. These are {\tt mode 0}, which is optimal if the
user is only interested in one-step first-order phase transitions;
{\tt mode 1} if one wants to ensure tracing coverage with a global
minimum check at phase endpoints for multi-step phase tracing; {\tt mode 2}
enforces global minimum tracing coverage explicitly. The {\tt auto}
mode automatizes {\tt mode 1} and {\tt mode 2} by running {\tt mode 2}
in case the global minimum check at phase endpoints fails for {\tt mode
  1}. \s

More details on all five implemented multi-step phase transition modes
are given in the following. In
Fig.~\ref{fig:illustrate_multi_step_modes} the respective tracing
algorithms are illustrated. 
\begin{labeling}{{\tt mode default}\quad}
    \item[{\tt mode default}] The default tracing mode provides a fast and customizable grid-checked way of tracking phases for points with multi-step phase transition.
        It starts at the global minimum at the user-defined $T_\text{high}$ and tries to trace it down to $T_\text{low}=\SI{0}{GeV}$. 
        When the currently traced phase ends, the new global minimum is traced subsequently until a phase existing down to $T_\text{low}=\SI{0}{GeV}$ is found. 
        The global minimum at $T_\text{low}=\SI{0}{GeV}$ is also traced up to $T_\text{high}$.
        The default mode then uses the global minima found at an equidistant temperature grid as additional seed points to check the completeness of the tracing.
        Each of these points are checked whether they are part of an already traced phase, and if not, are traced between $\{0,\,T_\text{high}\}$ and added as a new phase.
        By increasing the grid-size of equally-spaced intermediate checked points, by setting {\tt --num\_pts} to a value larger than the default value {\tt --num\_pts=10}, the user can fine-tune the tracing granularity.

        Note, that this mode returns success, {\tt status\_tracing=success}, if tracing \textit{coverage} is found, so if at least one phase is found for each temperature in the traced temperature interval.
        This does not necessarily indicate that the found phase structure contains the global minimum in the whole temperature range, which we call \textit{global minimum coverage}.
        Global minimum coverage, if not already achieved with the default settings of {\tt --multistepmode=default}, can be ensured by requesting a larger grid-size.

        In case temperature gaps between traced phases are identified,
        we try to {\it patch up} such gaps by explicitly choosing and
        tracing seed points inside the temperature gap. 
        Note, that we attempt to patch up gaps until $\Delta T < \SI{e-6}{GeV}$. 
        Gaps smaller than $\SI{e-6}{GeV}$ are no longer patched up. 
        In that case we cannot numerically find tracing coverage with
        {\tt mode default} and {\tt status\_tracing} is set to the
        error code {\tt no\_coverage}. 
    \item[{\tt mode 0}] \textbf{one-step phase transition mode}: This
      is a dedicated mode to exclusively look for one-step first-order
      phase transitions.  It only traces the global minima
      from $T_\text{low}=\SI{0}{GeV}$ towards
        $T_\text{high}$ and from $T_\text{high}$ towards 
        $T_\text{low}=\SI{0}{GeV}$, respectively. If they are found
        to overlap and the high-temperature phase is found to be the
        global minimum when the low-temperature phase ends at its
        highest temperature and vice versa, a valid one-step phase
        transition point was found and {\tt status\_tracing} reports
        {\tt success}. 

        The calculation for this mode reports an error code if the
        low-temperature and the high-temperature phase are not found
        to overlap ({\tt no\_coverage}) or no global minimum coverage
        was found ({\tt no\_glob\_min\_coverage}), indicating that no
        valid one-step phase transition can exist for this point. 
        If in this mode we cannot find a stable seed point for the
        two phases, an error code {\tt no\_mins\_at\_boundaries} is
        reported. 
    \item[{\tt mode 1}] \textbf{enforced tracing coverage mode}: This mode is specialized to deal with multi-step phase transitions, and will (as far as it is numerically possible) enforce coverage while checking for global minimum coverage in a performance-optimized way, similarly to the check done by {\tt mode 0}, as elaborated below.
        It therefore can deal best with points illustrated in Fig.~\ref{fig:illustrate_multi_step_modes} (second from right), which have multiple phases and multiple overlaps between them which only consist of exactly two phases at a time.
        Again, like in {\tt mode 0}, the initial low- and high-temperature phases are traced and if they are not found to coexist, then, at their respective phase-end-points, we determine the global minimum again and trace it up and down in temperature, until we reach full coverage by traced phases over the whole temperature region.
        Temperature gaps with no phases found larger than
        $\SI{e-6}{GeV}$ are patched up as described for {\tt mode default}.
        Global minimum coverage is ensured by checking at all temperatures coinciding with phase end points (from tracing up and down in temperature) whether the lower and higher temperature phases coincide with the global minimum.
        If global minimum coverage is achieved in this sense, we classify a valid multi-step phase transition point with {\tt status\_tracing=success}.
        In this case, using {\tt mode 1} and not {\tt mode 2}, can significantly save runtime.
        If for any intermediate overlap we do not find the global minimum as part of any of the already traced phases, we miss tracing the global minimum in some areas of the temperature region. 
        Then, rerunning the parameter point with {\tt --multistepmode=2} is recommended and minima tracing fails with {\tt status\_tracing=no\_glob\_min\_coverage}.
    \item[{\tt mode 2}] \textbf{enforced global minimum tracing coverage mode}: Finally, {\tt mode 2} has the strongest implemented check for global minimum coverage. 
        It can reliably deal with multi-step phase transition points
        with multiple overlaps between any number of phases as well as
        overlaps between phases that only coexist while both no longer
        include the global minimum, as illustrated in
        Fig.~\ref{fig:illustrate_multi_step_modes} (right). 
        It works similarly to {\tt mode 1}, but in addition we track how long a traced minimum is still found to be the global minimum. 
        At temperatures where the global minimum is no longer part of the traced phase, a new phase gets traced and added using the new global minimum as a seed point.
        The procedure is repeated until the whole temperature range is covered, making sure that the global minimum is a subset of all traced phases in the whole traced temperature range.
        In this mode we again patch up gaps as described for {\tt mode default} and it can only fail with {\tt no\_coverage} in case tracing coverage can numerically not be achieved.
    \item[{\tt mode auto}] \textbf{automatic mode}: This mode automatizes the choice between {\tt mode 1} and {\tt mode 2}.
        It first attempts to run {\tt mode 1} and switches to the more resource-intensive {\tt mode 2} in case of failure with {\tt no\_glob\_min\_coverage}.
        Note that {\tt mode auto} therefore relies in a first iteration on the global minimum coverage check that only takes into account phase end points, as described above, and only in case of failure, moves on to {\tt mode 2}.
\end{labeling}
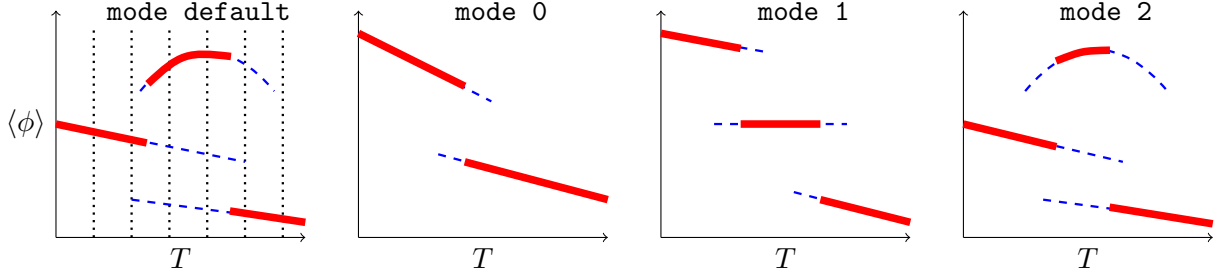
\begin{figure}[ht!]
  \centering
  \begin{tikzpicture}
    \def\w{\textwidth/4-20}
    \def\wmd{\w}
    \def\wmdg{\wmd+20}
    \def\wmnull{\wmdg+\w}
    \def\wmnullg{\wmnull+20}
    \def\wmone{\wmnullg+\w}
    \def\wmoneg{\wmone+20}
    \def\wmtwo{\wmoneg+\w}
    \def\textgap{40}
    \def\h{3}
    \def\gapbelow{-.5}
    \def\d{\gapbelow-1}
    \def\lwthick{1}
    \def\lwthin{.3}
    
    \node[align=center] (a) at (\wmd-\textgap,\h) {{\tt mode default}};
    \draw[->] (0,0) to node [left]{$\langle\phi\rangle$} (0,\h); 
    \draw[dotted, line width=\lwthin mm] (.5,0) to (.5,\h-.2);
    \draw[dotted, line width=\lwthin mm] (1,0) to (1,\h-.2); 
    \draw[dotted, line width=\lwthin mm] (1.5,0) to (1.5,\h-.2);  
    \draw[dotted, line width=\lwthin mm] (2,0) to (2,\h-.2);
    \draw[dotted, line width=\lwthin mm] (2.5,0) to (2.5,\h-.2);  
    \draw[dotted, line width=\lwthin mm] (3,0) to (3,\h-.2); 
    \draw[->] (0,0) to node [below]{$T$} (\wmd,0);
    \node[] (p1) at (1, 1.8) {};
    \node[] (p2) at (3, 1.8) {};
    \node[] (p3) at (1.08, 1.9) {};
    \node[] (p4) at (2.45, 2.38) {};
    \path[blue, dashed, line width=\lwthin mm, bend left=50, looseness=1.3] (p1) edge (p2);
    \path[red, line width=\lwthick mm, bend left=25, looseness=1.3] (p3) edge (p4);
    \draw[blue, dashed, line width=\lwthin mm] (0, 1.5) to (2.5, 1);
    \draw[red, line width=\lwthick mm] (0, 1.5) to (1.2, 1.25);
    \draw[blue, dashed, line width=\lwthin mm] (1, .5) to (\wmd, .2);
    \draw[red, line width=\lwthick mm] (2.3, .35) to (\wmd, .2);

    \node[align=center] (b) at (\wmnull-\textgap,\h) {{\tt mode 0}};
    \draw[->] (\wmdg,0) to (\wmdg,\h); 
    \draw[->] (\wmdg,0) to node [below]{$T$} (\wmnull,0);
    \draw[blue, dashed, line width=\lwthin mm] (\wmdg, \h-.3) to (\wmdg+50, \h-1.2);
    \draw[red, line width=\lwthick mm] (\wmdg, \h-.3) to (\wmdg+40, \h-1);
    \draw[blue, dashed, line width=\lwthin mm] (\wmdg+30, 1.1) to (\wmnull, 0.5);
    \draw[red, line width=\lwthick mm] (\wmdg+40, 1) to (\wmnull, 0.5);

    \node[align=center] (c) at (\wmone-\textgap,\h) {{\tt mode 1}};
    \draw[->] (\wmnullg,0) to (\wmnullg,\h); 
    \draw[->] (\wmnullg,0) to node [below]{$T$} (\wmone,0);
    \draw[blue, dashed, line width=\lwthin mm] (\wmnullg, \h-.3) to (\wmnullg+40, \h-.55);
    \draw[red, line width=\lwthick mm] (\wmnullg, \h-.3) to (\wmnullg+30, \h-.5);
    \draw[blue, dashed, line width=\lwthin mm] (\wmnullg+20, 1.5) to (\wmnullg+70, 1.5);
    \draw[red, line width=\lwthick mm] (\wmnullg+30, 1.5) to (\wmnullg+60, 1.5);
    \draw[blue, dashed, line width=\lwthin mm] (\wmnullg+50, .6) to (\wmone, .2);
    \draw[red, line width=\lwthick mm] (\wmnullg+60, .5) to (\wmone, .2);

    \node[align=center] (d) at (\wmtwo-\textgap,\h) {{\tt mode 2}};
    \draw[->] (\wmoneg,0) to (\wmoneg,\h); 
    \draw[->] (\wmoneg,0) to node [below]{$T$} (\wmtwo,0);
    \node[] (p1) at (\wmoneg+20, 1.8) {};
    \node[] (p2) at (\wmoneg+80, 1.8) {};
    \node[] (p3) at (\wmoneg+31, 2.28) {};
    \node[] (p4) at (\wmoneg+59, 2.48) {};
    \path[blue, dashed, line width=\lwthin mm, bend left=50, looseness=1.3] (p1) edge (p2);
    \path[red, line width=\lwthick mm, bend left=10, looseness=1.3] (p3) edge (p4);
    \draw[blue, dashed, line width=\lwthin mm] (\wmoneg, 1.5) to (\wmoneg+60, 1);
    \draw[red, line width=\lwthick mm] (\wmoneg, 1.5) to (\wmoneg+35, 1.2);
    \draw[blue, dashed, line width=\lwthin mm] (\wmoneg+30, .5) to (\wmtwo, .2);
    \draw[red, line width=\lwthick mm] (\wmoneg+55, .4) to (\wmtwo, .18);
\end{tikzpicture}
\caption{Illustration of the multi-step phase transition modes {\tt
    default}, {\tt 0}, {\tt 1}, {\tt 2} (from left to right) for the
  exemplary class of points where the respective mode performs
  best. Mode {\tt auto} attempts to run {\tt mode 1} and only if
  unsuccessful, immediately afterwards starts {\tt mode 2}. 
  All modes are described in detail in the text. The diagrams show the phases in
  the generic field coordinate $\langle\phi\rangle$ for each point as
  a function of the temperature $T$. Temperature regions in which the
  found minimum is the global minimum are marked by bold red lines,
  regions in which the phase contains only a non-global minimum are
  marked by thinner dashed blue lines. The vertical dotted black lines
  in the left-most diagram that illustrates {\tt mode default}
  represent the grid points that are used for additional tracing
  seeds.}\label{fig:illustrate_multi_step_modes} 
\end{figure}
All above-mentioned status codes for the minima tracing are logged in the {\tt status\_tracing} column.
Note that, as described above, even though the performance-optimized
modes suffice, {\tt success} in {\tt mode default}, {\tt mode 0} or
{\tt mode 1} (and {\tt mode auto}) can still mean that the global 
minimum is not part of the 
traced phases at every temperature inside the interval. 
 The {\tt mode 0} and {\tt mode 1} only check if
 the global minimum at  
 the endpoints of the traced phases is part of a different traced
 phase and therefore for complicated transition histories might miss
 phases.\footnote{ 
    The multi-step modes {\tt mode 1} (and {\tt mode 0}) would report
    {\tt success} even though the global minimum
    is missed in case of a phase (or an overlap of phases) that only
    for an intermediate region does not contain the global minimum. In
    this scenario, the global minimum moves to a new phase only in an
    intermediate temperature range of the initial phase that remains
    the global minimum for its lower and higher temperatures,
    therefore passing the global minimum coverage check that only
    relies on the check of the phase end points. Such a scenario,
    where {\tt mode 1} would falsely report success, is shown in Fig. 3
    (right) illustrating the overlap of two phases where the global
minimum is no longer contained in any of the two phases.}
The same applies to {\tt mode auto}, as it relies in a first iteration
on the reduced global minimum endpoint coverage check done by {\tt mode 1}. 
Full global minimum coverage, however, can be achieved in the {\tt default}
mode by increasing the size of the checked point-grid by setting {\tt
  --num\_pts} to a value larger than the default value 10.
For a reasonable choice of {\tt --num\_pts}, {\tt mode default} is as
accurate as {\tt mode 2} while being orders of magnitude faster. The
{\tt mode 2} ensures full global minimum tracing coverage for any
point independent of its phase structure at the expense of runtime. 

\FloatBarrier
\subsubsection{Electroweak Symmetry Restoration Check}\label{sec:ewsrcheck}
The loop-corrected effective potential at finite temperature $T$ as
function of the classical constant field configuration, generically
denoted by $\omega$, implemented in {\tt BSMPT} is given by  
\begin{align}
V(\omega, T)=V(\omega)+V^T(\omega, T) \equiv
  V^{(0)}(\omega)+V^{\mathrm{CW}}(\omega)+V^{\mathrm{CT}}(\omega)+V^T(\omega,
  T) \;,
\end{align}
where $V^{(0)}(\omega)$ is the tree-level potential,
$V^{\mathrm{CW}}(\omega)$ is the zero-temperature Coleman-Weinberg
potential, $V^{\mathrm{CT}}(\omega)$ is the counterterm potential and
$V^T(\omega, T)$ contains the thermal corrections at finite
temperature $T$. In the following, we derive the high-temperature
limit of the effective potential, which is obtained from $V^T
(\omega,T)$, that in the notation of~\cite{Camargo-Molina:2016moz} is given by
\begin{align}
V^T(\omega, T)=\sum_{X=S, G, F}(-1)^{2 s_X}\left(1+2 s_X\right) \frac{T^4}{2 \pi^2} J_{ \pm}\left(\Lambda_{(X)}^{x y} / T^2\right),
\end{align}
where
\begin{align}
	J_{\pm}\left(\Lambda_{(X)}^{x y}/T^{2}\right)=\mathrm{Tr}\left[\int_{0}^{\infty}\mathrm{d}\mathbf{k}\,k^{2}\log\left[1\pm\exp\left(-\sqrt{k^{2}+\Lambda_{(X)}^{x y}/T^{2}}\right)\right]\right],
\end{align}
is a function of the mass matrices $\Lambda_{(X)}^{x y}$ of scalars ($S$), gauge bosons ($G$) and fermions ($F$), respectively, that is evaluated via a trace $\mathrm{Tr}$. The spins are denoted by $s_X$ for the scalar, gauge and
fermion fields, respectively, $J_-$ is used for bosons and $J_+$
is used for fermions. Additionally, daisy corrections
\cite{PhysRevD.45.2933} $\Pi_{(S)}^{i j}$ and $\Pi_{(G)}^{a b}$ are
also considered, given by 
\begin{align}
\Pi_{(S)}^{i j}=& \frac{T^2}{12}\left[(-1)^{2 s_S}\left(1+2 s_S\right) \sum_{k=1}^{n_{\text {Higgs }}} L^{i j k k}+(-1)^{2 s_G}\left(1+2 s_G\right) \sum_{a=1}^{n_{\text {gauge }}} G^{aaij}\right. \nonumber\\
& \left.+ (-1)^{2 s_F}\left(1+2 s_F\right) \frac{1}{2} \sum_{I, J=1}^{n_{\text {fermion }}}\left(Y^{* I J i} Y_{I J}^j+Y^{* I J j} Y_{I J}^i\right)\right]\\
\Pi_{(G)}^{a b}=& T^{2}\frac{2}{3}\left(\frac{\tilde{n}_{H}}{8}+5\right)\frac{1}{\tilde{n}_{H}}\sum_{m=1}^{n_{\mathrm{Higgs}}}\Lambda_{(G)}^{a a m m}\delta_{a b}\,,
\end{align}
where only the longitudinal modes of the gauge bosons get the daisy corrections and $
\tilde{n}_H \leq n_\text{Higgs}$ is the number of Higgs fields 
coupling to the gauge bosons, and $n_{\text{fermion}}$ and
$n_{\text{gauge}}$ are the numbers of the fermion and gauge fields in
the theory, respectively. The definition of the tensors 
$L^{ijkk}$, $G^{aaij}$, $Y^{IJj}$, and $\Lambda^{aamm}_{(G)}$ can be
found in \cite{Basler:2018cwe}. There are two different approaches to
implement the temperature-corrected Daisy-resummed masses in the
effective potential. In the Arnold-Espinosa
approach~\cite{Arnold:1992rz} one makes the replacement
\begin{align}
V^T(\omega, T) & \rightarrow V^T(\omega, T)+V_{\text {daisy }}(\omega,
                 T) \label{eq:vtae} \\
V_{\text {daisy }}(\omega, T) & =-\frac{T}{12 \pi}\left[\sum_{i=1}^{n_{\text {Higgs }}}\left(\left(\overline{m}_i^2\right)^{3 / 2}-\left(m_i^2\right)^{3 / 2}\right)+\sum_{a=1}^{n_{\text {gauge }}}\left(\left(\overline{m}_a^2\right)^{3 / 2}-\left(m_a^2\right)^{3 / 2}\right)\right],
\end{align}
where $m_i^2$, $\overline m_i^2$, $m_a^2$, $\overline m_a^2$ are the
eigenvalues of $\Lambda_{(S)}^{i j},\Lambda_{(S)}^{i j}+\Pi_{(S)}^{i
  j},\Lambda_{(G)}^{a b},\Lambda_{(G)}^{a b}+\Pi_{(G)}^{a b}$,
respectively. The tensors $\Lambda^{ij}_{(S)}$ and
$\Lambda_{(G)}^{ab}$ are the coefficients of the Lagrangian terms
bilinear in the scalar and in the gauge fields, respectively. Remark,
that only the longitudinal modes of the gauge bosons get the thermal
corrections $\Pi^{(ab)}_{(G)}$. 
In the Parwani approach~\cite{Parwani:1991gq}, one replaces 
\begin{align}
	\Lambda_{(S)}^{i j}\rightarrow\Lambda_{(S)}^{i j}+\Pi_{(S)}^{i j},
\end{align}
and also
\begin{align}
	\Lambda_{(G)}^{a b}\rightarrow\Lambda_{(G)}^{a b}+\Pi_{(G)}^{a b},
\end{align}
for the longitudinal modes. Therefore the Debye corrected masses are
also used in the $V^\text{CW}$ potential. Since the high-temperature limit
of the potential depends on which of the two approaches is used, is
has to be analysed for the two schemes separately. \s

In the Arnold-Espinosa scheme, the thermal correction are contained in
$V^T(\omega,T)$ and $V_\text{daisy}(\omega,T)$. For high temperatures,
i.e.~$x^2 = m^2/T^2 \ll 1$, the thermal functions $J_\pm$ can be
approximated as
\begin{align}
J_{+}\left(x^2, n\right)= & -\frac{7 \pi^4}{360}+\frac{\pi^2}{24} x^2+\frac{1}{32} x^4\left(\log x^2-c_{+}\right)   \nonumber\\
& -\pi^2 x^2 \sum_{l=2}^n\left(-\frac{1}{4 \pi^2} x^2\right)^l
\frac{(2 l-3) ! ! \zeta(2 l-1)}{(2 l) ! !(l+1)}\left(2^{2 l-1}-1\right)  \\
J_{-}\left(x^2, n\right)= & -\frac{\pi^4}{45}+\frac{\pi^2}{12}
x^2-\frac{\pi}{6}\left(x^2\right)^{3 /2}-\frac{1}{32} x^4\left(\log
                            x^2-c_{-}\right) \nonumber\\  
& +\pi^2 x^2 \sum_{l=2}^n\left(-\frac{1}{4 \pi^2} x^2\right)^l
                                                                                                                                                                   \frac{(2 l-3) ! ! \zeta(2 l-1)}{(2 l) ! !(l+1)} \,,
\end{align}
where
\begin{align}
  c_{+}&={\frac{3}{2}}+2\log\pi-2\gamma_{E}
  \\ c_{-}&=c_{+}+2\log4 \,,
\end{align}
where $\gamma_E$ denotes the Euler-Mascheroni constant, $\zeta (x)$
the Riemann $\zeta$-function and $(x)!!$ the double factorial. Taking
into account the leading two terms in the high-temperature expansion,
the asymptotic behaviour of the $J_\pm$ is given by
\begin{align}
  J_{+}\left(x^2\right) \sim & -\frac{7 \pi^4}{360}+\frac{\pi^2}{24}
                               x^2 \\
  J_{-}\left(x^2\right)  \sim & -\frac{\pi^4}{45}+\frac{\pi^2}{12} x^2.
\end{align}
Inserting this high-temperature expansion in (\ref{eq:vtae}), we find
the asymptotic behaviour of $V^T(\omega,T)$ as 
\beq
V^T(\omega,T) &\sim& -T^4\frac{\pi^2}{720}\left(8 n_{\text{Higgs}} -
  (2 \cdot 7) n_{\text{fermion}} + (3 \cdot 8) n_{\text{gauge}}\right)
\nonumber \\
&+& \frac{T^2}{24} \left[\mathrm{Tr}\left(\Lambda_{(S)}^{x y}\right) -
  \mathrm{Tr}\left(\Lambda_{(F)}^{x y}\right) + 3
  \mathrm{Tr}\left(\Lambda_{(G)}^{x y}\right)  \right] \;.
\eeq
We do not expand the daisy corrections in the high-temperature limit,
but explicitly factor out their dependence on the temperature
as
\beq
\Pi_{(S)}^{x y} = T^2 \tilde\Pi_{(S)}^{x y}
\eeq
and
\beq
\Pi_{(G)}^{a b} = T^2\tilde\Pi_{(G)}^{a b} \,,
\eeq
where the tilde denotes that these matrices are explicitly temperature
independent. The eigenvalues of $\Pi_{(S)}^{x  y}$ and $\Pi_{(G)}^{a
  b}$ can be written as $T^2 \widetilde m_{i}^2$ and $T^2
\widetilde m_{a}^2$, respectively, where $\widetilde m_{i}^2$ and $\widetilde m_{a}^2$
are the eigenvalues of $\tilde\Pi_{(S)}^{x y}$ and $\tilde\Pi_{(G)}^{a
    b}$, respectively. They are temperature independent. At high
temperature, we expect $\overline m_{i}^2$ ($\overline m_{a}^2$) and
$T^2\widetilde m_{i}^2$ ($T^2\widetilde m_{a}^2$) to differ by a
perturbative effect induced by $\Lambda_{(S)}^{x y}$ ($\Lambda_{(G)}^{a
  b}$). Similar to perturbation theory in quantum mechanics and using
the fact that the $\Pi_{(S)}^{x y}$ and $\Pi_{(G)}^{a b}$ are
hermitian, the shift of the mass eigenvalues is given by 
%
\begin{align}
  \overline m^2_{i} &\sim T^2\widetilde m^2_{i} + \frac{\vec{\psi}_i \cdot \Lambda^{xy}_{(S)} \cdot \vec{\psi}_i}{(\vec{\psi}_{i})^2}  \\
\overline m^2_{a} &\sim T^2\widetilde m^2_{a} + \frac{\vec{\psi}_a
                    \cdot \Lambda^{ab}_{(G)} \cdot
                    \vec{\psi}_a}{(\vec{\psi}_{a})^2} \,,
\end{align}
where $\vec\psi_{i}$ ($\vec\psi_a$) are the eigenvectors of $\Pi_{(S)}^{x
  y}$ ($\Pi_{(G)}^{a b}$) associated with the eigenvalue $T^2\widetilde
m^2_{i}$ ($T^2\widetilde m^2_{a}$). With this, we can write the asymptotic
behaviour of the daisy-corrected potential as
\begin{align}
V_{\text {daisy }}(\omega, T)  \sim-\frac{T}{12
  \pi}\Bigg[&\sum_{i=1}^{n_{\text {Higgs
              }}}\left(\left(T^2\tilde{m}_i^2\right)^{3 / 2} +
              \frac{3}{2}T\left(\widetilde
              m_i^2\right)^{1/2}\frac{\vec{\psi}_i \cdot
              \Lambda^{xy}_{(S)} \cdot
              \vec{\psi}_i}{(\vec{\psi}_{i})^2}\right)+ \nonumber\\
& \sum_{a=1}^{n_{\text {gauge}}}\left(\left(T^2\tilde{m}_a^2\right)^{3 / 2} + \frac{3}{2}T\left(\widetilde m_a^2\right)^{1/2}\frac{\vec{\psi}_a \cdot \Lambda^{ab}_{(G)} \cdot \vec{\psi}_a}{(\vec{\psi}_{a})^2}\right)\Bigg]. 
\end{align}

In the Parwani scheme, the asymptotic behaviour of $V^T(\omega,T)$ can be written as
\begin{align}V^T(\omega, T)&=\sum_{X=S, G, F}(-1)^{2 s_X}\left(1+2 s_X\right) \frac{T^4}{2 \pi^2} J_{ \pm}\left(\frac{\Lambda_{(X)}^{x y}+\Pi^{xy}_{(X)}}{ T^2 }\right),\nonumber\\
&=\sum_{X=S, G, F}(-1)^{2 s_X}\left(1+2 s_X\right) \frac{T^4}{2 \pi^2} \sum_{i \in X }J_{ \pm}\left(\frac{\overline{m}^2_i}{T^2}\right),\nonumber\\&\sim\sum_{X=S, G, F}\left(1+2 s_X\right) \frac{(-1)^{2 s_X}}{2 \pi^2}\sum_{i \in X }\left[ T^4 J_{ \pm}\left(
\widetilde m_i^2 \right) + T^2 \frac{\vec{\psi}_i \cdot \Lambda_{(X)} \cdot \vec{\psi}_i}{(\vec{\psi}_{i})^2} J_{ \pm}'\left(
\widetilde m_i^2 \right)\right],\end{align}
where we used the same definitions as in the last section. In this scheme, the daisy corrections also affect the Coleman-Weinberg potential 
\begin{align}
  &V^{\mathrm{CW}}(\omega)\to V^{\mathrm{CW}}(\omega,T)\nonumber\\ &= \frac{\varepsilon}{4}\sum_{X=S.G.F}(-1)^{2s_{X}}\left(1+2s_{X}\right)\mathrm{Tr}\left[\left(\Lambda_{(X)}^{x y}+\Pi_{(X)}^{x y}\right)^{2}\left(\log\left(\frac{1}{\mu^{2}}\left(\Lambda_{(X)}^{x y}+\Pi_{(X)}^{x y}\right)\right)-k_{X}\right)\right],\nonumber\\ 
&\sim \frac{\varepsilon}{4}\sum_{X=S.G.F}(-1)^{2s_{X}}\left(1+2s_{X}\right)\left(T^2 \log T^2\right)\mathrm{Tr}\left[T^2\left(1+\frac{\log\tilde\Pi^{xy}_{(X)}}{\log T^2}\right)\left(\tilde\Pi^{xy}_{(X)}\right)^2+\left\{\Lambda_{(X)}^{x y},\tilde\Pi_{(X)}^{x y}\right\}\right].
\end{align}

Using these results, we can factor out the temperature dependence from
the effective potential in the two different approaches as ($AS
\equiv  \text{Arnold-Espinosa}$ $P \equiv \text{Parwani}$) and arrive
at 
\begin{align}
  \left(\frac{V_\text{eff}}{T^2}\right)_\text{Arnold-Espinosa}  & \sim \text{(const.)}_{AS} \cdot T^2 + V_{AS} + \vec G_{AS} \cdot\vec\omega + \vec\omega  \cdot \frac{H_{AS}}{2} \cdot\vec\omega\label{eq:arnold}\\
\left(\frac{V_\text{eff}}{T^2\log T^2}\right)_\text{Parwani}  &\sim
\text{(const.)}_{P,1} \cdot T^2+\text{(const.)}_{P,2}  \cdot \frac{T^2}{\log  T^2} +
V_{P}+ \vec G_{P} \cdot\vec\omega + \vec\omega\cdot\frac{H_{P}}{2}\cdot\vec\omega\;.\label{eq:parwani}
\end{align}
The rescaled potentials Eq.~(\ref{eq:potrescal}) for both schemes
have a field-independent temperature dependence. In the investigation of
  the electroweak symmetry restoration (EWSR) of the potential in the high-temperature
  limit we can therefore ignore the first
  term in Eq.~(\ref{eq:arnold}) and Eq.~(\ref{eq:parwani}), respectively. The remaining 
potential parameters $V_{AS},V_{P},\vec G_{AS},\vec
G_{P},H_{AS}$, and $H_{P}$ are field independent. The relevant 
part of the effective potential for our investigation is just a
quadratic function in the fields $\vec\omega$, which has a
field-independent Hessian. Therefore, if $H_{AS}/H_P$ has a 
negative eigenvalue this means that the potential has a concavity in one of
the directions. Therefore the minimum, if it exists, must be outside the region where the high temperature expansion holds. If the smallest eigenvalue is zero, then
there is an infinite number of degenerate VEVs and more orders 
  in the high-temperature expansion are  
needed in order to lift this degeneracy, which is not considered in
this paper. If the Hessian is positive definite then there exists a
single minimum in the region where the high temperature expansion is valid, and it is located at the VEV,
given by  
\begin{align}
  \langle\vec\phi\rangle^{AS}_{T\to\infty} &= H^{-1}_{AS}G_{AS}\quad \text{for the Arnold-Espinosa scheme}\\
\langle\vec\phi\rangle^P_{T\to\infty}&= H^{-1}_{P}G_{P}\quad \text{for the Parwani scheme.}\end{align}

The flag {\tt --checkewsr=} allows for the check of
electroweak symmetry restoration (EWSR) at high temperature.  
The results of this check are reported in the column {\tt
  status\_ewsr} that is added in the output file. For the EWSR
calculation we iteratively calculate the Hessian matrix of the
rescaled potential at the origin $\vec\omega_0 = \{0,\,\cdots,\,0\}$
until its behaviour is temperature independent, allowing us to
determine $\vec G_{AS}$ ($\vec G_{P}$) and $H_{AS}$ ($H_{P}$), respectively, and,
consequently, the shape of the potential. Iteratively means that we calculate the smallest eigenvalue
 of the Hessian matrix of $V_{\text{eff}}/T^2$, i.e.~the lowest mass value,
 at three different temperatures until the relative difference of the
 obtained mass values is $\lsim 10^{-6}$. \s

The four options that can be set for the flag\ {\tt --checkewsr=} are
\begin{labeling}{{\tt keep\_ewsr}\quad}
\item[{\tt on}] Enables the check and saves the result without removing any point.
\item[\tt keep\_pos\_def] Enables the check and removes all points
  that do not have a positive definite Hessian in the high temperature
  expansion.  (Only for {\tt CalcTemps} 
  and {\tt CalcGW}.) 
\item[{\tt keep\_ewsr}] Enables the check and removes all points with no
  electroweak symmetry restoration. (Only for {\tt CalcTemps} and {\tt
    CalcGW}.)
\item[{\tt off}] Disables the check. The {\tt status\_ewsr}-column in
  this case is filled with {\tt off}.  
\end{labeling}

The possible {\tt status\_ewsr} codes that can be reported in the
output file and their respective meaning are 
\begin{labeling}{{\tt ew\_sym\_non\_res}\quad}
\item[{\tt off}] The test was disabled.  
\item[{\tt failure}] The check failed, because the numerical precision
  was not sufficient. 
\item[{\tt non\_pos\_def}] The potential does not have a positive definite Hessian.
\item[{\tt flat\_region}] There is an infinite number of degenerate
  VEVs that minimise the rescaled potential.
\item[{\tt ew\_sym\_non\_res}] There is a single minimum at high temperature that does not restore the EW symmetry.
\item[{\tt ew\_sym\_res}] There is a single minimum at high temperature that restores the EW symmetry. 
\end{labeling}

It is important to note that our conclusions are only valid inside
  the region where the high-temperature expansion holds. If we
  predict that a parameter point restores the EW symmetry at high
  temperature, this does not rule out the possibility of another
  minimum with a lower energy than the EW-restoring minimum. The same
  reasoning applies to a parameter point with a non-positive definite
  Hessian which indicates that no minimum exists inside the region
  where the high-temperature expansion applies. Still, a minimum could
  exist outside this region. The conclusion that can definitely be drawn,
  however, is that if we find that the Hessian is not positive
  definite then there is no EWSR.

\FloatBarrier
\subsection{The Executable \texttt{CalcTemps}}\label{sec:CalcTemps}
Based on the information obtained from the tracing of the phases in a
temperature interval, we calculate characteristic temperatures for all
found coexisting phase pairs. 
The {\tt CalcTemps} executable is an interface to obtain these
temperature values directly and therefore extends the {\tt
  MinimaTracer} algorithm by additional steps to
solve the bounce equation and derive the critical, nucleation,
percolation and completion temperature, as described in Sec.~\ref{sec:characteristic_temperature_scales}.
Calling {\tt CalcTemps} without arguments, {\tt ./bin/CalcTemps}, or
with the {\tt --help} flag, {\tt ./bin/CalcTemps --help}, prints out the
following menu: 
\begin{lstlisting}
CalcTemps calculates characteristic temperatures for phase transitions
it is called by

	./bin/CalcTemps model input output firstline lastline

or with arguments

	./bin/CalcTemps [arguments]

with the following arguments, ([*] are required arguments, others are optional):

argument                   default   description
--help                               shows this menu
--model=                             [*] model name
--input=                             [*] input file (in tsv format)
--output=                            [*] output file (in tsv format)
--firstline=                         [*] line number of first line in input file
                                         (expects line 1 to be a legend)
--lastline=                          [*] line number of last line in input file
--thigh=                   300       high temperature [GeV]
--multistepmode=           default   multi-step PT mode
                                     default: default mode
                                     0: single-step PT mode
                                     >0 for multi-step PT modes:
                                     1: tracing coverage
                                     2: global minimum tracing coverage
                                     auto: automatic mode
--num_pts=                 10        intermediate grid-size for default mode
--vwall=                   0.95      wall velocity: >0 user defined
                                     -1: approximation
                                     -2: upper bound
--perc_prbl=               0.71      false vacuum fraction for percolation
--compl_prbl=              0.01      false vacuum fraction for completion
--checknlo=                on        check for NLO stability
                                     on: only keep NLO stable points
                                     off: check disabled
--checkewsr=               on        check for EWSR at high temperature
                                     on: perform check and add info
                                     keep_bfb: only keep BFB points
                                     keep_ewsr: only keep EWSR points
                                     off: check disabled
--maxpathintegrations=     7         number of solutions of 1D equation =
                                     number of path deformations + 1
--usegsl=                  true      use GSL library for minimization
--usecmaes=                true      use CMAES library  for minimization
--usenlopt=                true      use NLopt library for minimization
--usemultithreading=       false     enable multi-threading for minimizers
--json=                              use a json file instead of cli parameters
\end{lstlisting}
Again, the required flags to set are the name of the model to
investigate {\tt --model=}, the name of the input
file in tsv-format {\tt 
  --input=}, the name of the output file in
  tsv-format {\tt --output=}, and the line number of the first and
the last line in the input file, {\tt --firstline=} and {\tt --lastline=}, respectively. 
A minimal example call is
\begin{lstlisting}
./bin/CalcTemps --model=MODEL --input=input.tsv --output=output.tsv --firstline=2 --lastline=2
\end{lstlisting}
Optionally, it is again possible to specify the temperature range in which to trace the phases, whether or not the check for NLO vacuum stability or the check for electroweak symmetry restoration at high
temperature is enabled and which mode should be used to handle
multi-step phase transitions. Note, that contrary to {\tt
  MinimaTracer}, if {\tt --checknlo=on}, {\tt no\_nlo\_stability} acts
as an error code, and {\tt --checkewsr=keep\_bfb} or {\tt
  --checkewsr=keep\_ewsr} only keep points that are bounded-from-below
or restore the EW symmetry at high temperature, respectively. The wall
velocity can be set via the flag {\tt --vwall=}. The 
different options are:  
\begin{labeling}{{\tt >0}\quad}
    \item[{\tt >0}] If a value $\in (0,1)$ is given, the wall velocity is set to this value. By default, if no flag is provided, the wall velocity is set to \num{0.95}.
    \item[{\tt -1}] For {\tt --vwall=-1} the approximation, see Eq.~(\ref{eq:vwallapprox}), from Refs.~\cite{Lewicki2022,Ellis2023} is chosen.
    \item[{\tt -2}] For {\tt --vwall=-2} the upper bound, see Eq.~(\ref{eq:vwallupperbound}), defined in \cite{Athron:2023xlk} is chosen.
\end{labeling}
Additionally, it is possible to define the false vacuum fraction used
to define the percolation and the completion temperature via the flags
{\tt --perc\_prbl} and the {\tt --compl\_prbl}, respectively.
By default, the percolation false vacuum fraction is set to \SI{71}{\%}, {\tt --perc\_prbl=0.71}, and the completion false vacuum fraction to \SI{1}{\%}, {\tt --compl\_prbl=0.01}.
By setting the optional {\tt --maxpathintegrations=} flag one can
specify the number of solutions to the 1D equation which equals the
number of path deformations plus one. 
Note that the choice of the number of path integrations ideally finds
a good (model-dependent) balance between the number of attempts and
computational time. 
All other optional flags and the {\tt Logger} classes work in the same
way as for {\tt MinimaTracer}, cf.~Sec.~\ref{sec:MinimaTracer}.\s

A successful run of the {\tt CalcTemps} executable attaches the following columns to {\tt input.tsv} and creates and saves the output to {\tt output.tsv}.
The first columns report on several status codes whose output partially depends on the set flags:
\begin{labeling}{{\tt status\_nlo\_stability}\quad}
\item[{\tt status\_nlo\_stability}]
Information on whether the global minimum 
      of the loop-corrected effective potential at $T=0$~GeV
      coincides with the global minimum of the tree-level
      potential.
  {\tt success} if the point is found to be NLO stable when {\tt --checknlo=on}, if not {\tt no\_nlo\_stability} discards the point and {\tt off} indicates that the check is disabled with {\tt --checknlo=off}.
    \item[{\tt status\_ewsr}] Information on electroweak symmetry restoration at high temperature, all details can be found in Sec.~\ref{sec:ewsrcheck}.
    \item[{\tt status\_tracing}] Status of the minima tracing, see Sec.~\ref{sec:multistepmode}.
    \item[{\tt status\_coex\_pairs}] If the tracing is successful this
      column informs on whether ({\tt success}) or not ({\tt
        no\_coex\_pairs}) coexisting phases are found.
    \item[{\tt runtime}] Runtime of code in seconds.
\end{labeling}
More details on all status codes can be found in Sec.~\ref{sec:status_codes}. 
If pairs of coexisting phases can be identified, we then try to obtain a critical temperature for each pair {\tt i} of coexisting phases:
\begin{labeling}{{\tt omega\_X\_crit\_false\_i}\quad} 
\item[{\tt status\_crit\_i}] If for a phase pair coexisting in
  $\{T_{{\tt i},\,\text{high}},\,T_{{\tt i},\,\text{low}}\}$ we have
  $\Delta V(T_{{\tt i},\,\text{high}}) > 0$ and $\Delta V(T_{{\tt
      i},\,\text{low}}) < 0$ with $\Delta V(T) \equiv
  V_\text{true}(T) - V_\text{false}(T)$, the critical temperature is
  identified via binary search between $T_{{\tt i},\,\text{high}}$ and
  $T_{{\tt i},\,\text{low}}$ and {\tt success} is reported in the
  status column.  
        If $\Delta V < 0$ in the whole range of coexistence, the true
        phase is always the lower minimum. We then set $T_c = T_{ {\tt
            i},\,\text{high}}$, and the reported status is {\tt
          true\_lower}. 
        If $\Delta V > 0$ over the whole range of coexistence, the
        false phase is always the lower minimum and there is no
        critical temperature for this pair, the reported error is {\tt
          false\_lower}. 
        The identification of the critical temperature for a pair of
        (false, true) phases fails with {\tt failure} if the true phase
        starts as a lower minimum at $T_{ {\tt i},\,\text{high}}$ and
        the false phase ends as a lower minimum at $T_{ {\tt
            i},\,\text{low}}$. 
  \item[\smash{\tshortstack[l]{{\tt T\_crit\_i},\,\\{\tt omega\_X\_crit\_false\_i},\,\\{\tt omega\_X\_crit\_true\_i}}}] This set of columns for phase pair {\tt i} contains
  information about the critical temperature $T_c$ in [\si{GeV}], the
  coordinates of the false vacuum and the coordinates of the true
  vacuum at the critical temperature.   
\end{labeling}
If a critical temperature can be identified successfully for a coexisting phase pair {\tt i}, the next step is to solve the bounce equation and extend the output by the following columns:
\begin{labeling}{{\tt omega\_X\_nucl\_approx\_false\_i}\quad}
    \item[{\tt status\_bounce\_sol\_i}] If a bounce solution for pair
      {\tt i} can be calculated, the status is {\tt success}, and the
      derivation of the nucleation, percolation and completion
      temperatures is attempted. 
If the calculation of the bounce solution fails, due to e.g. too small
overlap, the status is {\tt failure}, and no nucleation, percolation and completion
temperature can be calculated for this transition. 
  \item[{\tt status\_nucl\_approx\_i}] {\tt success} if
      Eq.~(\ref{eq:nucl_approx}) can be met, {\tt 
      not\_met} if not.
\item[\smash{\tshortstack[l]{{\tt T\_nucl\_approx\_i},\,\\{\tt omega\_X\_nucl\_approx\_false\_i},\,\\{\tt omega\_X\_nucl\_approx\_true\_i}}}] Attached next are the columns for the approximate nucleation temperature
  $T_n$ obtained from Eq.~(\ref{eq:nucl_approx}) and the
  false and true phase coordinates at this temperature, respectively. 
\item[{\tt status\_nucl\_i}] {\tt success} if
  Eq.~(\ref{eq:nucl_exact}) can be met, {\tt not\_met} if not. 
\item[\smash{\tshortstack[l]{{\tt T\_nucl\_i},\,\\{\tt omega\_X\_nucl\_false\_i},\,\\{\tt omega\_X\_nucl\_true\_i}}}] Contains the nucleation temperature 
  $T_n$ derived from Eq.~(\ref{eq:nucl_exact}) and the
  false and true phase coordinates at $T_n$, respectively.\\
\item[{\tt status\_perc\_i}] {\tt success} if Eq.~(\ref{eq:perc}) with $P_f(T_p)$ optionally set by {\tt --perc\_prbl} can be met, {\tt not\_met} if not.
\item[\smash{\tshortstack[l]{{\tt T\_perc\_i},\,\\{\tt omega\_X\_perc\_false\_i},\,\\{\tt omega\_X\_perc\_true\_i}}}] Reports the percolation 
      temperature $T_p$ derived from Eq.~(\ref{eq:perc}) and the false
      and true phase coordinates at $T_p$, respectively.\\
\item[{\tt status\_compl\_i}] {\tt success} if Eq.~(\ref{eq:compl})
  with $P_f(T_f)$ optionally set by {\tt --perc\_prbl} can be met,
  {\tt not\_met} if not. 
\item[\smash{\tshortstack[l]{{\tt T\_compl\_i},\,\\{\tt omega\_X\_compl\_false\_i},\,\\{\tt omega\_X\_compl\_true\_i}}}] Informs on the completion 
  temperature $T_f$ derived from Eq.~(\ref{eq:compl}) and the false
  and true phase coordinates at $T_f$, respectively.\\
\end{labeling}
Note, that an error message {\tt not\_met} might indicate vacuum trapping.
The last added column, {\tt transition\_history}, reports on the history of transitions that likely took place for the point. For details, compare Sec.~\ref{subsec:transtracer} as well as see the examples in Sec.~\ref{sec:examples}.

\subsection{The Executable  \texttt{CalcGW}}\label{sec:CalcGW}
Based on the tracing of the phases in the temperature interval $T_\text{low}=\SI{0}{GeV}\leq T\leq T_\text{high}$, the identification of coexisting phase pairs and the determination of the characteristic temperatures, the {\tt CalcGW} executable provides the calculation of the spectrum of primordial gravitational waves sourced by sound waves and turbulence.
The used terminology is introduced in Sec.~\ref{sec:GW_spectrum}.
Running {\tt ./bin/CalcGW} or {\tt ./bin/CalcGW --help} prints the
following menu, specifying all required and optional arguments:
\begin{lstlisting}
CalcGW calculates the gravitational wave signal
it is called by

	./bin/CalcGW model input output firstline lastline

or with arguments

	./bin/CalcGW [arguments]

with the following arguments, ([*] are required arguments, others are optional):

argument                   default   description
--help                               shows this menu
--model=                             [*] model name
--input=                             [*] input file (in tsv format)
--output=                            [*] output file (in tsv format)
--firstline=                         [*] line number of first line in input file
                                         (expects line 1 to be a legend)
--lastline=                          [*] line number of last line in input file
--thigh=                   300       high temperature [GeV]
--multistepmode=           default   multi-step PT mode
                                     default: default mode
                                     0: single-step PT mode
                                     >0 for multi-step PT modes:
                                     1: tracing coverage
                                     2: global minimum tracing coverage
                                     auto: automatic mode
--num_pts=                 10        intermediate grid-size for default mode
--vwall=                   0.95      wall velocity: >0 user defined
                                     -1: approximation
                                     -2: upper bound
--perc_prbl=               0.71      false vacuum fraction for percolation
--compl_prbl=              0.01      false vacuum fraction for completion
--trans_temp=              perc      transition temperature, options are:
                                     nucl_approx: approx nucleation temperature
                                     nucl: nucleation temperature
                                     perc: percolation temperature
                                     compl: completion temperature
--epsturb=                 0.1       turbulence efficiency factor
                                     >0: user defined
                                     -1: upper bound
--pnlo_scaling=            1         1 -> N NLO pressure
                                     1: propto gamma
                                     2: propto gamma^2
--checknlo=                on        check for NLO stability
                                     on: only keep NLO stable points
                                     off: check disabled
--checkewsr=               on        check for EWSR at high temperature
                                     on: perform check and add info
                                     keep_bfb: only keep BFB points
                                     keep_ewsr: only keep EWSR points
                                     off: check disabled
--maxpathintegrations=     7         number of solutions of 1D equation =
                                     number of path deformations + 1
--usegsl=                  true      use GSL library for minimisation
--usecmaes=                true      use CMAES library  for minimization
--usenlopt=                true      use NLopt library for minimization
--usemultithreading=       false     enable multi-threading for minimizers
--json=                              use a json file instead of cli parameters
\end{lstlisting}
In addition to the previously described required and optional arguments, cf.~Secs.~\ref{sec:MinimaTracer}-\ref{sec:CalcTemps}, {\tt CalcGW} allows the user to set the transition temperature.
By default, it is set to $T_*=T_p$, and by specifying {\tt --trans\_temp=} one can choose:
\begin{labeling}{{\tt nucl\_approx}\quad}
    \item[{\tt nucl\_approx}] Nucleation temperature determined via the approximation of Eq.~(\ref{eq:nucl_approx}).
    \item[{\tt nucl}] Nucleation temperature determined via the condition, Eq.~(\ref{eq:nucl_exact}).
    \item[{\tt perc}] Percolation temperature evaluated via Eq.~(\ref{eq:perc}). Note that the false vacuum fraction used to determine the percolation temperature can be set optionally with {\tt --perc\_prbl}.
    \item[{\tt compl}] Completion temperature calculated via Eq.~(\ref{eq:compl}). Note that the false vacuum fraction used to determine the completion temperature can be set optionally with {\tt --perc\_compl}.
\end{labeling}
The {\tt --epsturb} flag allows the user to set the
  turbulence efficiency factor in case a value $>0$ is entered. Alternatively, for {\tt -1}, the upper bound of Eq.~\eqref{eq:eps_turb} can be chosen.
With the flag {\tt --pnlo\_scaling=1,2} the user can switch between
the NLO pressure with the linear $\gamma$ scaling of
Eq.~\eqref{eq:pnlo1} or with the $\gamma^2$ scaling of
Eq.~\eqref{eq:pnlo2}. By default, the linear $\gamma$ scaling is used.
A minimal example call can look like:
\begin{lstlisting}
./bin/CalcGW --model=MODEL --input=input.tsv --output=output.tsv --firstline=2 --lastline=2
\end{lstlisting}
The first columns added to {\tt input.tsv} in {\tt output.tsv} are
{\tt status} columns, compare again with Secs.~\ref{sec:MinimaTracer}-\ref{sec:CalcTemps} and Sec.~\ref{sec:status_codes} for a summary of all status codes.
Then for each identified coexisting phase pair {\tt i}, the columns containing information on the bounce solution and characteristic temperatures are added, cf.~Sec.~\ref{sec:CalcTemps}.
In addition, the information on the gravitational wave spectrum is
given out in the following columns with the respective contents: 
\begin{labeling}{{\tt SNR(LISA-3yrs)\_turb\_i}\quad}
    \item[{\tt status\_gw\_i}] Status of the gravitational wave
      calculation, {\tt success} if successful, {\tt failure} if an
      error was encountered. Possible encountered errors are that
      the requested transition temperature could not be calculated or
      that $\frac{\beta}{H} < 1$, c.f. Sec.~\ref{sec:GW_spectrum}. 
    \item[{\tt T\_star\_i}] Transition temperature $T_*$.
    \item[{\tt T\_reh\_i}] Reheating temperature $T_\mathrm{reh}$.
    \item[{\tt v\_wall\_i}] Wall velocity.
    \item[{\tt alpha\_PT\_i}] Strength of the phase transition, Eq.~(\ref{alpha}).
    \item[{\tt beta/H\_i}] Inverse time scale, Eq.~(\ref{betaH}).
    \item[{\tt kappa\_col\_i}] Efficiency factor for bubble collisions as defined in Eq.~\eqref{eq:kappa_coll}.
     \item[{\tt kappa\_sw\_i}] Sound-wave efficiency factor derived with the method from \cite{Giese:2020rtr,Giese:2020znk}, as described in Sec.~\ref{sec:GW_spectrum}.
    \item[{\tt eps\_turb\_i}] Efficiency factor for the
      turbulence contribution as defined in
      Eq.~\eqref{eq:kappa_turb}.
    \item[{\tt cs\_f\_i}] Sound velocity in the false vacuum as defined in Eq.~\eqref{eq:sound_speed}.
    \item[{\tt cs\_t\_i}] Sound velocity in the true vacuum as defined in Eq.~\eqref{eq:sound_speed}.
    \item[{\tt fb\_col\_i}] Characteristic frequency of the collision spectrum as defined in Eq.~\eqref{eq:fb}
    \item[{\tt h2Omegab\_col\_i}] Amplitude of the collision spectrum as defined in Eq.~\eqref{eq:Omegab}
    \item[{\tt f\_1\_X\_i}] First characteristic
      frequency break for the sound wave and turbulence source, with {\tt X = sw/turb}, as defined in Eqs.~\eqref{eq:f1_sw} and~\eqref{eq:f1_turb}.
    \item[{\tt f\_2\_X\_i}] Second characteristic
      frequency breaks for the sound wave and turbulence source, with {\tt X = sw/turb}, as defined in Eqs.~\eqref{eq:f2_sw} and~\eqref{eq:f2_turb}.
    \item[{\tt h2Omega\_2\_X\_i}] Amplitude $\Omega_2$
      for sound waves, {\tt X = sw}, and turbulence, {\tt X = turb},
      as defined in Eq.~\eqref{eq:Omega2_sw}
      and~\eqref{eq:Omega2_turb}, respectively.
  \item[{\tt SNR(LISA-3yrs)\_col\_i}] Signal-to-noise ratio (SNR) at {\tt LISA} with an acquisition period of three years, given by Eq.~(\ref{eq:SNRLISA}), for the collision contribution only.
    \item[{\tt SNR(LISA-3yrs)\_sw\_i}] Signal-to-noise
    ratio for the sound-wave contribution only.
    \item[{\tt SNR(LISA-3yrs)\_turb\_i}] SNR for the turbulence contribution only.
    \item[{\tt SNR(LISA-3yrs)\_i}] SNR for the collision, sound-wave and the
      turbulence contribution combined.
\end{labeling}

The last added column, {\tt transition\_history}, reports on the history of transitions that likely took place for the point. For details, compare Sec.~\ref{subsec:transtracer} as well as see the examples in Sec.~\ref{sec:examples}.


\subsection{The Executable \texttt{PotPlotter}}\label{sec:PotPlotter}
Visualizing the multi-dimensional effective potential  often is useful for understanding complicated minima landscapes.
The executable {\tt PotPlotter} provides an interface for extracting
(multi-dimensional) effective potential data grids that can be used to
generate different kinds of contour plots. If {\tt ./bin/PotPlotter
  --help} is called, its menu is printed: 
\begin{lstlisting}
PotPlotter calculates the effective potential on a user-specified field grid
it is called by

	./bin/PotPlotter [arguments]

with the following arguments, ([*] are required arguments, others are optional):

argument                   default   description
--help                               shows this menu
--model=                             [*] model name
--input=                             [*] input file (in tsv format)
--output=                            [*] output file (in tsv format)
--line=                              [*] line number of line in input file
                                         (expects line 1 to be a legend)
--temperature=                       [*] temperature [GeV]
--point=                   0,..,0    grid reference point
--npointsi=                0         number of points in direction i
                                     (with i = [1,..,6])
--lowi=                    0         lowest field value in direction i
                                     [* if npointsi > 0] (with i = [1,..,6])
--highi=                   0         highest field value in direction i
                                     [* if npointsi > 0] (with i = [1,..,6])
--slice=                   false     enable slice mode
--min_start=                         [* in slice mode] start minimum
--min_end=                           [* in slice mode] end minimum
--npoints=                 100       grid size in slice mode
--json=                              use a json file instead of cli parameters
\end{lstlisting}
The user has to specify the model, the input and output files and the
line number, as well as the temperature, at which the
contour is to be evaluated. 
Furthermore, one of the two different operation modes of {\tt
  PotPlotter} has to be chosen. The two modes work as follows:
    \begin{labeling}{{\tt slice mode}\quad}
    \item[{\tt grid mode}] The potential values are evaluated on a
      user-defined grid that lies along the field directions of the
      model, in which we (model-specifically) allow for the generation
      of a non-zero finite temperature VEV, called VEV directions in
      the following.  
        The user has to specify the number of grid points and the grid
        ranges in all VEV directions in which the grid should span by setting {\tt --npointsi=}, {\tt --lowi=} and {\tt --highi=} to the desired values.
        Note, that the index {\tt i} runs from {\tt 1} to {\tt n} with {\tt n} being the total number of VEV directions.
        The order of the VEV coordinates is set in the model file and can be read e.g. from the model-specific implementation of {\tt addLegendVEV()}.
        At the moment, up to six field dimensions are possible for a grid, for higher-dimensional VEV spaces, the algorithm needs to be extended.
        Optionally, the user can force the evaluation of the point grid with all VEV dimensions that are not axes of the grid set to the coordinates of a reference point.
        The reference point coordinates are supplied via {\tt --point=x1,..,xn}.
        If no reference point is specified, all VEV coordinates that are not varied in the grid are set to zero.
        This is useful if a user wants to display a lower-dimensional projection of a higher-dimensional VEV space.
    \item[{\tt slice mode}] The potential values are evaluated along a straight line between two user-defined points. 
        The result is a one-dimensional array of potential values along this one-dimensional path.
        In order to enable the {\tt slice} mode, the user has to set
        {\tt --slice=true} and specify the coordinates of the two
        points via {\tt--min\_start} and {\tt--min\_end}.
        Again, the order of the VEV coordinates is set in the model file and can be derived e.g. from the model-specific implementation of {\tt addLegendVEV()}.
        Optionally, the number of points along the straight line at
        which the potential gets evaluated, can be changed by setting
        {\tt --npoints=} to the requested number. By default, {\tt --npoints=100} are evaluated.
\end{labeling}
The output of {\tt PotPlotter} is then saved to the output file where each line corresponds to one grid or slice point. The columns are
\begin{labeling}{{\tt Veff(point,T)}\quad}
    \item[{\tt v\_X}] Field value of direction {\tt X} in \si{GeV} for one grid or slice point.
        The labels of the direction {\tt X} are model-specific and defined in {\tt addLegendVEV()} in the respective model file.
    \item[{\tt v\_X\_point}] (only in {\tt grid} mode) Coordinates of the reference point.
    \item[{\tt Veff(v,T)}] Value of the effective potential in \si{GeV} at the grid or slice point and temperature.
    \item[{\tt Veff(point,T)}] (only in {\tt grid} mode) Effective potential value in \si{GeV} at the reference point and temperature.
    \item[{\tt T}] Temperature in \si{GeV} at which the effective potential is evaluated.
\end{labeling}
Examples on how the output of {\tt PotPlotter} can be used for
visualizations can be found in
Figs.~\ref{fig:bp1_contour},\,\ref{fig:bp2_contour}, and
\ref{fig:bp3_contour}. The figures in the respective left columns
were made using the {\tt slice} mode, the ones in the respective middle and the
right columns were made with the {\tt grid} mode with the coordinates
of the global minimum chosen as the reference point of the
two-VEV-dimensional projection.  


\subsection{The Folder \texttt{standalone}}\label{sec:standalone}
In case the user wants to use some particular function or class of
\texttt{BSMPTv3}, such as those explained in the last sections, we
also provide a few examples on how to do so. They are placed
in the folder \texttt{standalone} and automatically compiled when
\texttt{BSMPTv3} is compiled. If new \texttt{.cpp} files are
created/moved into \texttt{standalone} then it is necessary to run
\texttt{CMake} and compile again so that all libraries are properly
linked. Three examples are already put inside \texttt{standalone}:
\begin{labeling}{{\tt GenericPotential.cpp}\quad}
    \item[{\tt CalculateAction.cpp}] Solves the bounce equation and
      calculates the Euclidean action. The user is expected to provide
      the initial guess path and the potential, the gradient is
      optional. 
    \item[{\tt GenericModel.cpp}] The user provides a potential
      $V(\vec\omega)$, the zero temperature VEV and the dimensionality of
      the VEV directions. This tracks the minima and calculates the
      GWs spectrum.
    \item[{\tt TunnellingPath.cpp}] Solves the bounce equation using
      the full \texttt{BSMPTv3} and prints the tunnelling path and the
      VEV profile in \texttt{Mathematica} and \texttt{Python}
      formats. 
\end{labeling}
Remark that the provided examples merely serve as 
demonstrations of how to use the classes.
In case some functionality is missing, the recommended way of 
extending {\tt BSMPT} is by adding these features in form of functions
which can be tested in unit tests. Our team welcomes suggestions. 
To ensure everything is still working fine, we recommend running the
unit tests during development.

\subsection{Summary on Status Codes}\label{sec:status_codes}
We summarize here all codes in text format to log the status of
several steps of the calculation. 
The new status code framework is used by all executables that were
added with the release of {\tt BSMPTv3}. For information on status
codes of the previous versions, cf.~\cite{Basler:2018cwe, Basler:2020nrq}. 
The status codes are listed and described in the following and
illustrated in Fig.~\ref{fig:logicflow}:
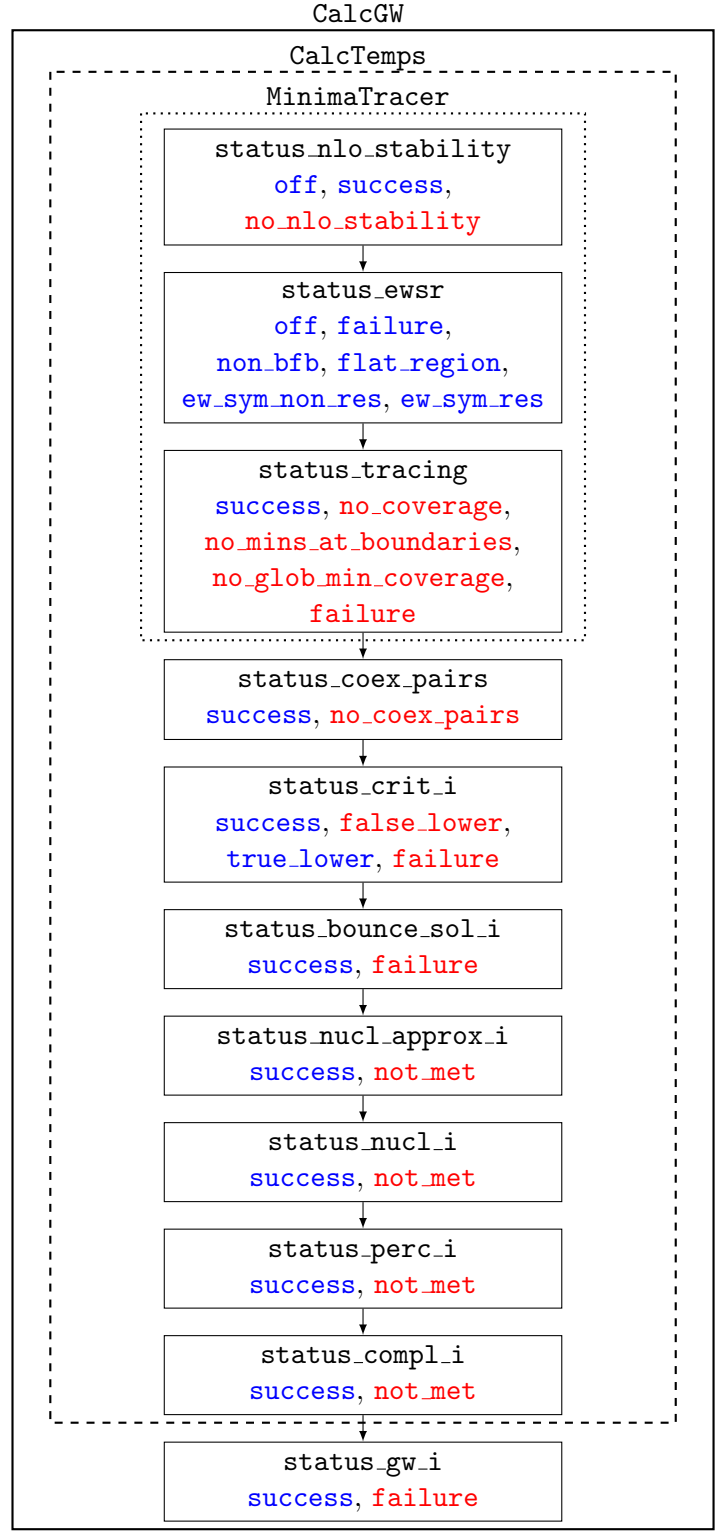
\begin{figure}[t!]
\vspace*{-0.8cm}  
  \centering
  \begin{tikzpicture}[node distance=3.5mm]
      \node[draw,align=center,text width=5cm] (a) at (5,15) {
          {\tt status\_nlo\_stability}\par
          {\color{blue}\tt off}, {\color{blue}\tt success}, {\color{red}\tt no\_nlo\_stability}
      };
      \node[draw,below=of a,align=center,text width=5cm] (b) {
          {\tt status\_ewsr}\par
          {\color{blue}\tt off}, {\color{blue}\tt failure}, {\color{blue}\tt non\_bfb}, {\color{blue}\tt flat\_region}, {\color{blue}\tt ew\_sym\_non\_res}, {\color{blue}\tt ew\_sym\_res}
      };
      \node[draw,below=of b,align=center,text width=5cm] (c) {
          {\tt status\_tracing}\par
          {\color{blue}\tt success}, {\color{red}\tt no\_coverage}, {\color{red}\tt no\_mins\_at\_boundaries}, {\color{red}\tt no\_glob\_min\_coverage}, {\color{red}\tt failure}
      };
      \node[draw,below=of c,align=center,text width=5cm] (d) {
          {\tt status\_coex\_pairs}\par
          {\color{blue}\tt success}, {\color{red}\tt no\_coex\_pairs}
      };
      \node[draw,below=of d,align=center,text width=5cm] (e) {
          {\tt status\_crit\_i}\par
          {\color{blue}\tt success}, {\color{red}\tt false\_lower}, {\color{blue}\tt true\_lower}, {\color{red}\tt failure}
      };
      \node[draw,below=of e,align=center,text width=5cm] (f) {
          {\tt status\_bounce\_sol\_i}\par
          {\color{blue}\tt success}, {\color{red}\tt failure}
      };
      \node[draw,below=of f,align=center,text width=5cm] (g) {
          {\tt status\_nucl\_approx\_i}\par
          {\color{blue}\tt success}, {\color{red}\tt not\_met}
      };
      \node[draw,below=of g,align=center,text width=5cm] (h) {
          {\tt status\_nucl\_i}\par
          {\color{blue}\tt success}, {\color{red}\tt not\_met}
      };
      \node[draw,below=of h,align=center,text width=5cm] (i) {
          {\tt status\_perc\_i}\par
          {\color{blue}\tt success}, {\color{red}\tt not\_met}
      };
      \node[draw,below=of i,align=center,text width=5cm] (j) {
          {\tt status\_compl\_i}\par
          {\color{blue}\tt success}, {\color{red}\tt not\_met}
      };
      \node[draw,below=of j,align=center,text width=5cm] (k) {
          {\tt status\_gw\_i}\par
          {\color{blue}\tt success}, {\color{red}\tt failure}
      };
      \node[black,align=center] (l1) at (5,16.2) {
          {\tt MinimaTracer}
      };
      \node[black,align=center] (l2) at (5,16.7) {
          {\tt CalcTemps}
      };
      \node[black,align=center] (l3) at (5,17.3) {
          {\tt CalcGW}
      };
      \draw[black,thick,dotted] ($(a.north west)+(-0.3,0.2)$) rectangle ($(c.south east)+(0.3,-0.1)$);
      \draw[black,thick,dashed] ($(a.north west)+(-1.5,0.75)$) rectangle ($(j.south east)+(1.5,-0.1)$);
      \draw[black,thick] ($(a.north west)+(-2,1.3)$) rectangle ($(k.south east)+(2,-0.1)$);
      \draw[-latex] (a) -- (b);
      \draw[-latex] (b) -- (c);
      \draw[-latex] (c) -- (d);
      \draw[-latex] (d) -- (e);
      \draw[-latex] (e) -- (f);
      \draw[-latex] (f) -- (g);
      \draw[-latex] (g) -- (h);
      \draw[-latex] (h) -- (i);
      \draw[-latex] (i) -- (j);
      \draw[-latex] (j) -- (k);
  \end{tikzpicture}
  \caption{Logical-flow diagram of {\tt BSMPTv3}. Status codes are marked in blue, error codes in red. If {\tt --checkewsr=keep\_bfb} ({\tt --checkewsr=keep\_ewsr}) the codes {\tt failure}, {\tt non\_bfb} and {\tt flat\_region} ({\tt ew\_sym\_non\_res}) for {\tt status\_ewsr} act as error codes.
  All possible error codes in {\tt status\_nlo\_stability}, {\tt status\_ewsr} and {\tt status\_tracing} only act as status codes for the executable {\tt MinimaTracer}. Codes are described in the text.}\label{fig:logicflow}
\end{figure}

\begin{labeling}{{\tt status\_nlo\_stability}\quad}
    \item[{\tt status\_nlo\_stability}] The NLO stability status is set
      to {\tt success} if the global minimum 
      of the loop-corrected effective potential at $T=\SI{0}{GeV}$
      coincides with the global minimum of the tree-level
      potential. It is set to {\tt no\_nlo\_stability} otherwise.
    \item[{\tt status\_ewsr}] Status of the EWSR
      check, described in Sec.~\ref{sec:ewsrcheck}. If the
  check is enabled, it will be filled with one of the following
  results: {\tt failure} if the test failed; {\tt non\_bfb} if the
  potential is not bounded from below at high temperature; {\tt
    flat\_region} if there is an infinite number of degenerate VEVs
  that minimise the rescaled potential; {\tt ew\_sym\_non\_res} if
  there is a single minimum at high temperature that does not restore
  the electroweak symmetry and {\tt ew\_sym\_res} if there is a single
  minimum at high temperature that restores the EW symmetry.  
\item[{\tt status\_tracing}] Status of the phase tracing
  algorithm. Successful tracing is logged with {\tt success}. The
  tracing fails, if no coverage is found or the global minimum is
  missed for some temperature regions, reported as {\tt no\_coverage}
  and {\tt no\_glob\_min\_coverage}, respectively. 
  If {\tt mode=0} is chosen, meaning that it is searched
  for a one-step first-order phase transition exclusively, an
  error code {\tt no\_mins\_at\_boundaries} indicates that we cannot
  identify a numerically stable local minimum at the edge temperatures
  0 or $T_\text{high}$. Successful tracing in the mode {\tt default} can still mean that the
  global minimum escapes tracing in some temperature regions. In this
  case or in the case of failure, increasing the 
  equidistant point grid size or using a different multi-step phase
  transition tracing mode might help, cf.~also 
  Sec.~\ref{sec:multistepmode}.
  The failure code {\tt failure} is reported if either no phases could be traced or the global minimum at $T=0$~GeV is found at too large field values.
\item[{\tt status\_coex\_pairs}] Status of the check for coexisting phase pairs. 
    If no coexisting phases are found for the point in the whole temperature range, this status is set to {\tt no\_coex\_pairs}, ending the calculation for this parameter point. As soon as at least one coexisting phase pair is identified, {\tt success} is reported.
\item[{\tt status\_crit\_i}] Status of the calculation of the critical temperature for a coexisting phase pair {\tt i}. If the false phase starts as the lower minimum at the upper temperature of the coexisting region and the true phase ends as the lower minimum at the lower temperature of the coexisting region, the critical temperature lies in between and the status is {\tt success}. 
    If the true phase is always the lower minimum, the critical temperature is located at the upper end of the overlap and the status is {\tt true\_lower}.
    If the false phase remains the lower minimum over the whole overlap region with the true phase, there is no critical temperature within the overlap and the error is {\tt false\_lower}. 
    If the false phase is found to be the lower minimum at the low
    temperature and the true phase is found to be the lower minimum at
    the high temperature, the reported error is {\tt failure} and
    there is no critical temperature for the identified pair of
    (false, true) phase.
\item[{\tt status\_bounce\_sol\_i}] Status of the bounce solution
  calculation for coexisting phase pair {\tt i}. {\tt success} if a
  bounce solution can be calculated in the temperature range of the
  phase pair overlap, {\tt failure} otherwise.
\item[{\tt status\_nucl\_approx\_i}] Status of the approximate
  nucleation temperature calculation for coexisting phase pair {\tt
    i}. {\tt success} if Eq.~(\ref{eq:nucl_approx}) can be fulfilled,
  {\tt not\_met} if not.
\item[{\tt status\_nucl\_i}] Status of the nucleation
  temperature calculation for coexisting phase pair {\tt i}. {\tt
    success} if Eq.~(\ref{eq:nucl_exact}) can be fulfilled, {\tt
    not\_met} if not.
\item[{\tt status\_perc\_i}] Status of the percolation temperature
  calculation for coexisting phase pair {\tt i}. {\tt success} if
  Eq.~(\ref{eq:perc}) can be fulfilled, {\tt not\_met} if not.
\item[{\tt status\_compl\_i}] Status of the completion temperature
  calculation for coexisting phase pair {\tt i}. {\tt success} if
  Eq.~(\ref{eq:compl}) can be fulfilled, {\tt not\_met} if not.
\item[{\tt status\_gw\_i}] Status of the gravitational wave calculation for coexisting phase pair {\tt i}. Set to {\tt failure} if the requested transition temperature could not be calculated or a $\frac{\beta}{H} < 1$ is identified, {\tt success} otherwise.
\end{labeling}
\FloatBarrier

\section{Examples and Comparison with {\tt
    CosmoTransitions} \label{sec:examples}}
This section illustrates the functionality and usage of {\tt BSMPTv3}
by discussing some sample parameter points and by performing a
comparison between {\tt BSMPTv3} and {\tt 
  CosmoTransitions}.\footnote{To the best of our
knowledge, {\tt CosmoTransitions}  is the only other code that is
capable of calculating the bounce solution and the critical as well as
the nucleation temperature, where in {\tt CosmoTransitions} the approximation of
Eq.~(\ref{eq:nucl_approx}) is used. } We start in
Sec.~\ref{sec:toymodel} with the comparison of 
the solutions provided by the two codes for the bounce equation in a
toy model. We then  
compare in Sec.~\ref{sec:examplesbps} the phases and phase
transitions for sample benchmark points. In Sec.~\ref{sec:comparison} a
comparison is performed on a broader basis by using a parameter point
sample obtained from a parameter scan in the 2HDM which takes into account all
relevant theoretical and experimental constraints. 

\FloatBarrier
\subsection{Comparison in a Toy Model \label{sec:toymodel}}
\begin{figure}[h!]
\centering
\includegraphics[width=0.8\textwidth]{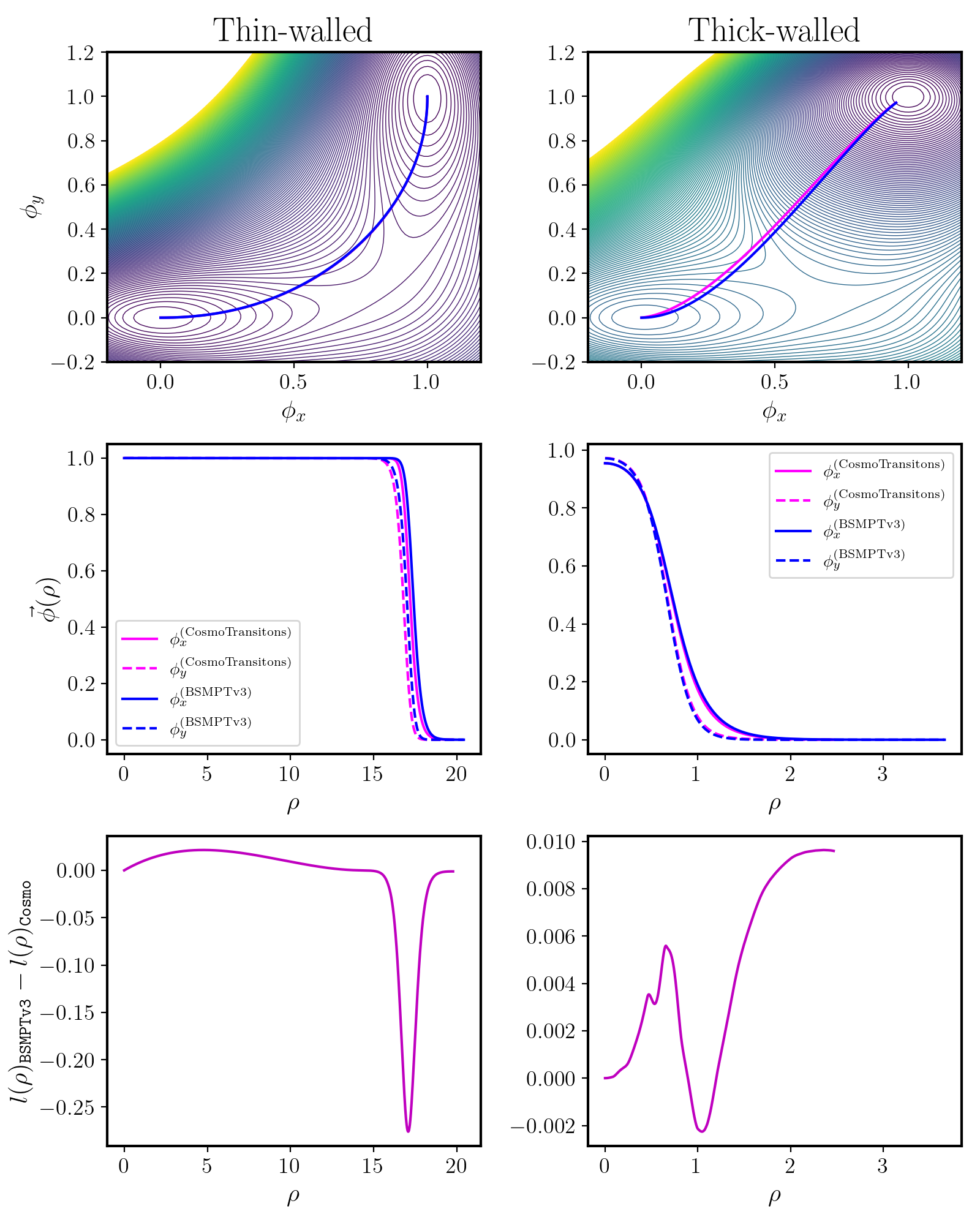}
\caption{Comparison between \texttt{CosmoTransitions} (red)
  and \texttt{BSMPTv3} (blue) for a toy model with $f_y = 2$ (left)
  and $f_y = 80$ (right). Upper: Potential contours in the
  $(\phi_y,\phi_x)$ plane. The color code denotes the potential
  values in arbitrary units. The lowest value is obtained at the true
  vacuum located at $(\phi_x,\phi_y)=(1,1)$. Middle: Tunnelling path as a function of $\rho$. Lower: Difference
    between the tunnelling path calculated by {\tt BSMPTv3} and {\tt
      CosmoTransitions} as a function of $\rho$.}
\label{fig:cosmotransitionscomparison}
\end{figure}

We compare the results for the bounce equation found by {\tt
  BSMPTv3} and {\tt CosmoTransitions} for the 
toy model provided by \texttt{CosmoTransitions} as an example. 
It is given by the potential
\begin{align}
	V(\phi_x,\phi_y) = \left(c\,(\phi_x-1)^2+(\phi_y-1)^2\right) \left(c\,\phi_y^2+\phi_x^2\right)+f_x
   \left(\frac{\phi_x^4}{4}-\frac{\phi_x^3}{3}\right)+f_y
   \left(\frac{\phi_y^4}{4}-\frac{\phi_y^3}{3}\right)\,.
   \end{align}
For the potential parameters we choose $c = 5$, $f_x = 0$ and
consider two cases $f_y = 2$ and $f_y =80$. For all cases, the true
vacuum of the potential sits at $(\phi_x,\phi_y) = (1,1)$ and the false vacuum
at $(\phi_x,\phi_y) = (0,0)$. The potential contours for the
  two cases are depicted in Fig.~\ref{fig:cosmotransitionscomparison}
  (upper). The middle plots shows the tunnelling path
  obtained by \texttt{CosmoTransitions} (red) and \texttt{BSMPTv3}
  (blue), respectively, as a 
function of the distance $\rho$ from the true vacuum. The lower plots
display the difference between the tunnelling path calculated by {\tt
  CosmoTransitions} and {\tt BSMPTv3}. The left plots are for $f_y=2$
  and the right plots for $f_y=80$. \s
 
In the case of $f_y = 2$, the vacuum phases are almost degenerate so
that the starting position is extremely close to the true
vacuum, i.e. $\vec{\phi}(\rho=0) \simeq \vec{\phi}_t$. The field
starts so close to the true vacuum that it stays near it across a large range
of $\rho$ before rolling down the inverted potential. This is
because, since the minima are almost generate, the drag term $\propto
1/\rho$ needs to have a small impact on the
dynamics. As can be inferred from the
middle plot, there is a small difference between the \texttt{CosmoTransitions}
solution (red) and the \texttt{BSMPTv3} solution (blue). This difference stems
from the fact that thin-walled solutions extremely depend on the
starting position which ultimately dictates when the field rolls
down. Nevertheless, since the bounce solution minimises the Euclidean
action and fulfils the Euler-Lagrange equations, we expect the action
to be insensitive to small variations of the correct solution, which is
indeed what we found. The relative error between both actions
that is less than $0.2\%$. And the profile of the two 
  solutions (lower plot) is very similar.  \s 

In the case of $f_y = 80$, the vacuum phases are far apart in energy
so that the starting position is not near the true vacuum,
i.e. $\vec{\phi}(\rho=0) \not\approx \vec{\phi}_t$. There are
differences in both the tunnelling paths (middle plot) and the profile solution
 (lower plot).\footnote{In {\tt CosmoTransitions} the
  tunnelling path stops at a lower $\rho$ value than in {\tt BSMPTv3}
  due to different termination conditions in the codes.} Although this
is the case, the relative difference between 
both actions is around $1\%$. \s

The determination of the bounce action is a
  challenge both from a mathematical and computational point of
  view. The solution is highly dependent on the boundary conditions,
  i.e. the starting position of the field configuration. Hence, the
  calculation entails a numerical instability which has to be treated
  carefully. So it is not surprising that both codes attacking this complex problem
numerically show some numerical discrepancies.

\FloatBarrier

\subsection{Benchmark Points}\label{sec:examplesbps}
For the purpose of illustrating {\tt BSMPTv3}, we present and discuss
in this section a few benchmark points. Benchmark points {\tt BP1}
and {\tt BP2} have been chosen from an earlier publication of members
of our group because they exhibit several vacuum directions and
multi-step phase transitions as well as the interesting case of an
intermediate charge-breaking phase at non-zero temperature. Benchmark
point {\tt BP3} has been chosen from the literature as example for
flat field directions. We also comment on
the results obtained with {\tt CosmoTransitions} for each point and on
differences in the results. If not explicitly stated
otherwise, all our runtimes were obtained on one core of an
\texttt{Apple M1 Pro} using {\tt Clang 14.0.3} and {\tt Python 3.9.13}. 

\subsubsection{The Models}
The presented benchmark points are points of the CP-conserving 2HDM
and CxSM. We briefly introduce the
models to set our notation. For further details, we refer to
\cite{Basler:2018cwe, Basler:2020nrq}.

\paragraph{The CP-Conserving 2HDM}
In the 2HDM \cite{Lee:1973iz,Branco:2011iw}, the Higgs sector consists
of two $SU(2)_L$ Higgs doublets $\Phi_1$ and $\Phi_2$. The tree-level
potential with a softly broken $\mathbb{Z}_2$ symmetry, under which
the doublets transform as $\Phi_1\rightarrow \Phi_1,\ \Phi_2
\rightarrow -\Phi_2$, is given by
\begin{align}
\begin{split}
V_{\text{tree}} &= m_{11}^2 \Phi_1^\dagger \Phi_1 + m_{22}^2
\Phi_2^\dagger \Phi_2 - \left[m_{12}^2 \Phi_1^\dagger \Phi_2 +
  \mathrm{h.c.} \right] + \frac{1}{2} \lambda_1 ( \Phi_1^\dagger
\Phi_1)^2 +\frac{1}{2} \lambda_2 (\Phi_2^\dagger \Phi_2)^2 \\
&\quad + \lambda_3 (\Phi_1^\dagger \Phi_1)(\Phi_2^\dagger\Phi_2) +
\lambda_4 (\Phi_1^\dagger \Phi_2)(\Phi_2^\dagger \Phi_1)
+ \left[ \frac{1}{2} \lambda_5 (\Phi_1^\dagger\Phi_2)^2  + \mathrm{h.c.} \right] \ .
\end{split}\label{eq:model_1}
\end{align}
The mass parameters $m_{11}^2$, $m_{22}^2$ and $m_{12}^2$ and the couplings
$\lambda_{1}\dots\lambda_{5}$ are real in the CP-conserving 2HDM.
Allowing in general for four VEV directions, given by the CP-even  
VEVs $\omega_{1,2}$ of the scalar components of the Higgs doublets,
the charge-breaking VEV $\omega_{\text{CB}}$ and the CP-breaking VEV
$\omega_{\text{CP}}$, they can be parametrised in terms of the real fields $\rho_{i}$,
$\eta_i$, $\zeta_i$, and $\psi_i$ ($i=1,2$), as
\begin{align}
    \Phi_1 = \frac{1}{\sqrt{2}}\begin{pmatrix}\rho_1+i\,\eta_1\\\zeta_1+\omega_1+i\,\psi_1\end{pmatrix}\,,\quad\Phi_2 = \frac{1}{\sqrt{2}}\begin{pmatrix}\rho_2+\omega_\text{CB}+i\,\eta_2\\\zeta_2+\omega_2+i\,\left(\psi_2+\omega_\text{CP}\right)\end{pmatrix}\,.
\end{align}
At zero temperature, phenomenology requires that  
\begin{eqnarray}
&& \left. \{\omega_{\text{CB}},\,\omega_1,\,\omega_2,\,\omega_\text{CP}\}\right|_{T=0}
    =  \{0,v_1,v_2,0 \}\,, \; \mbox{with } \nonumber \\
&& \left. \omega_{\text{EW}}\right|_{T=0} \equiv
   \left. \sqrt{\omega_1^2 + \omega_2^2 + \omega_{\text{CB}}^2 +
   \omega_{\text{CP}}^2}\right|_{T=0} =
   \sqrt{v_1^2 + v_2^2} \equiv v = 246 \mbox{ GeV} \;.
\end{eqnarray}
The ratio of the zero-temperature CP-even VEVs is given by the mixing
angle $\beta$ as
\beq
\tan\beta = \frac{v_2}{v_1} \;.
\eeq
After EWSB the Higgs spectrum consists of two scalar, $H_{1,2}$, and one
pseudoscalar, $A$, Higgs bosons as well as a charged Higgs pair,
$H^\pm$. By convention $H_1$ is to be taken as 
the lighter of the two CP-even Higgs bosons, i.e.~$m_{H_1} <
m_{H_2}$. In order to avoid tree-level flavour-changing neutral
currents, the $\mathbb{Z}_2$ symmetry is extended to the Yukawa sectors,
leading to four different types of 2HDM. The here presented benchmark
points are those of the 2HDM type 1, where the doublet $\Phi_1$
couples to all quarks and leptons. 

\paragraph{The CxSM}
The Higgs potential of the CxSM
\cite{Coimbra:2013qq,Costa:2015llh,Barger:2008jx,Gonderinger:2012rd,Muhlleitner:2017dkd,Chiang:2017nmu}
is based on the extension of the SM Higgs 
potential by a complex scalar singlet field $\mathbb{S}$. 
The tree-level potential with a
softly broken global $U(1)$ symmetry is given by
\begin{align}
  V = & \frac{m^2}{2} \Phi^\dagger \Phi + \frac{\lambda}{4}
        \left(\Phi^\dagger \Phi\right)^2 + \frac{\delta_2}{2}
        \Phi^\dagger \Phi \vert \mathbb{S} \vert^2 + \frac{b_2}{2}
        \vert \mathbb{S} \vert^2 + \frac{d_2}{4} \vert \mathbb{S}
        \vert^4 + \left( \frac{b_1}{4} \mathbb{S}^2 + a_1 \mathbb{S} + c.c. \right) \,, 
  \label{eq:CxSMPot}
\end{align}
where 
\beq
\mathbb{S} =  \frac{1}{\sqrt{2}} \left( S + i A \right) 
\eeq
is a hypercharge zero scalar field.
Because of the hermicity of the potential, all parameters in
Eq.~(\ref{eq:CxSMPot}) are real, except 
for $b_1$ and $a_1$. In our presented benchmark point,
the parameters of the soft-breaking terms, written in parenthesis, are set
to zero, $b_1=a_1=0$, so that the global $U(1)$ symmetry is exact. Denoting the
electroweak VEV by $\omega_{\text{EW}}$, and the VEVs of the CP-even
and CP-odd singlet field components by $\omega_s$ and $\omega_a$,
respectively, the doublet and singlet fields can be parametrised as
\begin{align}
  \Phi = & \frac{1}{\sqrt{2}} \begin{pmatrix}
    G^+ \\ \omega_{\text{EW}} + h + i G^0
  \end{pmatrix} \,,        \label{eq:gaugestates1}  \\
  \mathbb{S} =    & \frac{1}{\sqrt{2}} \left( s + \omega_s + i
                    \left( a + \omega_a \right) \right) \,, 
  \label{eq:gaugestates} 
\end{align}
where $G^{+}$ and $G^0$ denote the charged and neutral Goldstone
boson, respectively, and $h$ is identified with the discovered
SM-like Higgs boson. At $T=0$
\beq
\left. \{ \omega_{\text{EW}}, \, \omega_s, \, \omega_a \}\right|_{T=0}
= 
\{ v, \, v_s, \, v_a \} \;, \quad \mbox{with } v= 246 \mbox{ GeV} \;.
\eeq
The input parameters used by {\tt ScannerS}  are the SM VeV $v$, the
real and imaginary parts of the complex singlet VEVs, $v_s$ and
$v_a$, respectively, and the potential parameters $a_1$, $m^2$, $b_1$,
$b_2$, $\lambda$, $\delta_2$, $d_2$. \s

In Table~\ref{tab:examples} all results are summarized for each
benchmark point. In the following paragraphs, we show plots and discuss 
the points in detail.
\begin{center}
  \begin{table}[ht!]
      \centering
      \begin{tabular}{p{2.7cm}p{3.7cm}p{3.7cm}p{3.7cm}}\toprule
          & {\tt BP1}& {\tt BP2}& {\tt BP3}\\\midrule
          phases$_{\text{\tt BSMPT}}$& {\tt 0}: $\{216,\,400\}$\par{\tt 1}: $\{0,\,237\}$&{\tt 0}: $\{0,\,400\}$\par{\tt 1}: $\{0,\,264\}$&{\tt 0}: $\{118,\,400\}$\par{\tt 1}: $\{0,\,133\}$\\
          pairs$_{\text{\tt BSMPT}}$ & {\tt 0}: $[{\tt 0} \rightarrow {\tt 1}]\,\{216,\,237\}$ & {\tt 0}: $[{\tt 0} \rightarrow {\tt 1}]\,\{0,\,264\}$&{\tt 0}: $[{\tt 0} \rightarrow {\tt 1}]\,\{118,\,133\}$\\
          $t_{\tt MinimaTracer}$&$\SI{41.47}{s}$&$\SI{52.39}{s}$&$\SI{31.98}{s}$\\
          $T_{c}$&$226.3$&$231.0$&$127.0$\\
          $T_{n}$&$\{222.9,\,222.9\}$&$\{202.2,\,203.5\}$&$\{122.2,\,122.3\}$\\
          $T_{p}$&$222.6$&$199.0$&$121.8$\\
          $T_{f}$&$222.6$&$198.4$&$121.8$\\
          $t_{\tt CalcTemps}$&$\SI{6.87}{min}$&$\SI{3.58}{min}$&$\SI{1.45}{min}$\\
          history&${\tt 0}-({\tt 0})\rightarrow{\tt 1}$&${\tt 0}-({\tt 0})\rightarrow{\tt 1}$&${\tt 0}-({\tt 0})\rightarrow{\tt 1}$\\\midrule
          phases$_{\text{\tt Cosmo}}$&$\{0,\,206\}$&$\{0,\,212\}$&$\{0,\,135\}$\\
          $T_c$&$-$&$-$&$-$\\
          $T_n$&$-$&$-$&$-$\\
          $t_{\tt Cosmo}$&$\SI{3.95}{s}$&$\SI{5.44}{s}$&$\SI{2.07}{s}$\\
          \bottomrule
      \end{tabular}
      \caption{Results for the benchmark points {\tt BP1}-{\tt BP3}
        (input parameters given in the main text) when tracing phases
        in a temperature range $T\in\{0,\,T_\text{high}\}\,\si{GeV}$
        with {\tt MinimaTracer} and calculating characteristic
        temperatures with {\tt CalcTemps} as well as for {\tt
          CosmoTransitions}, here short-named {\tt Cosmo}. For all
        three benchmark points we set $T_\text{high}=\SI{400}{GeV}$.
          Indices of phases and phase pairs found by {\tt BSMPTv3} are given following the conventions of the output described in Sec.~\ref{subsec:transtracer}. 
          The indices of the phases that coexist in a phase pair are given in square brackets in the format $[{\tt i}_\text{false}\rightarrow{\tt i}_\text{true}]$. 
          Temperature ranges for the phases and pairs are noted in curly brackets, $\{T_\text{low}=\SI{0}{GeV},\,T_\text{high}\}$ in units of \si{GeV}. 
          Calculated characteristic temperatures are given for each phase pair, the nucleation temperature $T_n$ from {\tt BSMPTv3} is reported being calculated via Eq.~(\ref{eq:nucl_approx}) (first number) as well as Eq.~(\ref{eq:nucl_exact}) (second number).
          {\tt history} comments on the transition history of the point, specifying the hierarchy of transitions that take place for this point.
          We also show runtimes for {\tt MinimaTracer}, $t_{\tt
            MinimaTracer}$, and for {\tt CalcTemps}, $t_{\tt CalcTemps}$, as well
            as the runtime for {\tt CosmoTransitions}, $t_{\tt Cosmo}$ and the respective
          results. 
          Runtimes are measured on one core of an
        \texttt{Apple M1 Pro}. The
        timings for {\tt CosmoTransitions} cover
          the initialisation of the
        model and running {\tt findAllTransitions()}, where we
        decrease the epsilon used by numerical gradients {\tt
          x\_eps} in case the algorithm does not converge. The
        function also  determines all
          transitions, calculates their critical and (approximate)
        nucleation temperatures and stores them in {\tt self.TnTrans}.}\label{tab:examples}
  \end{table}
\end{center}
%
\subsubsection{Benchmark Points {\tt BP1} and {\tt BP2}: Multi-Step Phase Transitions with Four Field Directions}\par
\begin{figure}[t!]
  \includegraphics[width=\textwidth]{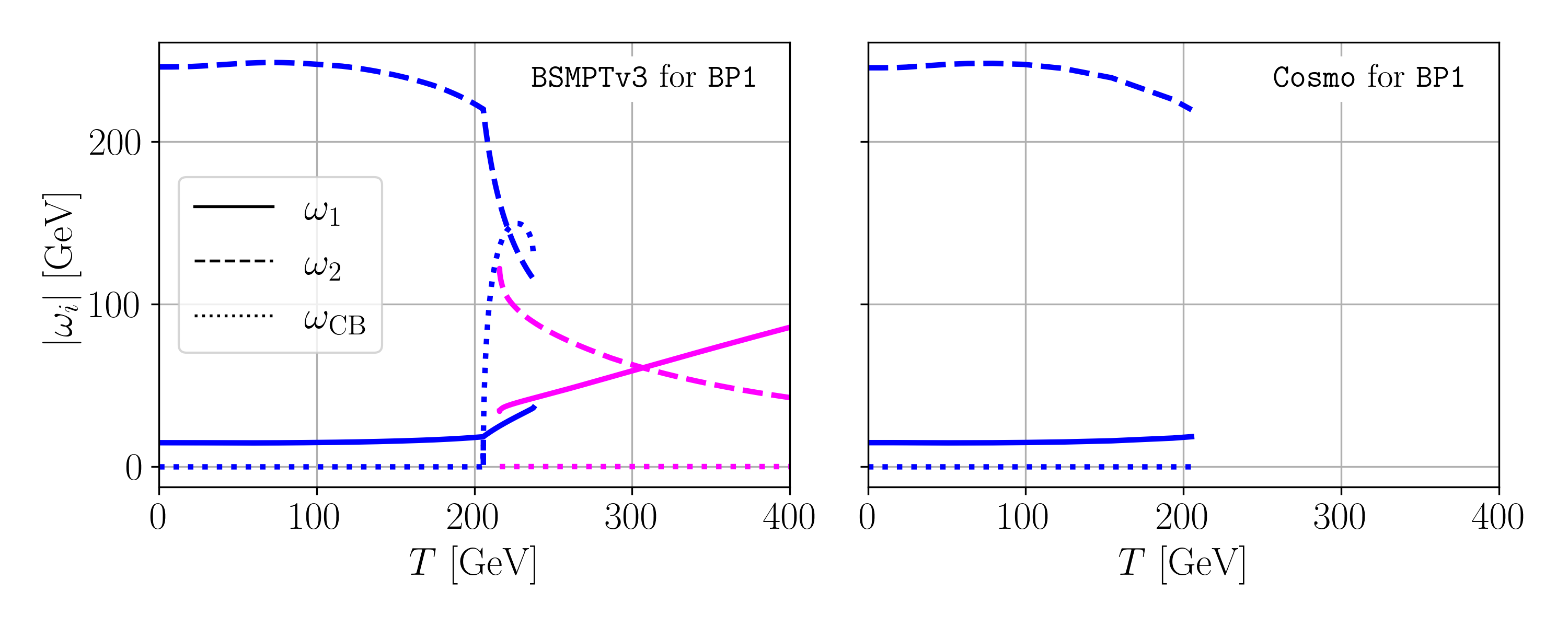}
  \caption{{\tt BP1:} Found phases as a function
      of the temperature $T\in\{0,\,400\}$~GeV with {\tt MinimaTracer}
      (left) and {\tt CosmoTransitions} (right). High-temperature phase (magenta)
      and low-temperature 
    phase (blue) for the three VEVs $\omega_1$ (solid),
    $\omega_2$ (dashed) and $\omega_\text{CB}$ (dotted). The forth VEV
    $\omega_\text{CP}$ is found to 
    remain zero for all found phases and temperatures.
    Inside the
    low-temperature phase (in blue) found by {\tt BSMPTv3} a
    second-order phase transition takes place into the 
    electroweak phase that contains the electroweak
    minimum $v = 246$~GeV at $T=\SI{0}{GeV}$.
}\label{fig:bp1}
\end{figure}

 
Our first two benchmark points {\tt BP1} and {\tt BP2} are taken from
\cite{Aoki:2023lbz} and are points of the CP-conserving 2HDM type 1.
For both presented benchmark points, we find a multi-step phase
structure in agreement with \cite{Aoki:2023lbz}, and moreover, we can
calculate a bounce solution and transition temperatures.
The benchmark point {\tt BP1} is defined by the following input
parameter set, 
\begin{eqnarray}
    \mbox{{\tt BP1}:} && \text{type} = \num{1}\,,\;\lambda_1=\num{6.931} \,, \;
    \lambda_2=\num{0.2631} \,, \; \lambda_3=\num{1.287} \,, \;
    \lambda_4=\num{4.772}\,, \; \lambda_5=\num{4.728} \,, \nonumber \\
    && m_{12}^2=\SI{1.893e4}{GeV^2}\,,\; \tan\beta=\num{16.578} \;.
\end{eqnarray}
As can be inferred from Fig.~\ref{fig:bp1}, it features a first-order
phase transition from a high-temperature neutral
(magenta) to a
charge-breaking (CB) phase (blue), that then transitions in a
second-order phase transition back into a neutral minimum. 
The nucleation, percolation and completion temperatures lie close
together slightly below $T=\SI{223}{GeV}$ and a transition history ${\tt 0}
\rightarrow {\tt 1}$ is reported, meaning that the universe will end up in phase {\tt 1} that contains the EW minimum
$v = \left.\sqrt{\omega_1^2 + \omega_2^2}\right|_{T=0} = \SI{246}{GeV}$ at $T=\SI{0}{GeV}$,
after the transition from the initial phase {\tt 0}. 
{\tt CosmoTransitions} agrees with the found low-temperature phase
until around $T=\SI{206}{GeV}$ and fails to trace any minima for
higher temperatures. The code then terminates after $t_{\tt
Cosmo}=\SI{3.95}{s}$ with no transitions found.  
By increasing the upper temperature by hand, {\tt CosmoTransitions}
might, however, successfully find a transition and reproduce the phase
structure found by {\tt BSMPTv3}. 
E.g. we find that for $T_\text{high}=\SI{900}{GeV}$ {\tt 
  CosmoTransitions} confirms the results of {\tt BSMPTv3}, however, with an increased
runtime by a factor of almost \num{17}, compared to the runtime of
{\tt CalcTemps} which also includes the calculation of the 
nucleation, percolation and completion
temperatures.\footnote{Note that {\tt
    CosmoTransitions} is numerically not stable enough to be able to
  consistently reproduce this result.}  \s

\begin{figure}[h!]
  \includegraphics[width=\textwidth]{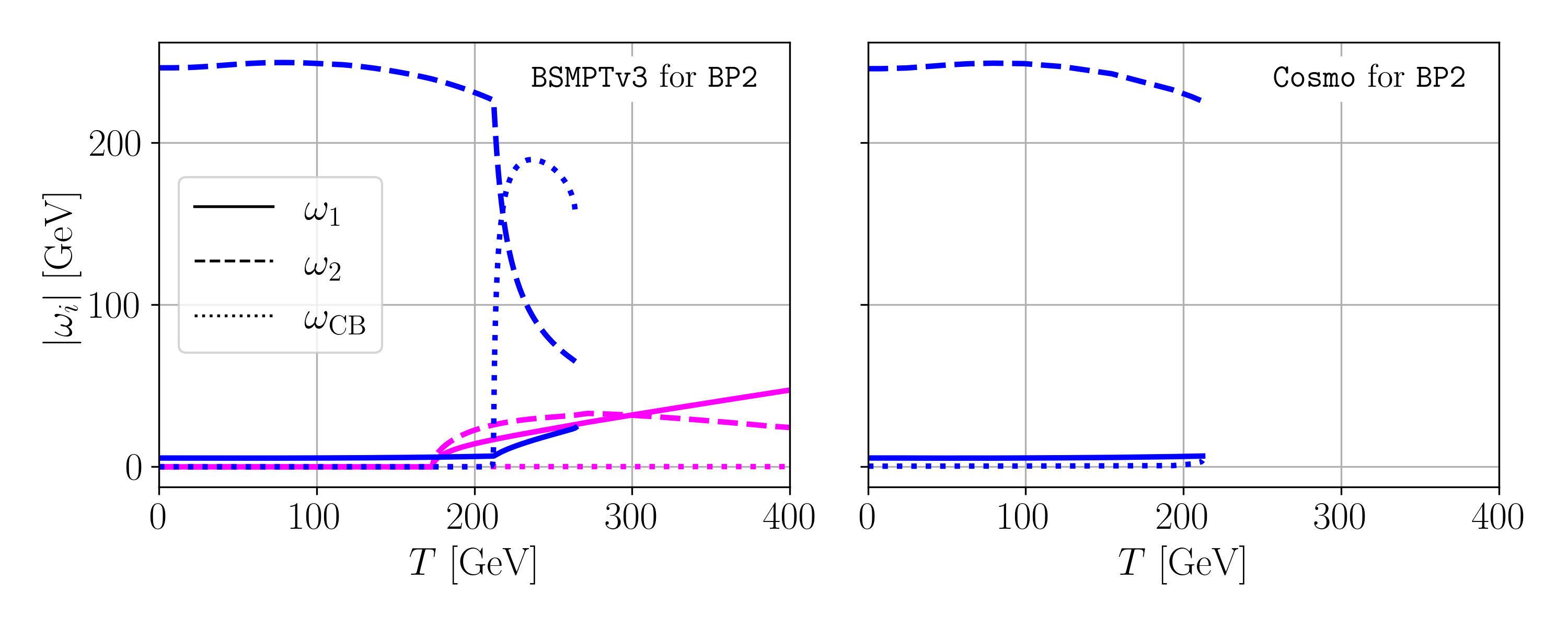}
  \caption{{\tt BP2:} Found phases as a function
      of the temperature $T\in\{0,\,400\}$~GeV with {\tt MinimaTracer}
      (left) and {\tt CosmoTransitions} (right). Colour/Line code same
      as in Fig.~\ref{fig:bp1}. Inside the low-temperature phase (in
      blue) found by {\tt BSMPTv3} a second-order phase transition
      takes place into the electroweak phase that contains the
    electroweak minimum $v = 246$~GeV at
    $T=\SI{0}{GeV}$. 
}\label{fig:bp2} 
\end{figure}
The second benchmark point {\tt BP2} is defined by 
\begin{eqnarray}
    \mbox{{\tt BP2}:} && \text{type} = \num{1}\,,\;\lambda_1=\num{6.846} \,, \;
    \lambda_2=\num{0.2589} \,, \; \lambda_3=\num{1.466} \,, \;
    \lambda_4=\num{4.498} \,, \; \lambda_5=\num{4.450} \,, \nonumber \\
    && m_{12}^2=\SI{6.630e3}{GeV^2}\,,\; \tan\beta= \num{45.320}\;,
\end{eqnarray}
%
As can be inferred from Fig.~\ref{fig:bp2}, we find a first-order
phase transition from the high-temperature phase 
(magenta lines) into the neutral low-temperature electroweak phase (blue lines) that
contains the electroweak minimum at $T=\SI{0}{GeV}$.  
This first-order phase transition happens around the same temperature
as the second-order phase transition from the CB to the neutral phase,
resulting in a transition history that unlike for {\tt BP1} can never
result in {\tt BP2} undergoing a CB intermediate phase. Due to
$T_p-T_c = \SI{30}{GeV}$, the true minimum cools
down during the CB phase and
enters the neutral phase before the phase transition happens. 
{\tt BP2}, similar to {\tt BP1} can also not be traced with {\tt
  CosmoTransitions} for $T_\text{high}=\SI{400}{GeV}$, but for a
choice of $T_\text{high}=\SI{600}{GeV}$ and a runtime of
$\SI{4.80}{min}$ it finds a transition with $T_{c}=\SI{233}{GeV}$ and $T_{n}=\SI{206}{GeV}$.  
These temperature results are then not only off by a few \si{GeV}s
from the {\tt BSMPTv3} results, but with this modified choice of
$T_\text{high}$, {\tt CosmoTransitions} also identifies three instead
of two phases in the range of $T\in\{0,\,400\}\,\si{GeV}$.  
Even though, compared to {\tt BP1}, {\tt CosmoTransitions}
reproducibly finds a transition for {\tt BP2} if
$T_\text{high}=\SI{600}{GeV}$, the phase tracing seems numerically
unstable: {\tt CosmoTransitions} is either observed to trace saddle
point directions, or cannot trace the
low-temperature phase around its second-order PT, resulting in it
finding two unconnected phases. \s
\begin{figure}[b!]
  \centering
  \includegraphics[width=\textwidth]{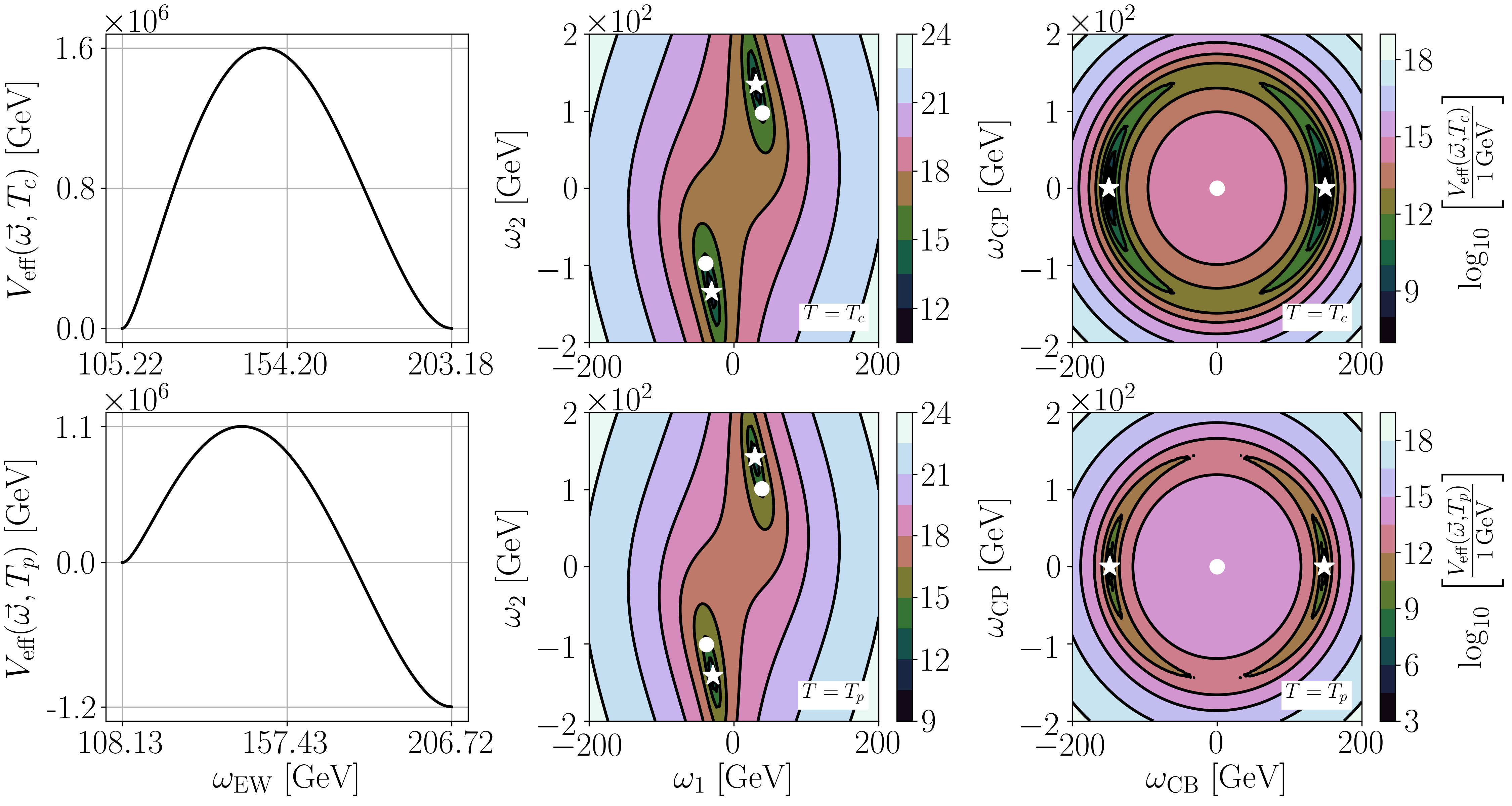}
  \caption{{\tt BP1:} Left: Slice of the effective potential from
    $\vec{\omega}_\text{false}$ to $\vec{\omega}_\text{true}$ at the
    critical temperature $T_c$, displayed via the coordinate of the EW
    VEV $\omega_\text{EW} \equiv
    \sqrt{\sum_{i=1,2,\text{CB},\text{CP}}\omega_i^2}$. Middle and
    right: Two-dimensional contour slices at $T_c$ in the
    $\omega_1-\omega_2$ (middle) and
    $\omega_\text{CB}-\omega_\text{CP}$ (right) planes.  
      The position of the false (true)
        minimum is denoted by a white dot (asterisk).
      Bottom: Same, but at the percolation temperature $T_p$.
    All contour plots are made with data-grids generated by {\tt
    PotPlotter}. The potential is shifted such that
  $V_\text{eff}(\vec{\omega}_\text{false},T)\equiv \SI{0}{GeV}$ (left
  column) as well as $V_\text{eff}(\vec{\omega}_\text{false},T)\equiv
  \SI{1}{GeV}$ (middle and right column), respectively. 
    The field directions which are not displayed in the
    two-dimensional contours are set to their global minimum
    coordinates. 
    }\label{fig:bp1_contour} 
\end{figure}

\begin{figure}[ht!]
  \centering
  \includegraphics[width=\textwidth]{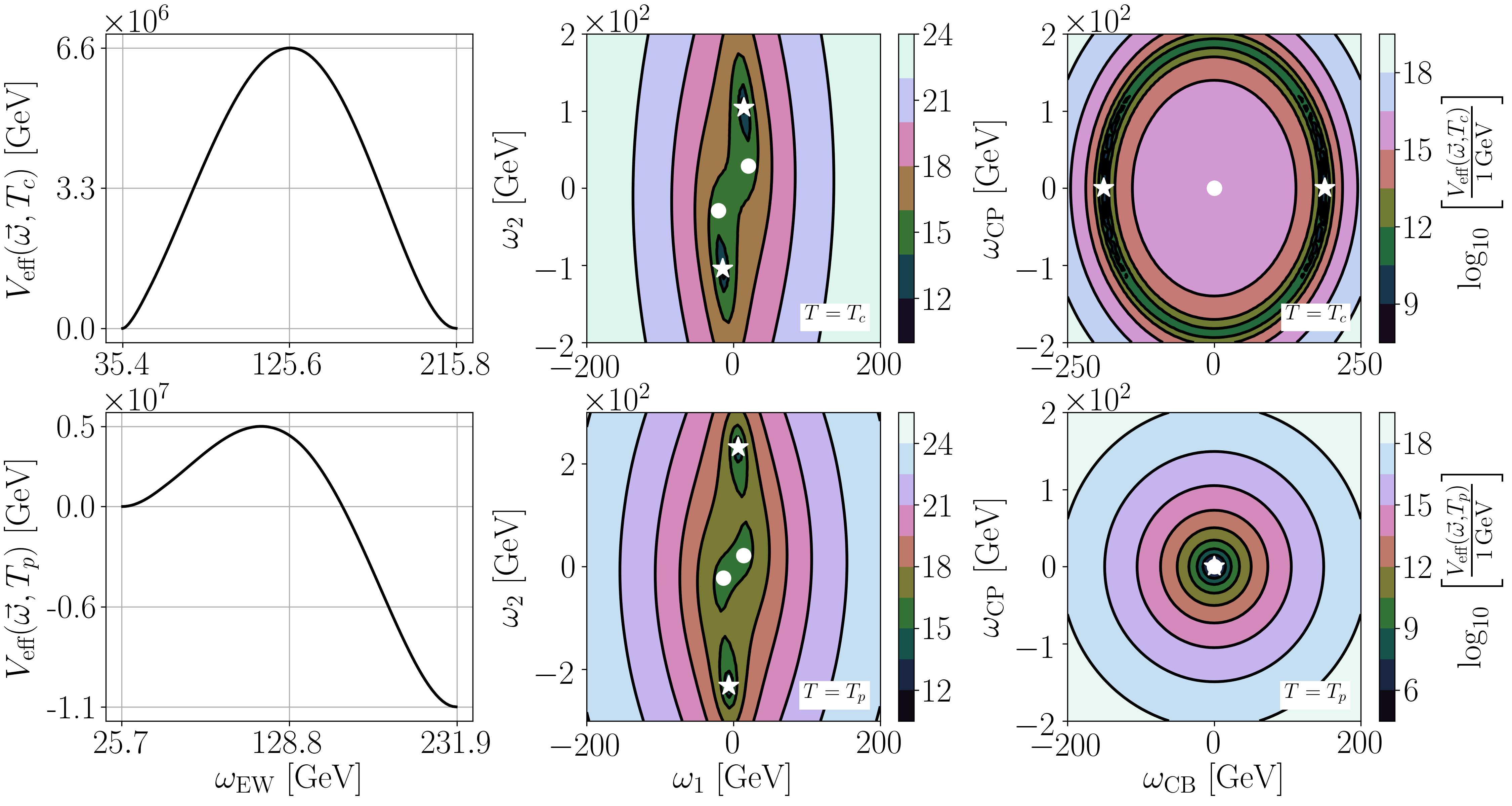}
  \caption{Same plot as Fig.~\ref{fig:bp1_contour}, but for {\tt BP2}.
  }\label{fig:bp2_contour}
\end{figure}

To further illustrate the benchmark points, in
Figs.~\ref{fig:bp1_contour} and \ref{fig:bp2_contour} we illustrate
selected potential contours at the critical and percolation
temperature $T_c$ and $T_p$, respectively, for {\tt BP1} and {\tt
  BP2}.  Note that because both 
points have $\lambda_4\approx\lambda_5$, the potential almost exhibits
an $SO(2)$ symmetry in the charge and CP-breaking VEV directions
$\{\omega_\text{CB},\,\omega_\text{CP}\}$, 
visible in Fig.~\ref{fig:bp1_contour} (right) as well as in
Fig.~\ref{fig:bp2_contour} (right) by the circle in the
$\{\omega_\text{CB},\,\omega_\text{CP}\}$-plane which is dented in the
$\omega_\text{CP}$-direction inducing a non-zero
$\omega_\text{CB}$ coordinate of the global minimum. \s

\FloatBarrier

\subsubsection{Benchmark Point {\tt BP3}: Dealing with Flat Field Directions in Three Field Directions} 
\begin{figure}[ht!]
  \includegraphics[width=\textwidth]{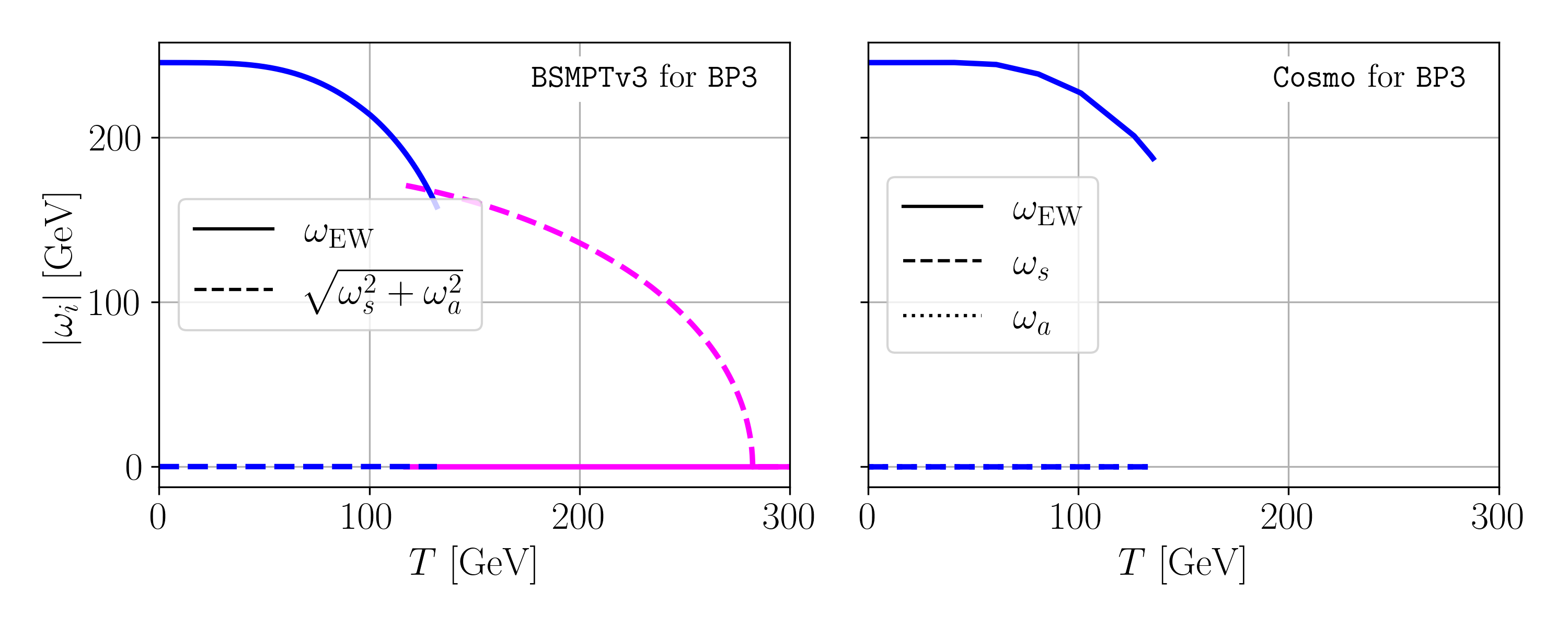}
  \caption{{\tt BP3:}  Phase structure $|\omega_i|$
      ($i=\mbox{EW},s,a$) as a function of the temperature $T$
identified with {\tt MinimaTracer} (left) and {\tt CosmoTransitions}
(right) for $T\in\{0,\,300\}$~GeV.  
      The low-temperature phase (blue) contains the electroweak
      minimum (solid line) at $T=\SI{0}{GeV}$; the high-temperature phase
      (magenta) contains the singlet phase (dashed) and is only found by {\tt
        MinimaTracer}.}
       \label{fig:bp3}
     \end{figure}

\begin{figure}[ht!]
  \centering
  \includegraphics[width=\textwidth]{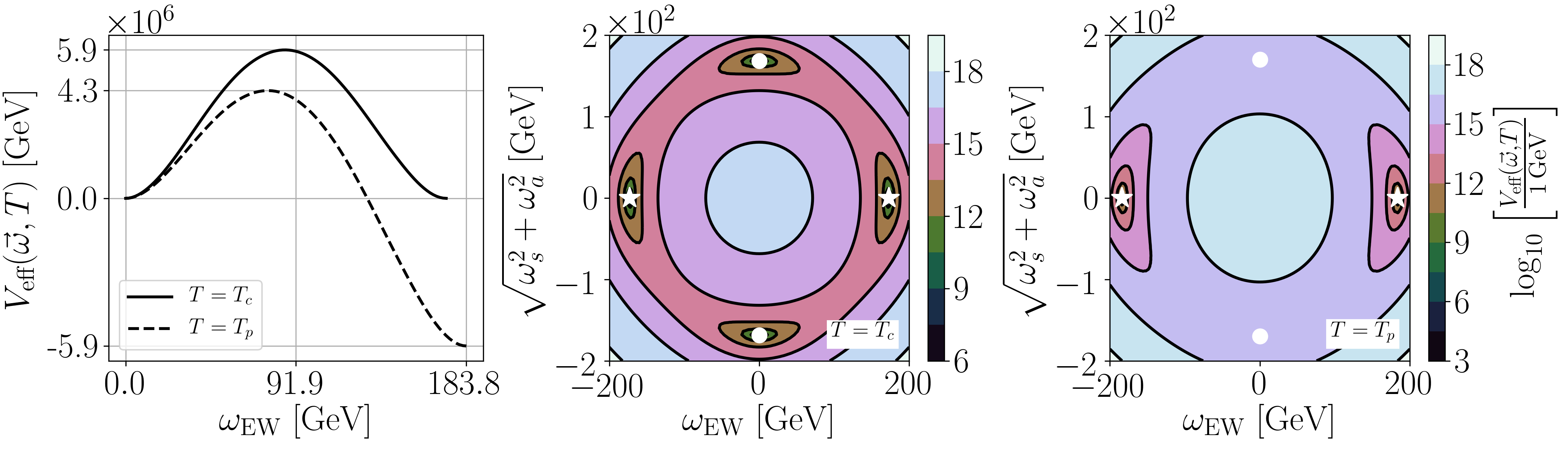}
  \caption{{\tt BP3:} Left: Slice of the effective potential
      from $\vec{\omega}_{\text{false}}$ to $\vec{\omega}_{\text{true}}$ 
      at $T_c$ (solid line) and $T_p$ (dashed line), displayed via
      $\omega_{\text{EW}}$. Middle and right: 
      Two-dimensional contours at $T_c$ (middle) and $T_p$
      (right) in the $\sqrt{\omega_s^2+\omega_a^2}-\omega_{\text{EW}}$ plane.
      The position of the false (true) minimum is denoted by a white
      dot (asterisk). The potential is shifted such that
      $V_\text{eff}(\vec{\omega}_\text{false},T)\equiv \SI{0}{GeV}$
      (left) as well as
      $V_\text{eff}(\vec{\omega}_\text{false},T)\equiv \SI{1}{GeV}$
      (middle and right), respectively. 
    }\label{fig:bp3_contour}
\end{figure}

For {\tt BP3} we illustrate a point of the complex singlet extension
of the SM (CxSM). In terms of the CxSM input parameters, the point is
defined by\footnote{We took this benchmark point from
  \cite{Chiang:2017nmu}, where it is benchmark point {\tt S2}.}
\begin{equation}
    \begin{split}
        \mbox{{\tt BP3}:}\quad & v = \SI{246.22}{GeV}\,,\; v_s = \SI{0}{GeV}\,,\; v_a = \SI{0}{GeV}\,,\; m^2 = \SI{-15650}{GeV^2}\,,\\
        & b_2 = \SI{-8859}{GeV^2}\,,\; \lambda = \num{0.52}\,,\; \delta_2 = \num{0.55}\,,\; d_2 = \num{0.5}\,,\\
        & a_1 = \SI{0}{GeV^3}\,,\; b_1 = \SI{0}{GeV^2}\;. 
    \end{split}
\end{equation} 
Since $b_1=a_1=0$, the global $U(1)$ symmetry is exact and the
potential is invariant under $v_s^2+v_a^2$, respectively 
$\omega_s^2+\omega_a^2$ at non-zero temperature. In the language of {\tt
  BSMPTv3}, this means that there is a flat 2-dimensional direction in
the potential. {\tt BSMPTv3} recognizes this flat direction and without loss
of generality sets $\omega_a=0$.  The resulting phase
structure is shown in Fig.~\ref{fig:bp3}, and the point is further
illustrated with contour slices in Fig.~\ref{fig:bp3_contour}.  
We find a first-order phase transition between a high-temperature
singlet phase (red) with $\sqrt{\omega_s^2+
    \omega_a^2} \ne 0$ and the electroweak VEV in
  the SM field direction $\omega_{\text{EW}}=0$ (corresponding to $v$ at
  $T=0$), and the low-temperature
electroweak phase (blue) with $\omega_{\text{EW}} \ne 0$ and
$\omega_s=\omega_a=0$. The corresponding critical temperature is given
by $T_c=\SI{127}{GeV}$ and the nucleation, percolation and completion temperatures lie close together at $\SI{122}{GeV}$.
{\tt CosmoTransitions} cannot identify flat directions and is
therefore forced to trace phases in all three dimensions. The code
fails to find any phase above $T=\SI{135}{GeV}$ in the requested range
of $T\in\{0,\,300\}$, cf.~Fig.~\ref{fig:bp3} (right). \s

We end this section by noting that here we of course compared only
three benchmark points, and in a broader comparison there may be
scenarios where the comparison of the performance of the two codes
{\tt BSMPTv3} and {\tt CosmoTransitions} may reveal different
features. To get a broader view, we therefore performed a comparison
based on a larger parameter sample, which we will present in the
following section.

\subsection{Parameter Scan}\label{sec:comparison}
For a broader comparison between {\tt BSMPTv3} and {\tt
  CosmoTransitions}, we performed a randomized parameter scan for the
real, i.e.~CP-conserving, 2HDM (R2HDM) type 1 by using {\tt
ScannerS-2.0.0} to check for theoretical and
experimental constraints. Details can be found 
in \cite{Abouabid:2021yvw}. Note, that for the check of the Higgs
constraints in {\tt ScannerS} the link has been updated to the
recently released program packages {\tt HiggsTools}
\cite{Bahl:2022igd}. For the scan, we chose the input parameters as
those allowed by the code. They are given by the masses of the five Higgs
states, the EW VEV $v$, the ratio of the CP-even VEVs, $\tan \beta=v_2/v_1$, the
coupling $c_{H_2 VV}$ of $H_2$ to two massive gauge bosons $V=W^\pm, Z$, and
the squared mass parameter $m_{12}^2$. The parameter ranges of our scan are given in
Tab.~\ref{tab:scan_ranges}. \s
\begin{table}[ht!]
    \centering
    \begin{tabular}{ccccccc}\toprule
        $m_{H_a}\,[\si{GeV}]$&$m_{H_b}\,[\si{GeV}]$&$m_A\,[\si{GeV}]$&$m_{H^\pm}\,[\si{GeV}]$&$c_{H_bVV}$&$\tan\beta$&$m_{12}^2\,[\si{GeV^2}]$\\\midrule
        $\num{125.09}$&$[\num{30},\,\num{1500}]$&$[\num{30},\,\num{1500}]$&$[\num{150},\,\num{1500}]$&$[\num{-.3},\,\num{.3}]$&$[\num{.8},\,\num{25}]$&$[\num{1e-3},\,\num{5e5}]$\\
        \bottomrule
    \end{tabular}
    \caption{Scan ranges for the CP-conserving 2HDM type 1 in the input parameters used by {\tt ScannerS}.}\label{tab:scan_ranges}
\end{table}

The thus obtained theoretically and
experimentally valid parameter points are then checked with respect to
their phase transitions with {\tt BSMPTv3} and independently with our
{\tt Python}-code that uses the methods of {\tt CosmoTransitions} and
traces the R2HDM potential in the full four-dimensional field space of
the R2HDM that is also used in {\tt BSMPTv3},
$\{\omega_\text{CB},\,\omega_1,\,\omega_2,\,\omega_\text{CP}\}$. \s

\begin{figure}[ht!]
    \centering
    \includegraphics[width=.5\textwidth]{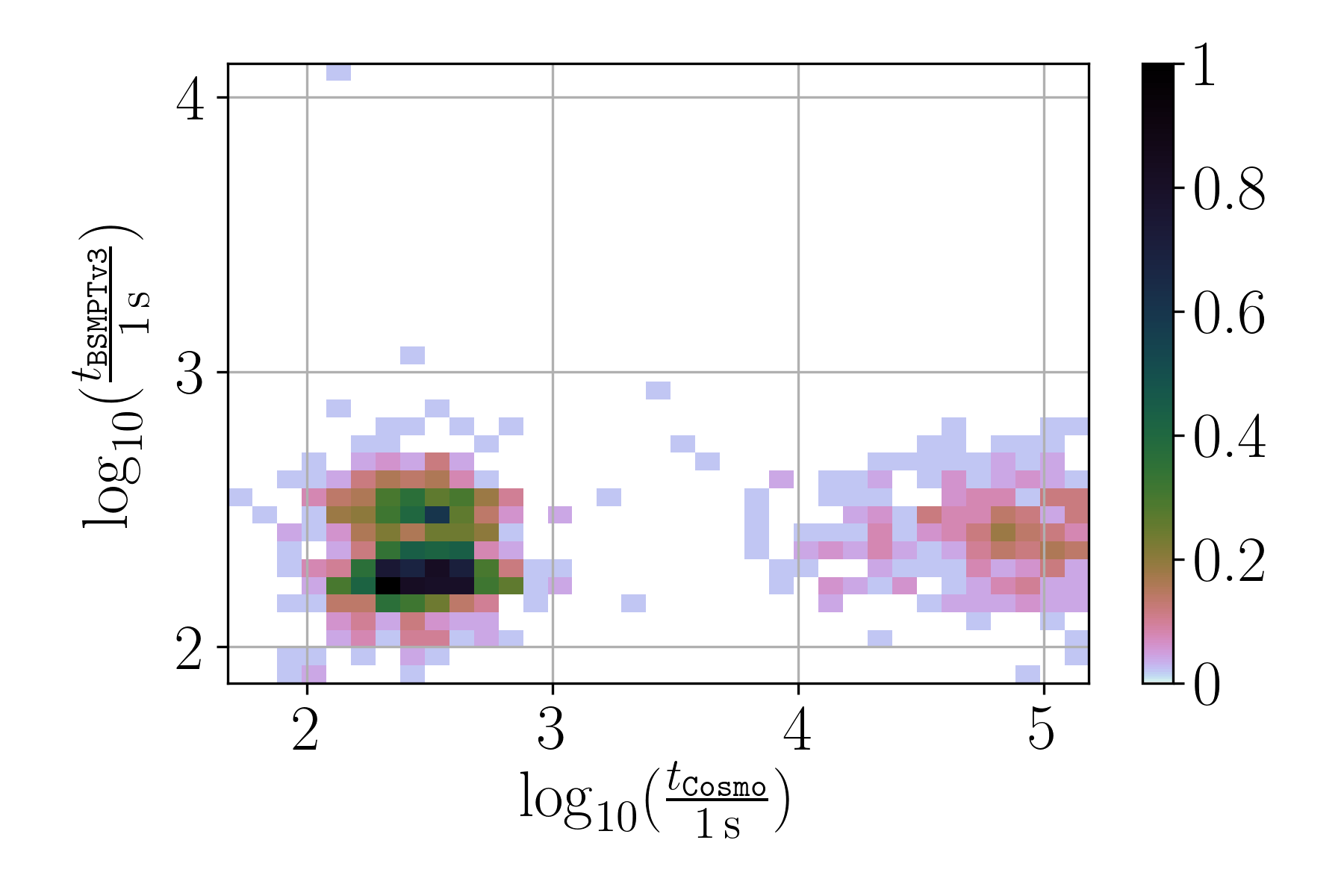}\hfill
    \includegraphics[width=.5\textwidth]{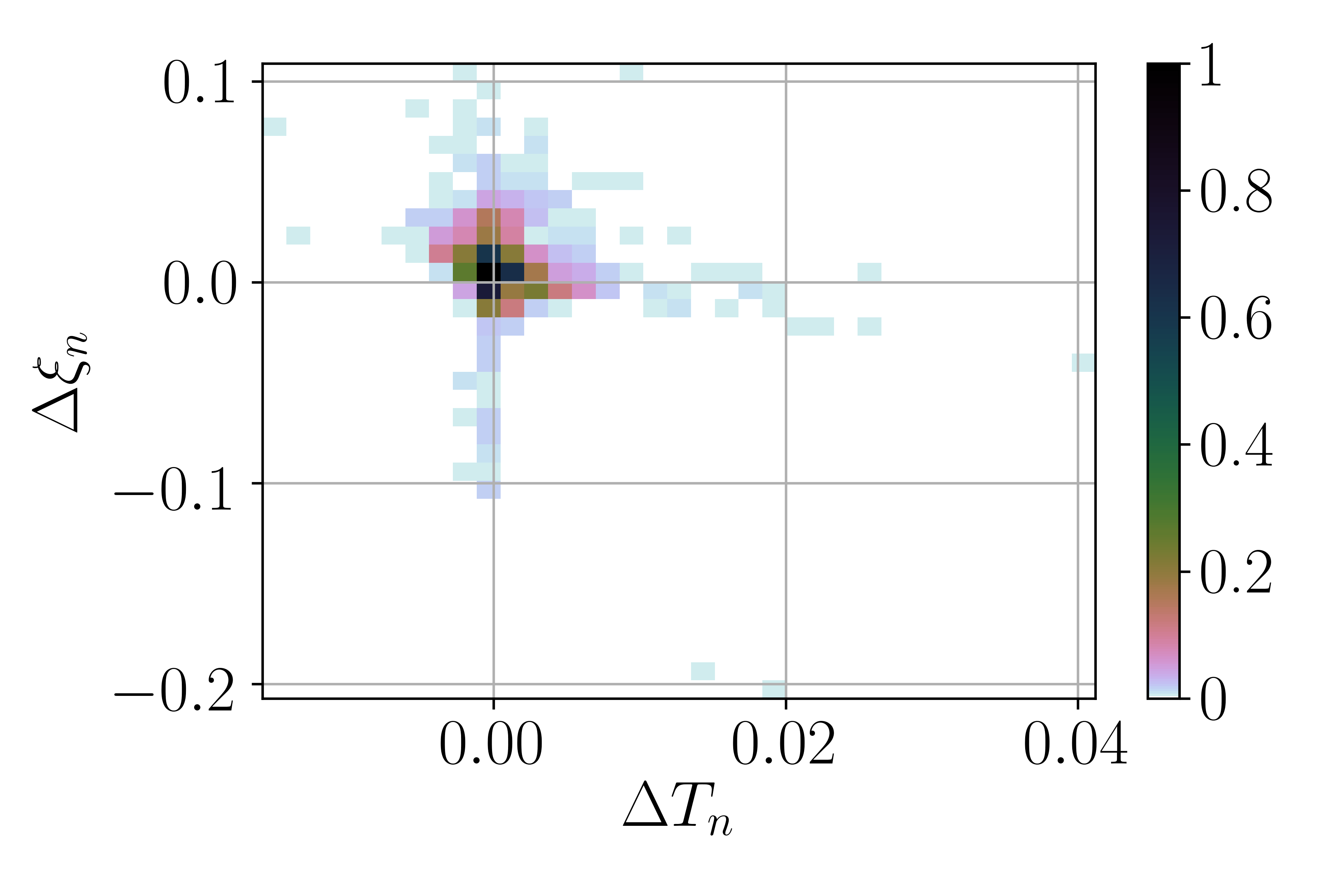}
    \caption{Left: Two-dimensional histogram showing the runtime of
      {\tt BSMPTv3} versus the runtime of {\tt  CosmoTransitions} for the sample for which both codes identify
      the same phase transitions. 
        Right: Two-dimensional histogram of the relative difference in $\xi$ at the
      nucleation temperature determined via the approximate condition
      of Eq.~(\ref{eq:nucl_approx}) versus the relative difference in
      the approximate nucleation temperature for the same sample.
        The colour of the bin indicates the proportion of the points
        falling into it.
    }\label{fig:scan_plot}
\end{figure}

In Fig.~\ref{fig:scan_plot} (left) we show a histogram of
the runtimes of {\tt BSMPTv3} versus {\tt CosmoTransitions}.  
The points taken into account are a subset of the full parameter sample, for which both codes find the same transitions.\footnote{
    If the number of found transitions differs between the two codes,
    the runtime comparison gets biased towards the code that finds
    less transitions. In that case, a direct runtime comparison would
    be biased towards the potentially less accurate code.}
    Runtimes are measured by running the codes on a mixture of {\tt
      Intel Xeon} and {\tt AMD EPYC} processors with {\tt Python
      3.6.15} for {\tt CosmoTransitions}.  
The runtime for {\tt BSMPTv3} is derived for running {\tt CalcTemps}
which traces all found phases and determines their critical
temperatures, bounce solutions, nucleation, percolation, and
completion temperatures for all found phase pairs. 
The runtime of the {\tt CosmoTransitions} routines is for initializing
the model and running the {\tt findAllTransitions()} method that
calculates the critical and approximate nucleation temperatures for
all found transitions. 
%
We find {\tt BSMPTv3} to be up to $10^3$ faster with a mean (median) runtime
of $\SI{4.15}{min}$ ($\SI{3.47}{min}$). For {\tt CosmoTransitions} we
find a mean (median) runtime of $\SI{41.46}{min}$ ($\SI{5.61}{min}$). 
If we only take into account points for which {\tt BSMPTv3} and {\tt
  CosmoTransitions} each only find one transition, their mean (median)
  runtimes are \SI{4.10}{min} (\SI{3.28}{min}) for {\tt BSMPTv3} and
  \SI{3.89}{h} (\SI{5.60}{min}) for {\tt CosmoTransitions}. \s 

While improvements of the runtime are of course desirable, the
determined temperatures of the phase transitions are the quantities
interesting for physics. In the following, we compare the values
of the temperatures and of the VEV-to-temperature ratios found by the
two codes, as well as the associated runtimes. This can be done in a meaningful
way only for parameter points where both codes find reliable
results. 
Defining the respective relative difference in the critical 
and approximate nucleation temperatures found by {\tt BSMPTv3} and
{\tt CosmoTransitions} as ($i=c, n$)
\begin{align}
    \Delta T_i = \frac{\left(T^{{\tt
  BSMPTv3}}_i - T^{{\tt Cosmo}}_i \right)}{T_i^{{\tt BSMPTv3}}}\,, 
  \label{eq:rel_diff_nucl_temp}
\end{align}
we find for the subset of points, in which both codes find the same
transitions, a maximal relative deviation of \SI{2.7}{\%} in the critical
temperature with mean (median) relative differences of
\SI{.07}{\%} (\SI{0.003}{\%}). We define the ratio between the electroweak VEV $v (T_i)$
at the temperature $T_i$ and the temperature $T_i$ as 
$\xi_i$,
\begin{eqnarray}
\xi_i = \frac{\sqrt{\sum_k \omega_k^2(T_i)}}{T_i}\quad\text{with}\quad \omega_k\in\{\omega_\text{CB},\,\omega_1,\,\omega_2,\,\omega_\text{CP}\}\,,
\end{eqnarray}  
and the relative difference $\Delta \xi_i$ in $\xi_i$ found by {\tt BSMPTv3}
and {\tt CosmoTransitions} as
\begin{align}
    \Delta \xi_i = \frac{\left(\xi_i^{{\tt BSMPTv3}} - \xi^{{\tt Cosmo}}_i\right)}{\xi_i^{{\tt BSMPTv3}}}\,.
\end{align}

In Fig.~\ref{fig:scan_plot} (right) we show the relative differences $\Delta \xi_{n}$
versus the relative differences $\Delta T_{n}$ in the found
approximate nucleation temperature. 
We find mean and median for both relative differences below
\SI{1}{\%}, however, we see outliers of up to \SI{4.1}{\%} in $\Delta
T_{n}$ as well as of up to \SI{-20.7}{\%} in $\Delta
\xi_{n}$. The outliers in $\Delta\xi_n$ are correlated with a rapidly
changing potential in a small temperature interval. Small $\Delta T_n$
in that case can lead to larger $\Delta\xi_n$ if the position of the
electroweak minimum changes significantly in small temperature
ranges.


\section{Conclusions}\label{sec:conclusions}
The detection of gravitational waves from first-order phase
transitions during the evolution of the universe combines
cosmological observation with particle physics in an exciting way that
may answer some of our most urgent open questions: What is the true theory
underlying nature? And how can we explain the observed
matter-antimatter asymmetry? For this to be meaningful, we need to
go through the whole chain from a particle physics model to the
possible detection of gravitational waves sourced by FOPTs at future
space-based interferometers like {\tt LISA}, taking into account the
state-of-the-art approaches to the various involved steps along this
way. At present, there exists no public code that is able to perform this
task. With the publication of the {\tt C++} code {\tt BSMPTv3} we
close this gap. It is the first publicly available code that performs
the whole chain from the particle physics model to the gravitational
wave spectrum. \s

The code {\tt BSMPTv3} is based on the extension of the previous
versions {\tt BSMPTv1} and {\tt v2}, which calculate the
loop-corrected effective potential at non-zero temperature in the
on-shell renormalization scheme, including
thermal masses, for extended Higgs sectors, to search for FOPTs. The
new release {\tt BSMPTv3} is able to trace vacuum phases as functions of
the temperature for complicated vacuum histories, involving also
multi-step PTs. It is able to treat multiple phase directions, discrete symmetries and
flat directions and to identify EW symmetry non-restoration at high
temperature. After tracing the minima, the bounce action is computed and the
bounce equation is solved for phase
pairs exhibiting a critical temperature. This then allows to evaluate the
tunnelling rate from the false to the true vacuum and to determine the
nucleation temperature, and thereby to decide if 
the universe is trapped in a vacuum or if a phase transition actually takes
place. In this case, the code also
calculates the percolation and the completion
temperature. Subsequently, the latent heat release and the inverse
time scale characteristic for a phase transition are evaluated at the
transition temperature, which by default is set to the percolation
temperature, but can also be chosen by the user. Together with the
wall velocity, for which various approximations are implemented among
which the user can choose, the thermal parameters are used to
calculate the GW spectrum sourced from bubble collisions and highly relativistic fluid shells, sound waves and
turbulence. Lastly, the signal-to-noise ratio at {\tt LISA} is
evaluated. \s

We compared our code with  {\tt CosmoTransitions} and found good
agreement between both codes, but showed that 
{\tt BSMPTv3} not only can be significantly faster, but also is more
powerful in dealing with higher-dimensional potentials. \s

The code is publicly available and can be downloaded at:

\centerline{\download\,.}

\noindent It will constantly be upgraded to include new developments in the field and
newly published improved calculations related to the various steps. \s

With the {\tt C++} code {\tt BSMPTv3} we provide an important
contribution to the reliable 
derivation of gravitational wave signals from FOPTs of BSM Higgs
sectors with several vacuum directions. Its application to the broad
new physics landscape will provide an exciting field for the
exploration and understanding of the Higgs vacuum structure and will 
advance our knowledge on the true model underlying nature.


\section*{Acknowledgements}\label{sec:acknowledgements}
LB wants to thank Thomas Biek\"otter and Christoph Borschensky for
helpful discussions about the code. Many thanks to Karo Erhardt and
Guilherme Monsanto for
testing early versions of {\tt BSMPTv3}. MM and LB acknowledge support
by the Deutsche Forschungsgemeinschaft (DFG, German Research
Foundation) under grant 396021762 - TRR 257. RS and JV are supported
by CFTC-UL under FCT contracts
https://doi.org/10.54499/UIDB/00618/2020,
https://doi.org/UIDP/00618/2020 as well as by the projects
CERN/FIS-PAR/0025/2021 and CERN/FIS-PAR/0021/2021. JV is also
supported by a PT-CERN PhD Grant, contract \\ PRT/BD/154191/2022.


\appendix
\section*{Appendix}
\section{Improvement of the bosonic thermal function
  $J_{-}(x^2)$} 

To solve the bounce equation, it is very important for the potential
and its gradient to behave properly, without any discontinuities
and/or unexpected behaviours. In the previous versions,
\texttt{BSMPTv1} and \texttt{BSMPTv2}, where all observables were
calculated without the use of the gradient, its behaviour was not
critical. \s

\begin{figure}[ht!]
\centering
\includegraphics[width=0.6\textwidth]{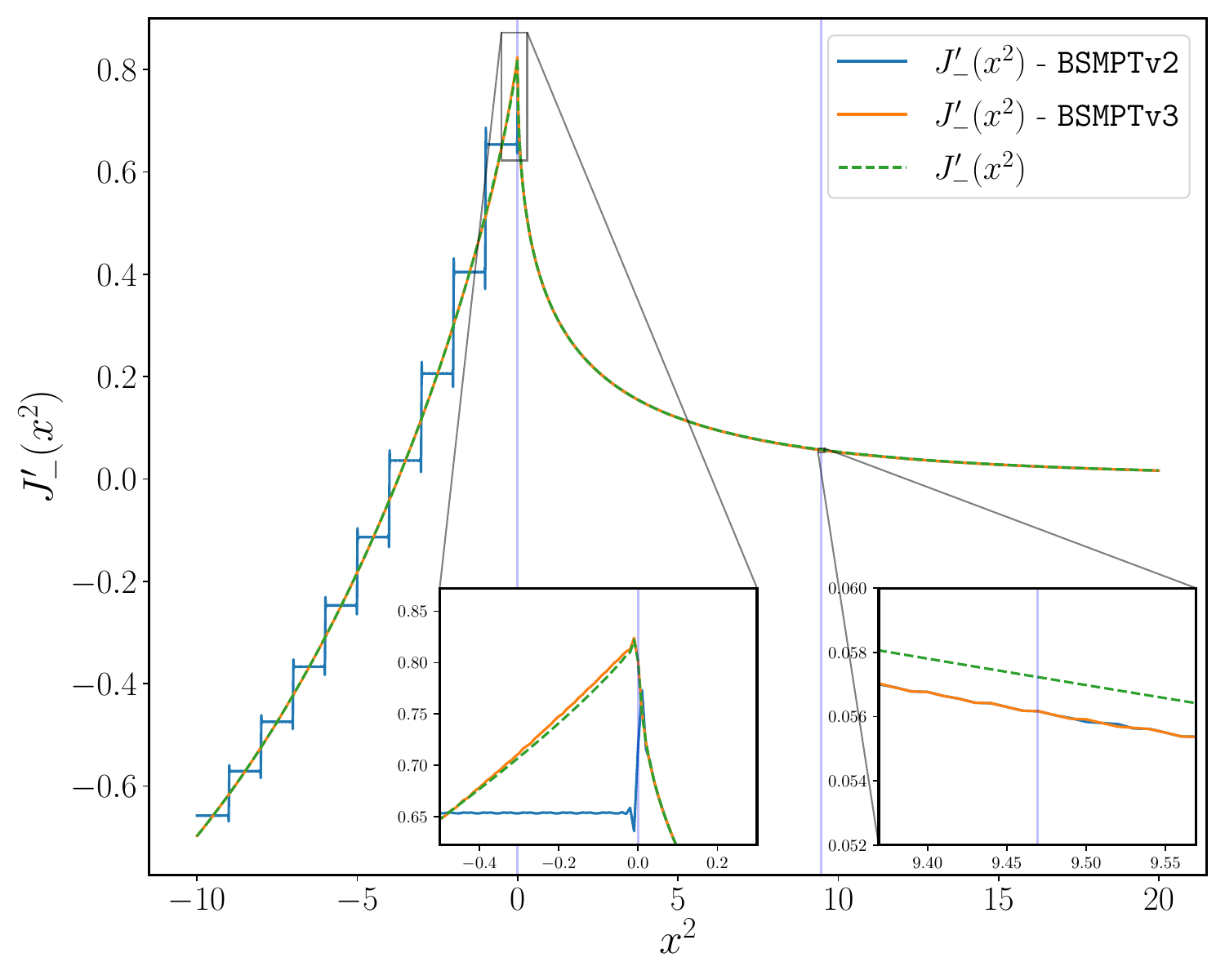}
\caption{Derivative $J_-'(x^2)$ of the bosonic thermal
  function as a function of 
  $x$. We compare the implementation of \texttt{BSMPTv2} (blue
  line) and \texttt{BSMPTv3} (orange) as well as a precise numerical
  evaluation of the integral (green dashed line). The blue vertical
  lines, at $x^2 = 0$ and $x^2 = -9.4692$, are the
  positions where we patch
  different implementations of $J_-(x^2)$ together,
  cf.~\cite{Basler:2018cwe} for more details.}
\label{fig:jminusfunction}
\end{figure}

The biggest problem arises from the evaluation of the bosonic thermal
function $J_-(x^2)$ at negative input values. In the previous
versions, the function values at $x^2 = \{0, -1, -2,
\cdots,-3000\}$ were hard-coded, and to calculate the function value at a
negative $x$-value a linear interpolation between the two
closest nodes was used. In the past, this was more than enough as the derivative
was never used and this interpolation produced a continuous $J_-(x^2)$
function. To solve the bounce equation we also need the derivative to
be well-behaved. Our solution for this is the construction of a
cubic spline using the same hard-coded function values shipped with
the previous versions. We also imposed that the spline derivative at
$x^2=0$ matches the analytical value of $J'_-(0) =
-\frac{\pi^2}{12}$. The result can be see in
Fig.~\ref{fig:jminusfunction} where we plot the derivative of
$J_-(x^2)$ at negative values for \texttt{BSMPTv2} (blue) and
\texttt{BSMPTv3} (orange) as well as a numerical derivative calculated with
high precision (green dashed). The new solution in {\tt BSMPTv3}
  approximates the derivative of the function at negative values much
  better. \s

It is important to check that this change does not completely alter
the results found in previous calculations. We therefore 
re-scanned some points of the R2HDM and its CP-violating version,
the C2HDM, applying both implementations of $J_-(x^2)$. We found that
the difference is a few per-cent only. Overall, the critical
temperature calculated before is above 
the critical temperature calculated in \texttt{BSMPTv3}. In the old
version, $\xi_c = \omega_{\text{EW}}(T_c) / T_c$ is slightly lower (by
at most \num{0.1}). The previously obtained results were hence a little bit more conservative.

\section{Comparison with Espinosa-Konstandin analytical solutions} 
Recently, a way of finding an analytical solution for
the bounce equation was developped in~\cite{Espinosa:2018szu,Espinosa:2023oml}.
The potentials/solutions used throughout this section assume an $\mathcal{O}(4)$ symmetry although similar expression can be constructed for  $\mathcal{O}(3)$-symmetric solutions. The method consists on defining the tunnelling potential 
\begin{align}
    V_t(\phi)\equiv V(\phi)-\frac{1}{2}\dot\phi_b^2\,,
\end{align}
where the subscript ``$b$'' indicates bounce, to clarify that the
tunnelling potential is only defined along the tunnelling path.
The tunnelling potential is simply the negative value of the energy that is not
  conserved, hence
\begin{align}
    \frac{d}{d\rho}\left[\frac{1}{2}\dot\phi_b^2-V(\phi_b)\right]=-\frac{3}{\rho}\dot\phi^2_b\le 0\,,
\end{align} 
so that the tunnelling potential describes how energy is dissipated by the drag term. The Euler-Lagrange equations provide a differential equation,
\begin{align}
    \left(4 V_t^{\prime}-3 V^{\prime}\right) V_t^{\prime}=6\left(V_t-V\right) V_t^{\prime \prime}\,,\label{eq:EK:euler}
\end{align}
where the prime denotes the derivative w.r.t.~$\phi$. The Euclidean action of the bounce solution can then be calculated as
\begin{align}
    S_E\left[V_t\right]=54\pi^2\int_{\phi_0}^{\phi_+}\frac{(V-V_t)^2}{(V'_t)^3}d\phi\,,
\end{align}
where $\phi_0\equiv \phi(\rho=0)$ and $\phi_+$ is the false
vacuum. The Euclidean action can have an analytical expression
provided that $V$ and $V_t$ are simple enough for the integral to be
solvable but, even if the integral is not solvable, it can usually be
calculated numerically with sufficient accuracy. \s

This method provides a way to produce analytical solutions. To solve the bounce equation one would use $V(\phi)$ to calculate $V_t(\phi)$, but we can start with $V_t(\phi)$ and calculate the $V(\phi)$ that solves Eq.~\eqref{eq:EK:euler}, something that is far easier to do. The potential can be calculated from the tunnelling potential as
\begin{align}
    V(\phi)=V_t(\phi)+\frac{V'_t(\phi)^2}{3}\int_{\phi_0}^\phi\frac{d\overline\phi}{V'_t(\overline\phi)}\,.
\end{align}
Therefore, given a tunnelling potential $V_t(\phi)$ and a wanted solution $\phi_0$ one could construct the potential $V(\phi)$ that has this solution. 
This method allows one to manufacture potentials, and their respective solutions, that then can be used to test other methods. 
With this in mind, we compared the examples provided in
Ref.~\cite{Espinosa:2023oml} to verify the accuracy of our code. The
results are provided here below.

\subsection{Example A}
We consider a tunnelling potential given by
\begin{align}
    V_t(\phi)=\phi^2(2\phi-3)\,, 
\end{align}
which produces the following potential
\begin{align}
    V(\phi)=\phi^2\left[2\phi-3+(1-\phi)^2\log\frac{(1-\phi)^2\phi_0^2}{(1-\phi_0)^2\phi^2}\right]\,, 
\end{align}
with $0<\phi_0<1$ and $\phi_0 \equiv \phi(\rho = 0)$ being the solution for the bounce. The value for the action is given by 
\begin{align}
    S(\phi_0)=-\frac{\pi^2}{3}\left[\phi_0+\text{Li}_2\left(\frac{\phi_0}{\phi_0-1}\right)\right]\,, 
\end{align}
where $\text{Li}_2(x)$ is the dilogarithm. For the value of $\phi_0 =
0.99$, the result of $\texttt{BSMPTv3}$ 
matches the analytical result with a $0.0007\%$ error. For the value of $\phi_0 = 0.50$, the result of $\texttt{BSMPTv3}$ matches the analytical result with a $0.07\%$ error.

\subsection{Example B}
We consider a tunnelling potential given by
\begin{align}
    V_t(\phi)=-\sin^2\phi\,, 
\end{align}
which produces the following potential
\begin{align}
    V=\left[-1+\frac{2}{3} \cos ^2 \phi \log \left(\frac{\tan \phi_0}{\tan \phi}\right)\right] \sin ^2 \phi\,.
\end{align}
with $0<\phi_0<\pi/2$. For the value of $\phi_0 = 1.4$, the result of $\texttt{BSMPTv3}$ matches the analytical result with a $0.002\%$ error. For the value of $\phi_0 = 1.54$, the result of $\texttt{BSMPTv3}$ matches the analytical result with a $0.0003\%$ error.

\subsection{Example C}
We consider a tunnelling potential given by
\begin{align}
    V_t(\phi)=\frac{\phi^2}{-1/2+\log\phi}\,, 
\end{align}
which produces the following potential
\begin{align}
    V(\varphi)=\frac{\phi^2}{-1 / 2+\log \phi}+\frac{8 \phi^2(1-\log \phi)^2}{3(1-2 \log \phi)^2}\left(2 \log ^2 \phi-2 \log ^2 \phi_0+\log \frac{1-\log \phi}{1-\log \phi_0}\right)\,.
\end{align}
with $0<\phi_0<\sqrt{e}$. For the value of $\phi_0 = 0.8$, the result of $\texttt{BSMPTv3}$ matches the analytical result with a $0.004\%$ error. 

\subsection{Example D}
We consider a tunnelling potential given by
\begin{align}
    V_t(\phi)=\text{Ei}(2\log\phi)\,, 
\end{align}
which produces the following potential
\begin{align}
    V(\varphi)=\text{Ei}(2\log\phi)+\frac{1}{6}\phi^2\left(1-\frac{\log^2\phi_0}{\log^2\phi}\right)\,.
\end{align}
with $0<\phi_0<1$. For the value of $\phi_0 = 0.8$, the result of $\texttt{BSMPTv3}$ does not match the analytical solution. Instead, we find a tunnelling solution around $\phi_0 \approx 0.5323$, that can be checked using the undershoot/overshoot algorithm around this point. From Ref.~\cite{Espinosa:2023oml}, the analytical solution for this potential also has an explicit form for the $\phi_0 = 0.8$ solution, which is given by 
\begin{align}
    \phi_B(\rho,\phi_0)=e^{-\sqrt{\rho^2/3+\log^2\phi_0}}\,.\label{eq:EK:exampleD:profile}
\end{align}
If we use the analytical solution starting at a different starting point $\tilde\phi_0$, the Euler-Lagrange equations give
\begin{align}
    \left(\log ^2(\phi_0)-\log ^2(\tilde\phi_0)\right)\frac{  \left(\sqrt{3 \log ^2(\tilde\phi_0)+\rho ^2}+\sqrt{3}\right)}{\left(3
    \log ^2(\tilde\phi_0)+\rho ^2\right)^{3/2}}e^{-\frac{\sqrt{3 \log ^2(\tilde\phi_0)+\rho ^2}}{\sqrt{3}}}=0\,,
\end{align}
which is only realized for all $\rho\ge0$ if
$\phi_0=\tilde\phi_0$.\footnote{There is another solution $1/\phi_0$
  that fulfills the Euler-Lagrange equation but it
  lies outside the range $0<\phi_0<1$.} 
This result indicates that there are two bounce solutions with
different functionals, i.e. only one of the solutions is given by
Eq.~\eqref{eq:EK:exampleD:profile}.

\subsection{Example E}
We consider a tunnelling potential given by
\begin{align}
    V_t=e^2 \operatorname{Ei}(-2+2 \log \phi)-e^3 \operatorname{Ei}(-3+3 \log \phi)\,, 
\end{align}
which produces the following potential
\begin{align}
    V(\phi)=e^2 \operatorname{Ei}(-2+2 \log \phi)-e^3 \operatorname{Ei}(-3+3 \log \phi)+\frac{(1-\phi)^2 \phi^2}{18(1-\log \phi)^2} r^2(\phi)\,, 
\end{align}
where $r^2(\phi)$ is the bounce profile implicitly given by
\begin{align}
    r^2(\phi)=3\left[2 \log \frac{(1-\phi) \phi_0}{\left(1-\phi_0\right) \phi}+\log ^2 \phi-\log ^2 \phi_0+2 \operatorname{Li}_2(1-\phi)-2 \mathrm{Li}_2\left(1-\phi_0\right)\right]\,.
\end{align}
For the value of $\phi_0 = 0.8$, the result of $\texttt{BSMPTv3}$ matches the analytical result with a $0.0006\%$ error.

\subsection{2-Dimensional Example}
There is a way to handle $n$-dimensional potentials.
The method is described in detail in Ref.~\cite{Espinosa:2023oml}. The
tunnelling potential, that we consider here is the same as in example A.
The $2$-dimensional potential is given by 
\begin{align}
    V_2\left(\phi_1, \phi_2\right)=V\left(\varphi\left(\phi_1\right)\right)+W\left(\phi_1\right)\left[\phi_2-\Phi_2\left(\phi_1\right)\right]+\frac{9}{2}\left(2-\phi_1\right)\left(\phi_2^2-\phi_1^2-1\right)^2\,, 
\end{align}
where $\varphi(\phi_1) = -i\operatorname{E}\left[i \operatorname{Arcsinh}(\phi_1),2\right]$ with $E[\phi,m]$ being the incomplete elliptic function of the second kind. The function $W(\phi_1$) is given by 
\begin{align}
    W(\phi_1)=V'\left(\varphi(\phi_1)\right)\tanh(\phi_1/\alpha)+\frac{2[V\left(\varphi(\phi_1)\right)-V_t\left(\varphi(\phi_1)\right)]}{\alpha\cosh(\phi_1/\alpha)^3}\,.
\end{align}
The parameter $\alpha$ is free and generates different tunnelling
paths, we set it to $\alpha=1/2$ to match
Ref.~\cite{Espinosa:2023oml}. For $\phi_0 = 0.8$, the result of $\texttt{BSMPTv3}$ matches the analytical result with a $1.5\%$ error.

\subsection{3-Dimensional Example}
The tunnelling potential is the same as in example A and in the
2-dimensional example. The $3$-dimensional potential is given by 
\begin{align}
    V_3\left(\phi_1, \phi_2, \phi_3\right)&= V\left(\varphi\left(\phi_1\right)\right)+W_2\left(\phi_1\right)\left(\phi_2-\Phi_2\left(\phi_1\right)\right)+W_3\left(\phi_1\right)\left(\phi_3-\Phi_3\left(\phi_1\right)\right) \\
    &\quad+ 25\left(\phi_2-\Phi_2\left(\phi_1\right)\right)^2+25\left(\phi_3-\Phi_3\left(\phi_1\right)\right)^2\,,
\end{align}
where $\Phi_2(\phi_1)$ and $\Phi_3(\phi_1)$ are given by
\begin{align}
    \Phi_2\left(\phi_1\right)=\rho \sin \left(\frac{\sqrt{1-\alpha^2}}{\rho \alpha} \phi_1\right), \quad \Phi_3\left(\phi_1\right)=\rho-\rho \cos \left(\frac{\sqrt{1-\alpha^2}}{\rho \alpha} \phi_1\right)\,,
\end{align}
and the $W_{1/2}$ functions are given by
\begin{align}
    W_2\left(\phi_1\right)&=V^{\prime}\left(\varphi\left(\phi_1\right)\right) \frac{\Phi_2^{\prime}\left(\phi_1\right)}{\varphi^{\prime}\left(\phi_1\right)}+2\left[V\left(\varphi\left(\phi_1\right)\right)-V_t\left(\varphi\left(\phi_1\right)\right)\right]\left[\frac{\Phi_2^{\prime \prime}\left(\phi_1\right)}{\varphi^{\prime}\left(\phi_1\right)^2}-\frac{\Phi_2^{\prime}\left(\phi_1\right) \varphi^{\prime
    \prime}\left(\phi_1\right)}{\varphi^{\prime}\left(\phi_1\right)^3}\right]\,, \\
    W_3\left(\phi_1\right)&=V^{\prime}\left(\varphi\left(\phi_1\right)\right) \frac{\Phi_3^{\prime}\left(\phi_1\right)}{\varphi^{\prime}\left(\phi_1\right)}+2\left[V\left(\varphi\left(\phi_1\right)\right)-V_t\left(\varphi\left(\phi_1\right)\right)\right]\left[\frac{\Phi_3^{\prime \prime}\left(\phi_1\right)}{\varphi^{\prime}\left(\phi_1\right)^2}-\frac{\Phi_3^{\prime}\left(\phi_1\right) \varphi^{\prime
    \prime}\left(\phi_1\right)}{\varphi^{\prime}\left(\phi_1\right)^3}\right]\,.
\end{align}
The parameters $\alpha$ and $\rho$ are free and generate different
tunnelling paths. We choose $\alpha=1/2$ and $\phi_0=0.999$ to match
Ref.~\cite{Espinosa:2023oml}. We chose two values of
$\rho$. When setting it to $\rho=\frac{1}{6}$, $\texttt{BSMPTv3}$
matches the analytical result with a $1.5\%$ error. If we set it to
$\rho=\frac{1}{8}$, the same value as in Ref.~\cite{Espinosa:2023oml},
then  $\texttt{BSMPTv3}$ converges to a different solution with an Euclidean action of $S\approx55.6$.

\section{Effective Degrees of Freedom for the Energy and Entropy}\label{sec:gstar}
\label{sec:gstar}
 At the early stages of the Universe, when the temperature was larger
than $\mathcal{O}(100\text{GeV})$, the cosmic fluid was dense and hot
enough that the interaction rate of the particles with the fluid was
much larger than the Hubble rate so that the Universe was fully
thermalised~\cite{Kierkla:2022odc}. Therefore, all
  particles fulfilled their respective equilibrium distribution
  functions, i.e.~the Bose-Einstein distribution for bosons and the
Fermi-Dirac distribution for fermions. The cosmic fluid energy
density and the entropy density can hence be calculated just by
knowing what type of particles is present in the cosmic fluid. For each
massless (relativistic) boson (b) and fermion (f) there is a
contribution (in natural units) to the energy ($\rho$) and entropy ($s$) density of
\begin{align}
	\rho_{\mathrm{b}}(T)&=g\frac{\pi^{2}}{30}T^{4}\,, & \rho_{\mathrm{f}}(T)&=\frac{7}{8}g\frac{\pi^{2}}{30}T^{4}\,,\\
	s_{\mathrm{b}}(T)&=g\frac{2\pi^2}{45}T^{3}\,, & s_{\mathrm{f}}(T)&=\frac{7}{8}g\frac{2\pi^2}{45}T^{3}\,,
\end{align} 
where $g$ are the internal degrees of freedom of the particle. To
obtain the complete energy and entropy density, respectively, one simply has to sum
over all particles in thermal equilibrium, 
\begin{align}
	\rho(T) &= \frac{\pi^2}{30}\left(\sum_{i \in \text{\{boson\}}} g^{(\rho)}_i + \frac{7}{8}\sum_{i \in \text{\{fermion\}}} g^{(\rho)}_i\right)T^4 = \frac{\pi^2}{30} g^{(\rho)}(T)  T^4 \,,\label{eq:energy_gstar}\\
	s(T) &= \frac{2\pi^2}{45}\left(\sum_{i \in \text{\{boson\}}} g_i^{(s)} + \frac{7}{8}\sum_{i \in \text{\{fermion\}}} g_i^{(s)}\right)T^3 = \frac{2\pi^2}{45} g^{(s)}(T) T^3\,.\label{eq:entropy_gstar}
\end{align}

\begin{figure}[ht!]
\centering
\includegraphics[width=12cm]{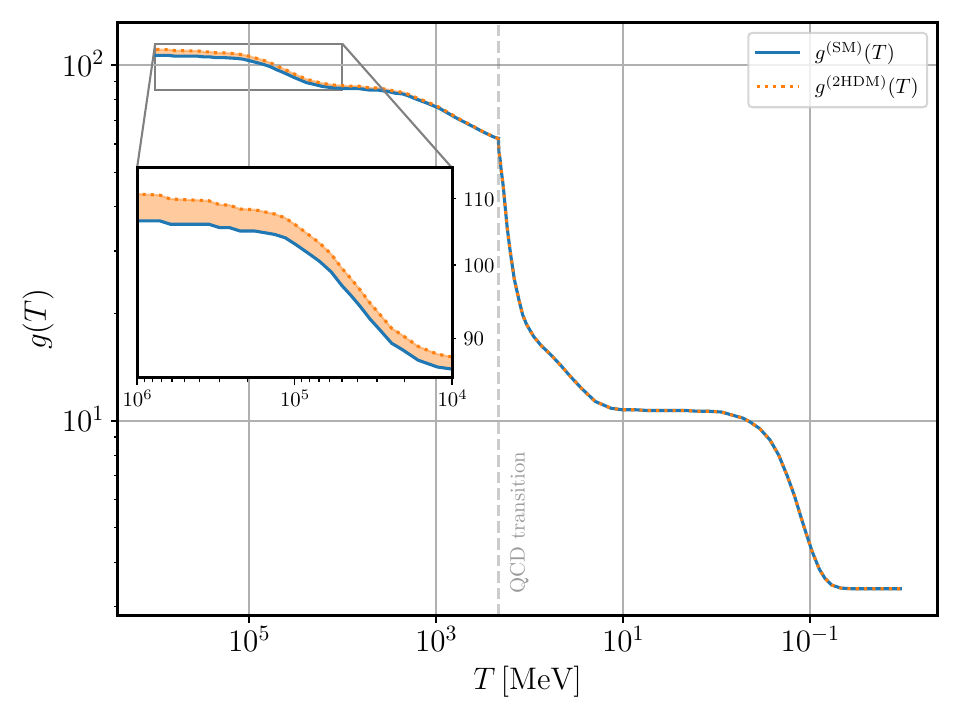}
    \caption{Approximation used by \texttt{BSMPTv3} to estimate the
  effective degrees of freedom $g(T)$ for a generic
  model, here the 2HDM. The 
  blue line represents the SM effective degrees of freedom
  $g^\text{(SM)}$ (with $g^\text{(SM)}(\infty) = 106.75$), the
  orange dotted line represents the 2HDM
  effective degrees of freedom $g^\text{(2HDM)}(T)$ (with
  $g^\text{(2HDM)}(\infty) = 110.75$), and the orange shaded region
  visualizes the change due to the additional degrees of freedom that the 2HDM has compared to
  the SM. The gray dashed vertical line at $T =
  214\,\text{MeV}$ is the temperature of the QCD
  phase transition where we assume
  that all additional particles have decoupled and we fall back to
    $g^\text{(SM)}$.}
\label{fig:gstar}
\end{figure}
As the Universe cools down and expands, particles gain mass which
decreases their contribution to the energy/entropy density of the
cosmic fluid. Additionally, their interaction rate with the fluid
decreases until the particle decouples from the cosmic fluid, which
further decreases the energy/entropy density of the fluid. For this
reason, it is customary to promote the $g_i^{(\rho/s)}$ in
Eq.~\eqref{eq:energy_gstar} and Eq.~\eqref{eq:entropy_gstar} to
temperature dependent quantities and call the sum the effective
degrees of freedom for the energy density and entropy density,
respectively, i.e.~$g^{(\rho)}(T)$ and $g^{(s)}(T)$. \s

For the scope of this paper, we make the approximation that the
effective number of degrees of freedom for energy and entropy is the
same, i.e. $g^{(\rho)}(T) \approx g^{(s)}(T)$, which we then denote as
$g(T)$. This should be a good
  approximation until $T =
100\,\text{keV}$~\cite{Kierkla:2022odc}. In the end, we have that
\begin{align}
	g(T) = \sum_{i \in \text{\{boson\}}} g^{(\rho/s)}_i(T) + \frac{7}{8}\sum_{i \in \text{\{fermion\}}} g^{(\rho/s)}_i(T)\,,
\end{align}
where we assumed that the sum is the same for the energy and entropy degrees of freedom. \s

The effective degrees of freedom $g(T)$ have been calculated for the
SM in detail, but for a generic model, the temperature dependence is
non-trivial to compute. At high temperature, when we assume that all
particles are thermalised and relativistic, we can calculate
$g(T\to\infty)\equiv g(\infty)$, and it is given by the sums in
Eq.~\eqref{eq:energy_gstar} and
\eqref{eq:entropy_gstar}. \s

To incorporate this in our code, $g(T)$
  is calculated using an approximation by interpolating the SM
  effective degrees of freedom $g^\text{(SM)}(T)$ and the correct
  value for a generic model in the relativistic limit
  $g(\infty)$. In our approximation, we assume that all additional
  fields decouple before the QCD transition at $T_\text{QCD} =
  214\,\text{MeV}$, and we make a smooth interpolation between the QCD
  transition and $T_\text{h} = 1\,\text{TeV}$, the temperature at which we assume all particles are relativistic and thermalised. Explicitly, the approximation is given by 
\begin{align}
	g(T) =   \begin{cases}
	g(\infty), & \text{for } T \geq T_\text{h} \\
        ({T}/{T_{\text{QCD}}})^{\log(g(\infty)/g^\text{(SM)}(T_\text{h}))/\log(T_\text{h}/T_{\text{QCD}})}g^\text{(SM)}(T), & \text{for } T_\text{h} > T > T_{\text{QCD}} \\
    g^\text{(SM)}(T) , & \text{for } T_{\text{QCD}} \geq T\,.
  \end{cases}
\end{align}
To show the effect of the additional degrees of
freedom of the 2HDM, which has four more effective degrees of freedom
at high temperature (one for each spin-zero boson), in comparison to
the SM, we plot both $g^\text{(SM/2HDM)}$ in Fig.~\ref{fig:gstar}
where the shaded area represents the change due to the additional
degrees of freedom of the additional fields.


\vspace*{0.5cm}
\printbibliography 

\end{document}